\documentclass[11pt,nohyper,twosided]{JHEPmod}
\usepackage{epsfig}
\usepackage{amsmath}
\usepackage{multirow}
\usepackage{color}
\usepackage{array}
\let\normalcolor\relax
\usepackage{cite}

\allowdisplaybreaks[2]

\title{\mbox{\hspace{-0.35cm}{\LARGE Tree-level Amplitudes in the Nonlinear Sigma Model}}}
\author{Karol Kampf$^{a}$, Ji\v{r}\'{\i} Novotn\'{y}$^{a}$ and Jaroslav
Trnka$^{a,b}$\\
{\it $^{a}$ Institute of Particle and Nuclear Physics, Faculty of
Mathematics and Physics,  Charles University in Prague, CZ-18000 Prague, Czech
Republic}\\
{\it $^{b}$ Department of Physics, Princeton University, Princeton,
New Jersey 08544, USA}}

\abstract{We study in detail the general structure and further properties of
the tree-level amplitudes in the $SU(N)$ nonlinear sigma model. We construct the
flavor-ordered Feynman rules for various parameterizations of the $SU(N)$ fields
$U(x)$,
write down the Berends-Giele relations for the semi-on-shell currents and discuss their efficiency for the amplitude calculation in comparison with those of renormalizable theories. We also present an explicit form of the partial amplitudes up to ten external particles.
It is well known that  the standard BCFW recursive relations cannot be used for reconstruction of the the on-shell amplitudes of effective theories like the $SU(N)$ nonlinear sigma model because of the inappropriate behavior of the deformed on-shell amplitudes at infinity.
We discuss possible generalization of the BCFW approach introducing ``BCFW formula with subtractions'' and with help of Berends-Giele relations we prove particular scaling
properties of the semi-on-shell amplitudes of the $SU(N)$ nonlinear sigma model under specific shifts of the external momenta.
These results allow us to define alternative deformation of the semi-on-shell amplitudes and derive
BCFW-like recursion relations. These provide a systematic and
effective tool for calculation of Goldstone bosons scattering
amplitudes and it also shows the possible applicability of on-shell
methods to effective field theories.
We also use these BCFW-like relations for the investigation of the Adler zeroes and double soft limit of the semi-on-shell amplitudes.}

\preprint{PUPT-2443}

\begin{document}

\newpage

\section{Introduction}

The chiral nonlinear sigma model is a widely used tool for description of many
phenomena in theoretical particle physics. It is based on a simple Lie Group
$G$ and the spontaneous symmetry breaking $G\times G\rightarrow G$ gives
rise to massless excitations - Goldstone bosons. For instance, in the theory
of strong interactions, the group $G$ is $SU(N_f)$ where $N_f=2,3$ is a
number of light quark flavors and Goldstone bosons are associated with the
triplet of pions (for $N_f=2$) or octet of pseudoscalar mesons $\pi $, $K$
and $\eta $ (for $N_f=3$). The interactions of these degrees of freedom
dominate the hadronic world at low energies. In this context, the leading
order nonlinear $U(3)\times U(3)$ chiral invariant effective Lagrangian,
the kinetic part of which corresponds to the chiral nonlinear $U(3)$ sigma
model, was constructed in the late sixties by Cronin \cite{Cronin:1967jq}
while the $SU(2)$ case was studied by Weinberg \cite{Weinberg:1966fm, Weinberg:1968de}, Brown \cite{Brown:1967qh} and Chang and G%
\"{u}rsey \cite{Chang:1967zza}.
Further generalization lead to the invention of Chiral Perturbation Theory
as a low energy effective theory of Quantum Chromodynamics by Weinberg \cite%
{Weinberg:1978kz} and by Gasser and Leutwyler \cite{Gasser:1983yg}, \cite%
{Gasser:1984gg}. Chiral Perturbation Theory became a very useful tool for
the investigation of the low energy hadron physics.

The focus of this paper is on scattering amplitudes of Goldstone bosons
within the $SU(N)$ nonlinear sigma model described by the leading order
Lagrangian. In principle, the standard Feynman diagram approach allows us to
calculate arbitrary amplitude. Because the model is effective, and the
Lagrangian contains an infinite tower of terms the calculation becomes very
complicated for amplitudes of many external Goldstone bosons even at
tree-level. It would be therefore desirable to find alternative
non-diagrammatic methods which could save the computational effort and
provide us with a tool to get the amplitudes more efficiently. In the past
an attempt to formulate the calculation of the tree-level without any
reference to the Lagrangian was made by Susskind and Frye \cite%
{Susskind:1970gf}. They postulated recursive procedure for pion amplitudes based on certain
algebraic duality assumptions supplemented with the requirement of Adler zero
condition which should have to  be satisfied separately for group-factor free kinematical functions recently known as the partial or stripped amplitudes. Such a condition had been proven in the special case of
pion amplitudes described by the $SU(2)$ nonlinear sigma model by Osborn
\cite{Osborn:1969ku}. In \cite{Susskind:1970gf} the authors successively
calculated the amplitudes up to eight pions and showed that these results  are equivalent  to the diagrammatic calculation
based on the $SU(2)$ nonlinear sigma model. The full equivalence for all amplitudes has been
proven by Ellis and Renner in \cite{Ellis:1970nt}.

Over the past two decades there has been a huge progress in
understanding scattering amplitudes using \textit{on-shell methods}
(for a  review see e.g. \cite{Mangano:1990by, Dixon:1996wi,
 Feng:2011np, Drummond:2010ep}). They do not use explicitly the
Lagrangian description of the theory and all on-shell quantities are
calculated using on-shell data only with no access to off-shell
physics (unlike virtual particles in Feynman diagrams). This has
lead to many new theoretical tools (e.g. unitary methods
\cite{Bern:1994zx, Bern:1994cg}, BCFW recursion relations for
tree-level amplitudes \cite{Britto:2004ap, Britto:2005fq} and the
loop integrand \cite{ArkaniHamed:2010kv}) as well as practical
applications of on-shell methods to LHC processes (for recent
results of the next-to-leading order QCD corrections for $W+4$-jets
see \cite{Berger:2010zx}). Most of the recent theoretical
developments have been driven by an intensive exploration of ${\cal
N}=4$ super Yang-Mills theory in the planar limit both at weak and
strong couplings (see e.g. \cite{Witten:2003nn,Bern:2005iz
,Alday:2007hr,Alday:2010vh,Drummond:2008vq,Drummond:2007cf,Drummond:2009fd,Bern:2008qj,ArkaniHamed:2009dn,Mason:2009qx,Mason:2010yk,ArkaniHamed:2012nw}).

There have been several attempts to extend some of these methods to other
theories. The most natural starting point are the recursion relations for
on-shell tree-level amplitudes, originally found by Britto, Cachazo, Feng
and Witten for Yang-Mills theory \cite{Britto:2004ap}, \cite{Britto:2005fq}
and later also for gravity \cite{Bedford:2005yy}, \cite{Cachazo:2005ca} .
The main idea is to perform a complex shift on external momenta and
reconstruct the amplitude recursively using analytic properties of the
S-matrix. More recently, this recursive approach was extended to Yang-Mills
and gravity theories coupled to matter, as well as more general class of
renormalizable theories \cite{Cheung:2008dn}. 

In this paper, we find the new
recursion relations for all on-shell tree-level amplitudes of Goldstone
bosons within $SU(N)$ nonlinear sigma model. This shows that on-shell
methods can be applied also for effective field theories and it gives new
computational tool in this model. Using these recursion relations we are
also able to prove more properties of tree-level amplitudes that are
invisible in the Feynman diagram approach.

The paper is organized as follows: In section 2 we discuss $SU(N)$
nonlinear sigma model, introduce \emph{stripped amplitudes} and using \emph{%
minimal} parametrization (the convenient properties of which has been
discussed in \cite{Ellis:1970nt}) we calculate tree-level amplitudes up to
10 points. In section 3 we review BCFW recursion relations and their
generalization to theories that do not vanish at infinity at large momentum
shift. Section 4 is the main part of the paper, we first introduce
semi-on-shell amplitudes, ie. amplitudes with $n-1$ on-shell and one off-shell external legs. Then
we prove scaling properties under particular momentum shifts which allows us
to construct BCFW-like recursion relations. Finally, we show explicit 6pt
example. In section 5 we use previous results to prove Adler zeroes and
double-soft limit formula for stripped amplitudes. Additional results and technical details are
postponed to appendices: In Appendix \ref{parametrization_appendix}, we
describe the general parametrization of the $SU(N)$ nonlinear sigma model.
In Appendix \ref{explicit_amplitudes_appendix} we give the results of the
amplitudes up to 10p. Appendix \ref{graph_numbers} is devoted to the
counting of flavor-ordered Feynman graphs needed for the calculations of the
amplitudes in nonlinear sigma models and other theories. In Appendix \ref%
{extra_scaling} we present additional scaling properties of the
semi-on-shell amplitudes. In Appendix \ref{double_soft_appendix}, we study
the double soft-limit for more general class of spontaneously broken
theories for complete (not stripped) amplitudes.

\section{Nonlinear sigma model}

\subsection{Leading order Lagrangian}

Let us first assume a most general case of the principal chiral nonlinear
sigma model based on a simple compact Lie group $G$. Such a model
corresponds to the spontaneous symmetry breaking of the chiral group $%
G_{L}\times G_{R}\ $where $G_{L,R}=G$ to its diagonal subgroup $G_{V}=G$,
i.e. to the subgroup of the elements $h=(g_{L},g_{R})$ where $g_{L}=g_{R}$.
The vacuum little group $G_{V}$ is invariant with respect to the involutive
automorphism $(g_{L},g_{R})\rightarrow (g_{R},g_{L})$ and the homogeneous
space $G_{L}\times G_{R}/G_{V}$ is a symmetric space which is isomorphic to
the group space $G$. A canonical realization of such an isomorphism is via
restriction of the mapping%
\begin{equation}
(g_{L},g_{R})\rightarrow g_{R}g_{L}^{-1}\equiv U
\end{equation}%
(which is constant on the right cosets of $G_{V}$ in $G_{L}\times G_{R}$) to
$G_{L}\times G_{R}/G_{V}$. Provided we induce the action of the chiral group
on $G_{L}\times G_{R}/G_{V}$ by means of the left multiplication, the
transformation $\ $\ of $U$ under general element $(V_{L},V_{R})$ of the
chiral group is linear%
\begin{equation}
U\rightarrow V_{R}UV_{L}^{-1}.  \label{V_RUV_L}
\end{equation}%
This can be used to construct the most general chiral invariant leading
order effective Lagrangian in general number $d$ of space-time dimensions
describing the dynamics of the Goldstone bosons corresponding to the
spontaneous symmetry breaking $G_{L}\times G_{R}\rightarrow G_{V}$ as%
\begin{equation}
\mathcal{L}^{(2)}=\frac{F^{2}}{4}\langle \partial _{\mu }U\partial ^{\mu
}U^{-1}\rangle =-\frac{F^{2}}{4}\langle (U^{-1}\partial _{\mu
}U)(U^{-1}\partial ^{\mu }U)\rangle ,  \label{L2}
\end{equation}%
where $F$ is a constant\footnote{%
The decay constant of the Goldstone bosons.} with the canonical dimension $%
d/2-1$. Here and in what follows we use the notation $\langle \cdot \rangle =%
\mathrm{Tr}(\cdot )$ and the trace is taken in the defining representation
of $G$. The overall normalization factor is dictated by the form of the
parametrization of the matrix $U$ in terms of the Goldstone boson fields $%
\phi ^{a}$ which we write for the purposes of this subsection\footnote{%
In what follows we will use also more general parametrization of $U$.} as
\begin{equation}
U=\exp \left( \sqrt{2}\frac{\mathrm{i}}{F}\phi \right)  \label{Uexp}
\end{equation}%
where $\phi =\phi ^{a}t^{a}$ and $t^{a}$, $a=1,\ldots ,\dim G$ are
generators of $G$ satisfying%
\begin{eqnarray}
\langle t^{a}t^{b}\rangle &=&\delta ^{ab}  \label{group_relation_1} \\
\lbrack t^{a},t^{b}] &=&\mathrm{i}\sqrt{2}f^{abc}t^{c}.
\label{group_relation_2}
\end{eqnarray}%
Here $f^{abc}$ are totally antisymmetric structure constants of the group $G$%
. According to (\ref{V_RUV_L}), the fields $\phi ^{a}$ transform linearly
under the little group $G_{V}$ as the vector in the adjoint representation
of $G$ while the general chiral transformations of $\phi ^{a}$ are
nonlinear.

The Lagrangian $\mathcal{L}^{(2)}$ can be rewritten in terms of the
Goldstone boson fields as follows. We have%
\begin{equation}
U^{-1}\partial _{\mu }U=-\frac{\exp \left( -\sqrt{2}\frac{\mathrm{i}}{F}%
\mathrm{Ad}(\phi )\right) -1}{\mathrm{Ad}(\phi )}\partial _{\mu }\phi =-%
\frac{1}{\sqrt{2}}t\cdot \frac{\exp \left( -\frac{2\mathrm{i}}{F}D_{\phi
}\right) -1}{D_{\phi }}\cdot \partial \phi
\end{equation}%
where
\begin{equation}
\mathrm{Ad}(\phi )\partial _{\mu }\phi =[\phi ,\partial _{\mu }\phi ]=\sqrt{2%
}t^{a}D_{\phi }^{ab}\partial _{\mu }\phi ^{b}\equiv \sqrt{2}t\cdot D_{\phi
}\cdot \partial \phi ,
\end{equation}%
the matrix $D_{\phi }^{ab}$ is given as
\begin{equation}
D_{\phi }^{ab}=-\mathrm{i}f^{cab}\phi ^{c}
\end{equation}%
and the dot means contraction of the indices in the adjoint representation.
Inserting this in (\ref{L2}) we get finally%
\begin{equation}
\mathcal{L}^{(2)}=\frac{F^{2}}{4}\partial \phi ^{T}\cdot \frac{1-\cos \left(
\frac{2}{F}D_{\phi }\right) }{D_{\phi }^{2}}\cdot \partial \phi =-\partial
\phi ^{T}\cdot \left( \sum_{n=1}^{\infty }\frac{(-1)^{n}}{(2n)!}\left( \frac{%
2}{F}\right) ^{2n-2}D_{\phi }^{2n-2}\right) \cdot \partial \phi .
\label{L2adjoint}
\end{equation}

\subsection{General properties of the tree-level scattering amplitudes}

Note that, the only group factors which enter the interaction vertices are
the structure constants $f^{abc}$. In any tree Feynman diagram each $f^{abc}$
is contracted either with another structure constant within the same vertex
or via propagator factor $\delta ^{ab}$ with some structure constant
entering next vertex. Therefore, using the standard argumentation for a
general tree graph \cite{Mangano:1990by}, i.e. expressing any $f^{abc}$ as a
trace $f^{abc}=-\langle \mathrm{i}[t^{a},t^{b}]t^{c}\rangle /\sqrt{2}$ and
then successively using the relations like $f^{cde}t^{c}=-\mathrm{i}%
[t^{d},t^{e}]/$ $\sqrt{2}$ in order to replace the contracted structure
constants with the commutators of the generators inside the single trace, we
can prove that any tree level on-shell amplitude has a simple group
structure, namely%
\begin{equation}
\mathcal{M}^{a_{1}a_{2}\ldots a_{n}}(p_{1},p_{2},\ldots ,p_{n})=\sum_{\sigma
\in S_{n}/Z_{n}}\langle t^{a_{\sigma (1)}}t^{a_{\sigma (2)}}\ldots
t^{a_{\sigma (n)}}\rangle \mathcal{M}_{\sigma }(p_{1},\ldots ,p_{n}).
\label{M_stripped}
\end{equation}%
Here all the momenta treated as incoming \ and the sum is taken over the
permutation of the $n$ indices $1,2,\ldots ,n$ modulo cyclic permutations.
As a consequence of the cyclicity of the trace we get%
\begin{equation}
\mathcal{M}_{\sigma }(p_{1},p_{2}\ldots ,p_{n})=\mathcal{M}_{\sigma
}(p_{2},\ldots ,p_{n},p_{1})
\end{equation}%
Due to the Bose symmetry, the kinematical factors $\mathcal{M}_{\sigma
}(p_{1},,\ldots ,p_{n})$ has to satisfy%
\begin{equation}
\mathcal{M}_{\sigma\circ \rho }(p_{1},,\ldots ,p_{n})=\mathcal{M}_{\sigma
}(p_{\rho (1)},p_{\rho (2)},\ldots ,p_{\rho (n)})
\end{equation}%
(where $\sigma\circ \rho $ is a composition of permutations) and therefore%
\begin{equation}
\mathcal{M}_{\sigma }(p_{1},\ldots ,p_{n})=\mathcal{M}(p_{\sigma
(1)},p_{\sigma (2)},\ldots ,p_{\sigma (n)})
\end{equation}%
where we have denoted $\mathcal{M\equiv M}_{\mathrm{id}}$ (here ${\mathrm{id}}$ is identical permutation). The amplitudes $%
\mathcal{M}(p_{1},\ldots ,p_{n})$ are called the \emph{stripped} or \emph{%
partial} amplitudes. Note that the same arguments can be used also for the
Feynman rules for the interaction vertices, the general form of which can be
written as%
\begin{equation}
V_{n}^{^{a_{1}a_{2}\ldots a_{n}}}(p_{1},p_{2},\ldots ,p_{n})=\sum_{\sigma
\in S_{n}/Z_{n}}\langle t^{a_{\sigma (1)}}t^{a_{\sigma (2)}}\ldots
t^{a_{\sigma (n)}}\rangle V_{n}(p_{\sigma (1)},p_{\sigma (2)},\ldots
,p_{\sigma (n)}).  \label{complete_vertex}
\end{equation}%
After some algebra we get explicitly (see Appendix \ref%
{parametrization_appendix} for details) $V_{2n+1}(p_{1,}\ldots ,p_{2n+1})=0$
and
\begin{equation}
V_{2n}(p_{1,}\ldots ,p_{2n})=\frac{(-1)^{n}}{(2n)!}\left( \frac{2}{F^{2}}%
\right) ^{n-1}\sum_{k=1}^{2n-1}(-1)^{k-1}\left(
\begin{array}{c}
2n-2 \\
k-1%
\end{array}%
\right) \sum_{i=1}^{2n}(p_{i}\cdot p_{i+k}).
\end{equation}%
Let us note that besides (\ref{L2}), (\ref{Uexp}) we need not to use any
algebraic relations specific for the concrete group $G$ when deriving this
formula and it is therefore valid for general $G$.  In the general case we
can therefore define the stripped amplitudes and stripped vertices, however,
their relation is not straightforward and may depend on the group $G$. In what
follows we will concentrate on the case $G=SU(N)$.

\subsection{Tree-level amplitudes for $G=SU(N)$}

\subsubsection{Flavor ordered Feynman rules}

The standard way of calculation of the tree-level amplitudes $\mathcal{M}%
^{a_{1}\ldots a_{n}}(p_{1},\ldots ,p_{n})$ is to evaluate the contributions
of all tree Feynman graphs with $n$ external legs build form the complete
vertices (\ref{complete_vertex}) and propagators $\Delta _{ab}=\mathrm{i}%
\delta _{ab}/p^{2}$. \ This includes rather tedious group algebra which is
specific for each group $G.$\ In the special case of $G=SU(N)$ the
calculations can be further simplified. Because we have the completeness
relations for the generators $t^{a}$ in the form%
\begin{equation}
\sum_{a=1}^{N^{2}-1}\langle Xt^{a}\rangle \langle t^{a}Y\rangle =\langle
XY\rangle -\frac{1}{N}\langle X\rangle \langle Y\rangle ,  \label{complet}
\end{equation}%
we can simply merge the traces from the vertices of any tree Feynman graphs
in one single trace preserving at the same time the order of the generators $%
t^{a_{j}}$ inside the trace. Note that the \textquotedblleft
disconencted'' $1/N$ terms have to cancel in the sum in
order to produce the single trace in (\ref{M_stripped})\footnote{%
As we shall see in what follows, this fact can be understood as a
consequence of the decoupling of the $U(1)$ Goldstone boson in the
nonlinear $U(N)$ sigma model.}. This enables us to formulate simple
\textquotedblleft flavor ordered Feynman rules'' directly
for the stripped amplitudes $\mathcal{M}$ completely in terms of the
stripped vertices $V_{n}$. The general recipe is exactly the same as in the
more familiar case of $SU(N)$ Yang-Mills theory, i.e. the tree graphs built
form the stripped vertices and propagators are decorated with cyclically
ordered external momenta and the corresponding ordering of the momenta
inside the stripped vertices are kept.

Let us note that such a simple way of the calculation of the stripped
amplitudes might not be possible for general group $G$. For instance for $G=SO(N)$
we have the following completeness relations%
\begin{equation}
\sum_{a=1}^{N(N-1)/2}\langle Xt^{a}\rangle \langle t^{a}Y\rangle =\frac{1}{2}%
\left(\langle XY\rangle -\langle XY^{T}\rangle\right)
\end{equation}%
the second term of which reverses the order of the generators in the merged
vertex and the aforementioned simple argumentation leading to the flavor ordered Feynman rules  has to be
modified.

The $SU(N)$ case has also another useful feature. As a consequence of the
completeness relations (\ref{complet}) for the group generators of $SU(N)$
and the analogous relation%
\begin{equation}
\sum_{a=1}^{N^{2}-1}\langle Xt^{a}Yt^{a}\rangle =\langle X\rangle \langle
Y\rangle -\frac{1}{N}\langle XY\rangle
\end{equation}%
it can be proved \cite{Mangano:1990by} that the traces $\langle t^{a_{\sigma
(1)}}t^{a_{\sigma (2)}}\ldots t^{a_{\sigma (n)}}\rangle $ and $\langle
t^{a_{\rho (1)}}t^{a_{\rho (2)}}\ldots t^{a_{\rho (n)}}\rangle $ are
orthogonal in the leading order of $N$ in the sense that%
\begin{equation}
\sum_{a_{1},a_{2},\ldots ,a_{n}}\langle t^{a_{\sigma (1)}}t^{a_{\sigma
(2)}}\ldots t^{a_{\sigma (n)}}\rangle \langle t^{a_{\rho (1)}}t^{a_{\rho
(2)}}\ldots t^{a_{\rho (n)}}\rangle ^{\ast }=N^{n-2}(N^{2}-1)\left( \delta
_{\sigma \rho }+O\left( \frac{1}{N^{2}}\right) \right)
\label{N_orthogonality}
\end{equation}%
where $\delta _{\sigma \rho }=1$ for $\rho =\sigma $ modulo cyclic
permutation and zero otherwise. This relation is enough to uniquely
determine the coefficients $\mathcal{T}_{\sigma }$ in the general expansion
of the form%
\begin{equation}
\mathcal{T}^{a_{1}a_{2}\ldots a_{n}}=\sum_{\sigma \in S_{n}/Z_{n}}\langle
t^{a_{\sigma (1)}}t^{a_{\sigma (2)}}\ldots t^{a_{\sigma (n)}}\rangle
\mathcal{T}_{\sigma },
\end{equation}%
(provided the coefficients $\mathcal{T}_{\sigma }$ are $N-$independent) as
the leading in $N$ terms of the \textquotedblleft scalar
product''
\begin{equation}
\sum_{a_{1},a_{2},\ldots ,a_{n}}\mathcal{T}^{a_{1}a_{2}\ldots a_{n}}\langle
t^{a_{\sigma (1)}}t^{a_{\sigma (2)}}\ldots t^{a_{\sigma (n)}}\rangle ^{\ast
}=N^{n-2}(N^{2}-1)\left( \mathcal{T}_{\sigma }+O\left( \frac{1}{N^2}\right)
\right)
\end{equation}%
Because the stripped amplitudes and vertices by construction do not depend
on $N$, the coefficients at the individual traces in the representation (\ref%
{M_stripped}) are unique a therefore the stripped amplitudes and vertices
are unique.

\subsubsection{Dependence on the parametrization}

Up to now we have identified the Goldstone boson fields $\phi ^{a}$ using
the exponential parametrization (\ref{Uexp}) of the group elements $U(\phi
^{a})$. However, according the equivalence theorem, the amplitudes $\mathcal{%
M}^{a_{1}a_{2}\ldots a_{n}}(p_{1},p_{2},\ldots ,p_{n})$ are the same for any
other parametrization $U(\widetilde{\phi }^{a})$ where
\begin{equation}
\widetilde{\phi }^{a}=\phi ^{a}+F^{a}(\phi )
\end{equation}%
where $F^{a}(\phi )=O(\phi ^{2})$ is at least quadratic in the fields $\phi $%
. Therefore, according to the aforementioned uniqueness, the stripped
amplitudes for the nonlinear $SU(N)$ sigma model do not depend on the
parametrization. Note, however, that this is not true for the stripped
vertices which do depend on the parametrization because the complete
vertices $V_{n}^{^{a_{1}a_{2}\ldots a_{n}}}(p_{1},p_{2},\ldots ,p_{n})$ do.

As far as the on-shell tree-level amplitudes are concerned, in various
calculations we are thus free to use the most suitable parametrization and
consequently the most useful form of the corresponding stripped vertices for
a given purpose. We shall often take advantage of this freedom in what
follows.

A wide class of parameterizations for the chiral nonlinear sigma model with
$G=U(N)$ and $G=SU(N)$ has been discussed in \cite{Cronin:1967jq}. The
general form of such a parameterizations reads
\begin{equation}
U=\sum_{k=0}^{\infty }a_{k}\left( \sqrt{2}\frac{\mathrm{i}}{F}\phi \right)
^{k}  \label{Ugenpar}
\end{equation}%
where $a_0=a_1=1$ and the remaining real coefficients $a_k$ are constrained
by the requirement $U U^+=1 $. The exponential parametrization (\ref{Uexp})
corresponds to the choice $a_{n}=1/n!$.  In fact, as was proved in \cite%
{Cronin:1967jq}, for $SU(N)$ nonlinear sigma model with $N>2$, the exponential parametrization is the only admissible choice
within the above class of parameterizations (\ref{Ugenpar}) compatible with
the nonlinearly realized symmetry with respect to the $SU(N)$ chiral
transformations (\ref{V_RUV_L}). On the other hand, for $SU(2)$ and for the extended chiral
group $G=U(N)$ with arbitrary $N$, the parameterizations of the form (\ref{Ugenpar}) represent
an infinite-parametric class. The more detailed discussion can be found in
Appendix \ref{parametrization_appendix}.

\subsubsection{Interrelation of the cases $G=U(N)$ and $G=SU(N)$}

Let us note, that the $SU(N)$ and $U(N)$ chiral nonlinear sigma models are
tightly related. Within the exponential parametrization we can write in the $%
U(N)$ case%
\begin{equation}
U=\exp \left( \frac{\mathrm{i}}{F}\sqrt{\frac{2}{N}}{\phi }^{0}\right)
\widehat{U}
\end{equation}%
where $\widehat{U}\in SU(N)$ and ${\phi }^{0}$ is the additional $U(1)$
Goldstone boson corresponding to the $U(1)$ generator $t^{0}=\mathbf{1}/%
\sqrt{N}$. We get then%
\begin{equation}
U^{-1}\partial _{\mu }U=\frac{\mathrm{i}}{F}\sqrt{\frac{2}{N}}\partial _{\mu
}{\phi }^{0}+\widehat{U}^{+}\partial _{\mu }\widehat{U}
\end{equation}%
and as a consequence,
\begin{equation}
\mathcal{L}^{(2)}=\frac{1}{2}\partial {\phi }^{0}\cdot \partial {\phi }^{0}+%
\frac{F^{2}}{4}\langle \partial _{\mu }\widehat{U}\partial ^{\mu }\widehat{U}%
^{-1}\rangle .
\end{equation}%
Therefore ${\phi }^{0}$ completely decouples. This means that for the
on-shell amplitudes in this model%
\begin{equation}
\mathcal{M}^{a_{1}a_{2}\ldots a_{n}}(p_{1},p_{2},\ldots ,p_{n})=0
\label{decoupleM}
\end{equation}%
whenever at least one $a_{j}=0$. Note that this statement does not depend on
the parametrization. We can therefore reproduce the on-shell amplitudes of
the $SU(N)$ chiral nonlinear sigma model from that of the $U(N)$ one simply
by assigning to the indices $a_{i}$ the values corresponding the $SU(N)$
Goldstone bosons. Keeping this in mind, in what follows we will freely
switch between the $U(N)$ and $SU(N)$ case and use the general
parameterizations (\ref{Ugenpar}) also in the context of the $SU(N)$ chiral
nonlinear sigma model.

The fact that the $U(1)$ Goldstone boson decouples gives also a nice
physical explanation why the ``disconnected`` $1/N$ term can be omitted in
the relation (\ref{complet}) when summing over virtual states in the
tree-level Feynman graphs for the $SU(N)$ nonlinear sigma model. This term
can be interpreted as the subtraction of the extra $U(1)$ virtual state
contained in the first ``connected`` part. However, because this state
decouples, no such correction is in fact needed.

The decoupling of the $U(1)$ Goldstone boson is an effect analogous to the
decoupling of the $U(1)$ component of the gauge field in the case of the $%
U(N)$ Yang-Mills theory. For the tree-level amplitudes (and the
corresponding stripped amplitudes) we get as a consequence a set of
identities constraining their form. For instance taking only one $a_{j}=0$
(say $a_{1}$) in (\ref{decoupleM}), we get the \textquotedblleft dual Ward
identity'' (or the $U(1)$ decoupling identity)
\begin{equation}
\mathcal{M}(p_{1},p_{2},p_{3},\ldots ,p_{n})+\mathcal{M}(p_{2},p_{1},p_{3},%
\ldots ,p_{n})+\ldots +\mathcal{M}(p_{2},p_{3},\ldots ,p_{1},p_{n})=0
\end{equation}%
exactly as in the Yang-Mills case (see e.g. \cite{Mangano:1990by} and
references therein).

\subsection{Explicit examples of $SU(N)$ on-shell amplitudes}

Using (\ref{M_stripped}) we can reconstruct the complete amplitude $\mathcal{%
M}^{a_{1}\ldots a_{n}}(p_{1},\ldots ,p_{n})$ just from a single stripped
amplitude $\mathcal{M}(p_{1},\dots p_{n})$ which is given by the sum of
Feynman diagrams with ordered external legs $\{1,2,\dots n\}$. Though the
aim of this paper is not to calculate scattering amplitudes using the
Feynman diagram approach, in this section we provide explicit examples for
diagrammatic calculation of the stripped 4pt and 6pt amplitudes of the
chiral nonlinear $SU(N)$ sigma model (the 8pt and 10pt amplitudes we
postpone to the Appendix \ref{explicit_amplitudes_appendix}) as the
reference result for the recursive formula given in section 4.

We can easily see that the only poles in the stripped amplitude are of the
form $1/s_{i,j}$ where
\begin{equation}
s_{i,j}=p_{i,j}^{2}\qquad \mbox{with}\qquad p_{i,j}=\sum_{k=i}^{j}p_{k}
\label{s_ij_definition}
\end{equation}%
(Obviously $s_{i,j}=s_{j+1,i-1}$ due to momentum conservation). The
variables $s_{i,j}$ are therefore well suited for presentation of the
amplitudes.

As we have discussed above, the $SU(N)$ stripped amplitudes are essentially
the same as those for the $U(N)$ case and, as we have discussed above, they
are independent on the parametrization of the unitary matrix $U$ in (\ref{L2}%
). The most convenient one for diagrammatic calculation of on-shell
scattering amplitudes is the \emph{minimal} parametrization \cite%
{Ellis:1970nt}%
\begin{equation}
U=\sqrt{2}\frac{\mathrm{i}}{F}\phi +\sqrt{1-2\frac{\phi ^{2}}{F^{2}}}=1+%
\sqrt{2}\frac{\mathrm{i}}{F}\phi -2\sum_{k=1}^{\infty }\left( \frac{1}{2F^{2}%
}\right) ^{k}C_{n-1}\phi ^{2k}
\label{minimal_parametrization_text}
\end{equation}%
where $C_{n}$ are the Catalan numbers (\ref{catalan}). The stripped Feynman
rules for vertices can be written in terms of $s_{i,j}$ as follows (see
Appendix \ref{parametrization_appendix} for details)
\begin{equation}
V_{2n+2}(s_{i,j})=\left( \frac{1}{2F^{2}}\right) ^{n}\frac{1}{2}%
\sum_{k=0}^{n-1}C_{k}C_{n-k-1}\sum_{i=1}^{2n+2}s_{i,i+2k+1}
\end{equation}%
Note that within this parametrization the stripped vertices do not depend on
the off-shellness of the momenta entering the vertex and when expressed in
terms of the variables $s_{i,j}$ they are identical taken both on-shell or
off-shell. This rapidly speeds up the calculation, because there are no
partial cancelations between the numerators and propagator denominators
within the individual Feynman graphs and it allows us to find the final
expressions for the amplitudes in very compact form.

The four-point amplitude is directly given by the Feynman rule in the simple
parametrization,
\begin{equation}
2F^{2}\mathcal{M}(1,2,3,4)=s_{1,2}+s_{2,3}.  \label{4p_tamplitude}
\end{equation}%
Note that for $n$-point amplitude $\sum_{k=1}^{n}p_{k}=0$ and this can be
used to systematically eliminate $p_{n}$ or equivalently $s_{\cdot ,n}$.

\begin{figure}[h]
\begin{center}
\epsfig{width=0.4\textwidth,figure=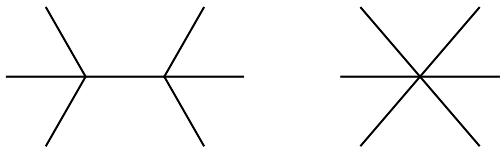}
\end{center}
\caption{Graphical representation of the 6-point amplitude (\protect\ref%
{6pt_amplitude}) with cycling tacitly assumed.}
\label{figure_6pt}
\end{figure}

The six-point amplitude is given by diagrams in Fig.~\ref{figure_6pt}. The
explicit formula reads
\begin{align}
& 4F^{4}\mathcal{M}(1,2,3,4,5,6)=  \notag \\
& =-\frac{(s_{1,2}+s_{2,3})(s_{1,4}+s_{4,5})}{s_{1,3}}-\frac{%
(s_{1,4}+s_{2,5})(s_{2,3}+s_{3,4})}{s_{2,4}}-\frac{%
(s_{1,2}+s_{2,5})(s_{3,4}+s_{4,5})}{s_{3,5}}  \notag \\
& \phantom{=}\,\,+(s_{1,2}+s_{1,4}+s_{2,3}+s_{2,5}+s_{3,4}+s_{4,5})
\label{6pt_amplitude}
\end{align}%
This can be rewritten as
\begin{equation*}
4F^{4}\mathcal{M}(1,2,3,4,5,6)=-\frac{1}{2}\frac{%
(s_{1,2}+s_{2,3})(s_{1,4}+s_{4,5})}{s_{1,3}}+s_{1,2}+\text{cycl}\,,
\end{equation*}%
with `cycl' defined for $n$-point amplitude as
\begin{equation}
A[s_{i,j},\ldots ,s_{m,n}]+\text{cycl}\equiv
\sum_{k=0}^{n-1}A[s_{i+k,j+k},\ldots ,s_{m+k,n+k}]\,,
\end{equation}%
which will quite considerably shorten the 8- and 10-point formulae. These
are postponed to Appendix \ref{explicit_amplitudes_appendix}.

\section{Recursive methods for scattering amplitudes}

Feynman diagrams are completely universal way how to calculate scattering
amplitudes in any theory (that has Lagrangian description). However, it is
well-known that in many cases they are also very ineffective. Despite the
expansion contains many diagrams each of them being a complicated function
of external data, most terms vanish in the sum and the result is
spectacularly simple. The most transparent example is Parke-Taylor formula
\cite{Parke:1986gb} for all tree-level Maximal-Helicity-Violating amplitudes
\footnote{%
Scattering amplitudes of gluons where two of them have negative helicity and
the other ones have positive helicity.}. The simple structure of the result
is totally invisible in the standard Feynman diagrams expansion.

Several alternative approaches and methods have been discovered in last
decades, let us mention e.g. the Berends-Giele recursive relations for the
currents \cite{Berends:1987me} and the more recent BCFW (Britto, Cachazo,
Feng and Witten) recursion relations for on-shell tree-level amplitudes that
reconstruct the result from its poles using simple Cauchy theorem \cite%
{Britto:2004ap}, \cite{Britto:2005fq}.

\subsection{BCFW recursion relations}

For concreteness let us consider tree-level stripped on-shell amplitudes of $%
n$ massless particles in $SU(N)$ Yang-Mills theory (\textquotedblleft
gluodynamics'' ).\footnote{%
The recursion relations can be also formulated for more general cases and
also for massive particles. See \cite{Badger:2005zh} for more details.} The
partial amplitude $\mathcal{M}_{n}$ is a gauge-invariant rational function
of external momenta and additional quantum numbers $h$ (helicities in case
of gluons)
\begin{equation}
\mathcal{M}_{n}\equiv \mathcal{M}_{n}(p_{1},p_{2},\dots
p_{n};h_{1},h_{2},\dots h_{n}).
\end{equation}%
The external momenta are generically complex but if we are interested in
physical amplitudes we can set them to be real in the end. Let us pick two
arbitrary indices $i$, $j$ and perform following shift.
\begin{equation}
p_{i}\rightarrow p_{i}(z)=p_{i}+zq,\qquad p_{j}\rightarrow p_{j}(z)=p_{j}-zq
\label{BCFWshift}
\end{equation}%
such that the momentum $q$ is orthogonal to both $p_{i}$ and $p_{j}$, ie. $%
q^{2}=(q\cdot p_{i})=(q\cdot p_{j})=0$ and the shifted momenta remain
on-shell. Let us note that such $q$ can be found only for the case of
spacetime dimensions $d\geq 4$. The amplitude becomes a meromorphic function
$\mathcal{M}_{n}(z)$ of complex parameter $z$ with only simple poles. The
original expression corresponds to $z=0$. If $\mathcal{M}_{n}(z)$ vanishes
for $z\rightarrow \infty $ we can use the Cauchy theorem to reconstruct $%
\mathcal{M}_{n}=\mathcal{M}_{n}(0)$,
\begin{equation}
0=\frac{1}{2\pi \mathrm{i}}\int_{C(\infty )}\frac{dz}{z}\mathcal{M}_{n}(z)=%
\mathcal{M}_{n}(0)+\sum_{k}\frac{\mathrm{Res}\left( \mathcal{M}%
_{n},z_{k}\right) }{z_{k}}
\end{equation}%
where $C(\infty )$ is closed contour at infinity. $\mathcal{M}_{n}$ can be
then expressed as
\begin{equation}
\mathcal{M}_{n}=-\sum_{k}\frac{\mathrm{Res}\left( \mathcal{M}%
_{n},z_{k}\right) }{z_{k}}  \label{Cauchy}
\end{equation}%
where $k$ is sum of all residues of $\mathcal{M}_{n}(z)$ in the complex $z$%
-plane. Residues of $\mathcal{M}_{n}(z)$ can be straightforwardly calculated
for the following reason: the only poles of $\mathcal{M}_{n}$ are $%
p_{a,b}^{2}=0$ where $p_{a,b}=(p_{a}+p_{a\hspace{0.5pt}\text{{\small +}}%
\hspace{-0.5pt}1}+\dots p_{b})$. The poles of $\mathcal{M}_{n}(z)$ have
still the same locations just shifted, namely $p_{a,b}^{2}(z)=0$ where $i\in
(a,a\hspace{0.5pt}\text{{\small +}}\hspace{-0.5pt}1,\dots b)$ or $j\in (a,a%
\hspace{0.5pt}\text{{\small +}}\hspace{-0.5pt}1,\dots b)$. If none of the
indices $i$, $j$ or both of them are in this range, the dependence on $z$ in
$p_{a,b}(z)$ cancels and it is not pole in $z$ anymore. It is easy to
identify all locations of the corresponding poles $z_{ab}$. Suppose that
particle $i\in (a,a\hspace{0.5pt}\text{{\small +}}\hspace{-0.5pt}1,\dots b)$%
,
\begin{equation}
p_{a,b}^{2}(z)=\left( p_{a}+\dots p_{i\mathrm{\rule[2.4pt]{6pt}{0.65pt}}%
1}+(p_{i}+zq)+p_{i\hspace{0.5pt}\text{{\small +}}\hspace{-0.5pt}1}+\dots
p_{b}\right) ^{2}=0\quad \Rightarrow \quad z_{a,b}=-\frac{p_{a,b}^{2}}{%
2(q\cdot p_{a,b})}
\end{equation}%
In the original amplitude $\mathcal{M}_{n}$ the residue on the pole $%
p_{a,b}^{2}=0$ is given by unitarity: on the factorization channel with
given helicity the amplitude factorizes into two sub-amplitudes, and
therefore
\begin{equation}
\mathrm{Res}\left( \mathcal{M}_{n},z_{a,b}\right) =\sum_{h_{ab}}\mathcal{M}%
_{L}(z_{a,b})^{-h_{ab}}\frac{\mathrm{i}}{2(q\cdot p_{a,,b})}\mathcal{M}%
_{R}^{h_{ab}}(z_{a,b})  \label{residue_schem}
\end{equation}%
where the summation over the helicities $h_{ab\text{ }}$of the one-particle
intermediate state is taken. The \textquotedblleft left''
and \textquotedblleft right'' sub-amplitudes $\mathcal{M}%
_{L,R}^{\pm h_{ab}}(z_{a,b})$ are
\begin{eqnarray}
\mathcal{M}_{L}^{-s_{ab}}(z_{a,b}) &=&\mathcal{M}_{b-a+2}(p_{a},\dots
,p_{i}(z_{a,b}),\dots p_{b},-p_{a,b}(z_{a,b});h_{a},\ldots ,-h_{ab}) \\
\mathcal{M}_{R}^{s_{ab}}(z_{a,b}) &=&\mathcal{M}%
_{n-(b-a)}(p_{a,b}(z_{a,b}),p_{b+1},\ldots ,p_{j}(z_{a,b}),\ldots
,p_{a-1};h_{ab},\ldots ,h_{a-1}).
\end{eqnarray}%
The amplitude $\mathcal{M}_{n}$ can be then written as
\begin{equation}
\mathcal{M}_{n}=\sum_{ab,h_{ab}}\mathcal{M}_{L}^{-h_{ab}}(z_{a,b})\frac{%
\mathrm{i}}{p_{a,b}^{2}}\mathcal{M}_{R}^{h_{ab}}(z_{a,b})
\label{phys_general_BCFW}
\end{equation}%
It is convenient to choose $i$ and $j$ to be adjacent because it eliminates
the number of factorization channels we have to consider.

\subsection{Reconstruction formula with subtractions\label{subtracted_BCFW}}

The BCFW recursion relations discussed above are very generic and applicable
for a large class of theories. The main restriction is the requirement of
large $z$ behavior: $\mathcal{M}_{n}(z)\rightarrow 0$ for $z\rightarrow
\infty $. However, this behavior is not guaranteed in general and there
exist examples when it is broken no matter which pair of momenta $p_{i}$ and
$p_{j}$ is chosen to be shifted. In such a case, an additional term (dubbed
\emph{boundary term}) is present on the right hand side of eq. (\ref%
{phys_general_BCFW}). The boundary term, which is hard to obtain in general
case, has been studied by various methods in the series of papers \cite%
{Feng:2009ei}, \cite{Feng:2010ku} and \cite{Feng:2011twa}, however no
general solution is still available. Sometimes this problem can be cured by
means of considering more general approach when all the external momenta $%
p_{k}$ are deformed (such an \emph{all-line shift} has been introduced in
\cite{Risager:2005vk}, see also \cite{Cohen:2010mi})
\begin{equation}
p_{k}\rightarrow p_{k}(z)=p_{k}+zq_{k}.
\end{equation}%
where $z$ is a complex parameter and $q_{k}$ are appropriate vectors
compatible with the requirements of the momentum conservation and on-shell
constraint for $p_{k}(z)$, ie. $p_{k}\cdot q_{k}=q_{k}^{2}=0$. The on-shell
amplitude
\begin{equation}
\mathcal{M}_{n}(z)\equiv \mathcal{M}_{n}(p_{1}(z),p_{2}(z),\ldots ,p_{n}(z))
\end{equation}%
become again meromorphic function of the variable $z$ the only singularities
of which are simple poles and the residue at these poles have the simple
structure (\ref{residue_schem}) dictated by unitarity. In some cases the
desired behavior $\mathcal{M}_{n}(z)\rightarrow 0$ for $z\rightarrow \infty $
can be achieved in this way. However, in general case the behavior of $%
\mathcal{M}_{n}(z)$ for $z\rightarrow \infty $ is power-like with
non-negative power of $z$. This fact requires some modification of the
reconstruction procedure.

\begin{figure}[t]
\begin{center}
\epsfig{width=0.4\textwidth,figure=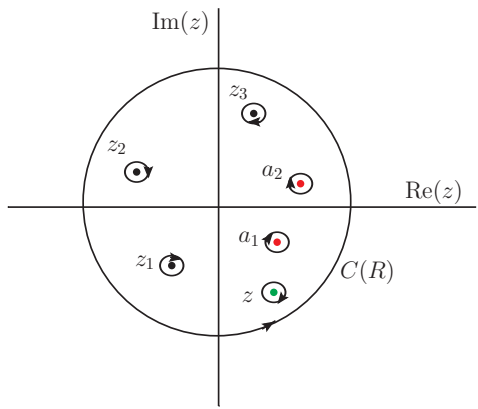}
\end{center}
\caption{Illustration of the contour used for the derivation of the subtracted Cauchy formula (\ref{subtracred_Cauchy}) with $k=1$ and $n_{C(R)}=3$. }
\label{contour_CR}
\end{figure}

This can be done as follows. Let us suppose that we have made any (linear)
deformation of the external momenta $p_{k}\rightarrow p_{k}(z)$ in such a
way that the deformed amplitude $\mathcal{M}_{n}(z)$ is a meromorphic
function the only singularities of which are simple poles and let us assume
the following asymptotic behavior
\begin{equation}
\mathcal{M}_{n}(z)\approx z^{k}  \label{asymptotic}
\end{equation}%
when $z\rightarrow \infty $. Let us denote the poles of $\mathcal{M}_{n}(z)$
as $z_{i}$, $i=1,2,\ldots n$. Assume $a_{j}$, $j=1,2,\ldots ,k+1$ to be
complex numbers satisfying $|a_{j}|<R$ different form the poles $z_{i}$.
Then we can write for $z\neq a_{j}$ inside the disc $D(R)$ (i.e. inside the
domain $|z|<R$  the boundary of which is a circle $C(R)$ of the radius $R$)
the following \textquotedblleft $k+1$ times subtracted Cauchy
formula'' (see Fig.\ref{contour_CR})
\begin{eqnarray}
&&\frac{1}{2\pi \mathrm{i}}\int_{C(R)}\mathrm{d}w\frac{\mathcal{M}_{n}(w)}{%
w-z}\prod_{j=1}^{k+1}\frac{1}{w-a_{j}}  \notag \\
&=&\mathcal{M}_{n}(z)\prod_{j=1}^{k+1}\frac{1}{z-a_{j}}+\sum_{j=1}^{k+1}%
\frac{\mathcal{M}_{n}(a_{j})}{a_{j}-z}\prod_{l=1,l\neq j}^{k+1}\frac{1}{%
a_{j}-a_{l}}+\sum_{i=1}^{n_{C(R)}}\frac{\mathrm{Res}\left( \mathcal{M}%
_{n};z_{i}\right) }{z_{i}-z}\prod_{j=1}^{k+1}\frac{1}{z_{i}-a_{j}}.
\label{subtracred_Cauchy}
\end{eqnarray}%
Here $z_{1},z_{2},\ldots ,z_{n_{C(R)}}$ are the poles inside $D(R)$ and $%
\mathrm{Res}\left( \mathcal{M}_{n};z_{i}\right) $ are corresponding
residues. 
In the limit $R\rightarrow \infty $ the integral 
vanishes due to (\ref{asymptotic}) and $D(\infty )$ will contain all $n$
poles. As a result we get a reconstruction formula with $k+1$ subtractions
\begin{equation}
\mathcal{M}_{n}(z)=\sum_{i=1}^{n}\frac{\mathrm{Res}\left( \mathcal{M}%
_{n};z_{i}\right) }{z-z_{i}}\prod_{j=1}^{k+1}\frac{z-a_{j}}{z_{i}-a_{j}}%
+\sum_{j=1}^{k+1}\mathcal{M}_{n}(a_{j})\prod_{l=1,l\neq j}^{k+1}\frac{z-a_{l}%
}{a_{j}-a_{l}}.  \label{generalized_BCFW}
\end{equation}%
This is the desired generalization of the usual prescription. In order to
reconstruct the amplitude with the asymptotic behavior (\ref{asymptotic})
from its pole structure, we need therefore along with the residues at the
poles $z_{i}$ (which are fixed by unitarity) also supplementary information,
namely the $k+1$ values $\mathcal{M}_{n}(a_{j})$ of the amplitude at the
points $a_{j}$. Such a additional information is the weakest point of the
relations (\ref{generalized_BCFW}): there exists no universal recipe how to
get the values $\mathcal{M}_{n}(a_{j})$ for a general theory. This corresponds
to the well known analogous situation of $k+1$ subtracted dispersion
relations, which allow to reconstruct a general amplitude from its
discontinuities uniquely up to the $k+1$ generally unknown subtraction
constants. Note that, provided we choose $a_{j}$ in such a way that $%
\mathcal{M}_{n}(a_{j})=0$ (i.e. $a_{j}$ are the roots of the deformed
amplitude $\mathcal{M}_{n}(z)$), we can reproduce the formula
\begin{equation}
\mathcal{M}_{n}(z)=\sum_{i=1}^{n}\frac{\mathrm{Res}\left( \mathcal{M}%
_{n};z_{i}\right) }{z-z_{i}}\prod_{j=1}^{k+1}\frac{z-a_{j}}{z_{i}-a_{j}}
\end{equation}%
first written in this context by Benincasa a Conde \cite{Benincasa:2011kn}
and further discussed by Bo Feng, Yin Jia, Hui Luo a Mingxing Luo in \cite%
{Feng:2011jxa}.

\section{BCFW-like relations for semi-on-shell amplitudes}

The straightforward application of the BCFW reconstruction procedure is not
possible for the $SU(N)$ nonlinear sigma model because the amplitudes $%
\mathcal{M}_{n}(z)$ do not have appropriate asymptotic behavior for $%
z\rightarrow \infty $.  The reason is that due to the derivative coupling
of the Goldstone bosons the interaction vertices are quadratic in the
momenta. Therefore after the BCFW shift the vertices along the
\textquotedblleft hard'' $z-$dependent line of the Feynman
graph are in general linear in $z$ and the linear large $z$ behavior of the
propagators cannot compensate for it. \ For instance, under the shift\footnote{Under the all-line (anti)holomorphic BCFW shift the large $z$ behavior is the same. Here we can use the general formulae derived in \cite{Cohen:2010mi}
which relate the number $n$ of external particles, the sum $H$ of their helicities and the overall dimension $c$ of the couplings to the asymptotics of the amplitude under the all-line holomorphic ($O(z^a)$) and anti-holomorphic ($O(z^s)$) shift. These  formulae reads $2s=4-n-c+H$ and $2a=4-n-c-H$. In our case $H=0$ and the only coupling constant is $F^{-1}$, therefore $c=2-n$, therefore in general case $a=s=1$ independently on $n$.}(\ref%
{BCFWshift}) with $i=1$, $j=2$ we get for the 6pt amplitude (\ref%
{6pt_amplitude}) for $z\rightarrow \infty $
\begin{eqnarray}
\mathcal{M}_{6}(z) &=&-2z\left( \frac{\left( q\cdot p_{2,3}\right)
(s_{1,4}+s_{4,5}-s_{1,3})}{s_{1,3}}+\frac{\left( q\cdot p_{2,5}\right)
\left( q\cdot p_{2,3}\right) }{\left( q\cdot p_{2,4}\right) }+\frac{\left(
q\cdot p_{2,5}\right) (s_{3,4}+s_{4,5}-s_{3,5})}{s_{3,5}}\right)  \notag \\
&&+O(z^{0}).
\end{eqnarray}%
and analogously $\mathcal{M}_{n}(z)=O(z)$ for general\footnote{%
The general statement can be derived by induction from Brends-Giele
recursive relations discussed in the next subsection.} $n$. As discussed in
the previous section, in order to reconstruct such an amplitude from its
pole structure, it would be sufficient to know the values of $\mathcal{M}%
_{n}(z)$ for two fixed values of $z$. However, such an information is
difficult to gain solely from the Feynman graph analysis restricted only to
the amplitudes $\mathcal{M}_{n}$. It is therefore useful to take into
account also more flexible objects, namely the semi-on-shell amplitudes,
which unlike the on-shell amplitudes depend on the parametrization of the
matrix $U~$\ and from which the on-shell amplitudes can be straightforwardly
derived. As we would like to show in this section, appropriate choice of
parametrization together with suitable way of BCFW-like deformation of the
semi-on-shell amplitudes allows to substitute for the missing information on
the amplitudes $\mathcal{M}_{n}$ and to construct generalized BCFW-like
relations for them.

\subsection{Semi-on-shell amplitudes and Berends-Giele relations}

The semi-on-shell amplitudes $J_{n}^{a_{1}a_{2}\ldots
a_{n}}(p_{1},p_{2},\ldots ,p_{n})$ (or \emph{currents} in the terminology of
the original paper \cite{Berends:1987me}, where they were introduced for QCD
and more generally for the $SU(N)$ Yang-Mills theory) can be defined in our
case as the matrix elements of the Goldstone boson field $\phi ^{a}(0)$
between vacuum and the $n$ \ Goldstone boson states $|\pi
^{a_{1}}(p_{1})\ldots \pi ^{a_{n}}(p_{n})\rangle $%
\begin{equation}
J_{n}^{a,a_{1}a_{2}\ldots a_{n}}(p_{1},p_{2},\ldots ,p_{n})=\langle 0|\phi
^{a}(0)|\pi ^{a_{1}}(p_{1})\ldots \pi ^{a_{n}}(p_{n})\rangle .
\label{currentJ}
\end{equation}%
Here the momentum $p_{n+1}$ attached to $\phi ^{a}(0)$
\begin{equation}
p_{n+1}=-\sum_{j=1}^{n}p_{j}.
\end{equation}%
is off-shell. Note that $J_{n}^{a,a_{1}a_{2}\ldots a_{n}}(p_{1},p_{2},\ldots
,p_{n})$ has a pole for $p_{n+1}^{2}=0.$

In complete analogy with the on-shell amplitudes, at the tree level the
right hand side of (\ref{currentJ}) can be expressed in terms of the
flavor-stripped semi-on-shell amplitudes $J_{n}(p_{1},p_{2},\ldots ,p_{n})$
in the form
\begin{equation}
\langle 0|\phi ^{a}(0)|\pi ^{a_{1}}(p_{1})\ldots \pi ^{a_{n}}(p_{n})\rangle
|_{\mathrm{tree}}=\sum_{\sigma \in S_{n}}\mathrm{Tr}(t^{a}t^{a_{\sigma
(1)}}\ldots t^{a_{\sigma (n)}})J_{n}(p_{\sigma (1)},p_{\sigma (2)},\ldots
,p_{\sigma (n)}).
\end{equation}%
Let us note that, at higher orders in the loop expansion the group structure
contains also  multiple trace terms. We normalize
the one particle states according to%
\begin{equation}
J_{1}(p)=1.
\end{equation}%
In this section the above semi-on-shell flavor-stripped amplitudes $%
J_{n}(p_{1},p_{2},\ldots ,p_{n})$ will be the main subject of our interest.
The on-shell stripped amplitudes $\mathcal{M}(p_{1},p_{2},\ldots ,p_{n+1})$
can be extracted from them by means of the Lehmann-Symanzik-Zimmermann (LSZ) formulas
\begin{equation}
\mathcal{M}(p_{1},p_{2},\ldots ,p_{n+1})=-\lim_{p_{n+1}^{2}\rightarrow
0}p_{n+1}^{2}J_{n}(p_{1},p_{2},\ldots ,p_{n}).
\end{equation}%
\begin{figure}[t]
\begin{center}
\epsfig{width=0.6\textwidth,figure=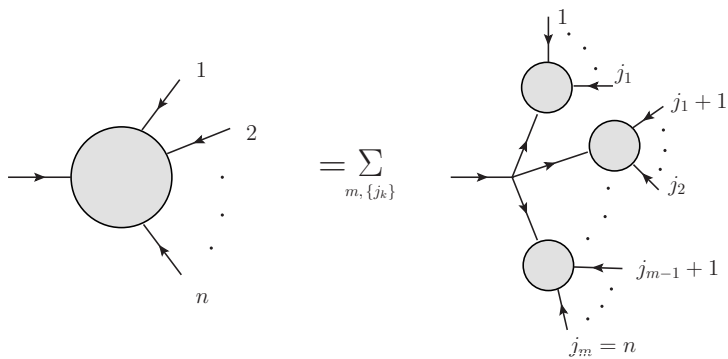}
\end{center}
\caption{Graphical representation of the Berends-Giele recursive relations}
\label{figure_0}
\end{figure}
The main advantage of the semi-on-shell amplitudes $J_{n}(p_{1},p_{2},\ldots
,p_{n})$ (in what follows we also use short-hand notation $J(1,2,\ldots ,n)$%
) is that they allow to abandon the Feynman diagram approach using
appropriate recursive relation. The latter has been first formulated by
Berends and Giele in the context of QCD \cite{Berends:1987me} and proved to
be very efficient for the calculation of the tree-level multi-gluon
amplitudes. For the $U(N)$ nonlinear sigma model the generalized recurrent
relations of Berends-Giele type can be written in the form (see Fig.\ref%
{figure_0})%
\begin{equation}
J(1,2,\ldots ,n)=\frac{\mathrm{i}}{p_{1,n}^{2}}\sum_{m=2}^{n}%
\sum_{\{j_k\}}\mathrm{i}%
V_{m+1}(p_{1,j_{1}},p_{j_{1}+1,j_{2}},\ldots
p_{j_{m-1}+1,n},-p_{1,n})\prod\limits_{k=1}^{m}J(j_{k-1}+1,\ldots ,j_{k})
\label{Berends_Giele_relations}
\end{equation}%
where the sum is over all splittings of the ordered set $\{1,2,\dots ,n\}$
into $m$ non-empty ordered subsets $\{j_{k-1}+1,j_{k-1}+2,\dots ,j_{k}\}$,
(here $j_{0}=0$ and $j_{m}=n$)\footnote{%
Explicitly%
\begin{equation*}
\sum_{\{j_{k}\}}\equiv
\sum_{j_{1}=1}^{n-m+1}\sum_{j_{2}=j_{1}+1}^{n-m+2}\cdots
\sum_{j_{m-1}=j_{m-2}+1}^{n-m+(m-1)}.
\end{equation*}%
}, $V_{m+1}$ is  the flavor-stripped Feynman
rule for vertices with $m+1$ external legs
and $p_{i,k}=\sum_{j=i}^{k}p_{j}$ as above.

Let us note that, because the Lagrangian of the nonlinear sigma model
includes infinite number of vertices with increasing number of fields, the
above Berends-Giele relation for $J_{n}$ have to contain vertices up to $n+1$
legs, i.e. much more terms than in the case of power-counting renormalizable
theories like QCD where the number of vertices is finite\footnote{The number of terms on the right hand side of (\ref{Berends_Giele_relations}) grows exponentially with increasing $n$ in contrast to the polynomial growths typical for the renormalizable theories. See Appendix \ref{graph_numbers} for details.}. This fact rather
reduces the efficiency of these relation for the calculations of the
amplitudes. We illustrate this in the Tab. 1, where the number of terms on
the right hand side of the Berends-Giele relation (\ref%
{Berends_Giele_relations}) written for $J_{2n+1}$ (denoted as $t(2n+1)$) and
the total number of terms necessary for the calculation of the same
semi-on-shell amplitude using the Berends-Giele recursion (denoted as $%
b(2n+1)$) is compared with the total number $f(2n+1)$ of the flavor ordered
Feynman graphs contributing to $J_{2n+1}$ and with the same numbers valid
for the theory with only quadrilinear vertices (\textquotedblleft $\phi ^{4}$
theory '' ) denoted with subscript \textquotedblleft $4$%
'' . See Appendix \ref{graph_numbers} for more details and
for derivation of the explicit formulae for these and other related cases.

\begin{table}[t]
{\small {\
\begin{tabular}{|c|r|r|r|r|r|r|r|r|r|r|}
\hline
$n$ & 2 & 3 & 4 & 5 & 6 & 7 & 8 & 9 & 10 & 11 \\ \hline
$t(2n+1)$ & 4 & 12 & 33 & 88 & 232 & 609 & 1 596 & 4 180 & 10 945 & 28 656
\\
$b(2n+1)$ & 5 & 17 & 50 & 138 & 370 & 979 & 2 575 & 6 755 & 17 700 & 46 356
\\
$f(2n+1)$ & 4 & 21 & 126 & 818 & 5 594 & 39 693 & 289 510 & 2 157 150 & 16
348 960 & 125 642 146 \\
$t_4(2n+1)$ & 3 & 6 & 10 & 15 & 21 & 28 & 36 & 45 & 55 & 66 \\
$b_4(2n+1)$ & 4 & 10 & 20 & 35 & 56 & 84 & 120 & 165 & 220 & 286 \\
$f_{4}(2n+1)$ & 3 & 12 & 55 & 273 & 1 428 & 7 752 & 43 263 & 246 675 & 1 430
715 & 8 414 640 \\ \hline
\end{tabular}%
}}
\caption{A comparison of the number $t$ of the terms on the right hand side
of the Berends-Giele recursive relation with the total number $b$ of terms
needed for the Berends-Giele recursive calculation of the amplitude $%
J(1,2,\ldots ,2n+1)$ and with the total number $f$ of flavor ordered Feynman
graphs contributing to the same amplitude. In the last three row we compare
these numbers with the analogous ones for the case of \textquotedblleft $%
\protect\phi ^{4}$ theory''.}
\end{table}

On the other hand, as we will see in what follows, the Berends-Giele
relations can be used as a very suitable tool for the investigation of the
general properties of the semi-on-shell amplitudes. Let us mention e.g. the
following simple relations valid for $J(1,2,\ldots ,n)$%
\begin{eqnarray}
J(1,2,\ldots ,2n) &=&0  \label{odd_zero} \\
J(1,2,\ldots ,n) &=&J(n,n-1,\ldots ,2,1).  \label{reversed_J}
\end{eqnarray}%
These relation are valid independently on the field redefinition. However,
as we shall see in what follows, some properties of the semi-on-shell
amplitudes are not valid universally and are tightly related to a given
parametrization.

\subsection{Cayley parametrization}

Unlike the on-shell amplitudes $\mathcal{M}^{a_{1}\dots
a_{n}}(p_{1},p_{2},\ldots ,p_{n})$, which are physical observables and do
not depend on the choice of the field variables provided the different
choices are related by means of admissible (generally nonlinear)
transformations, the concrete form of $J_{n}^{a,a_{1}\dots
a_{n}}(p_{1},p_{2},\ldots ,p_{n})$ as well as the flavor-stripped amplitudes
$J_{n}(p_{1},p_{2},\ldots ,p_{n})$ depends on the parametrization of the $%
U(N)$ nonlinear sigma model.\ In what follows we will almost exclusively use
the so called Cayley parameterizations
\begin{equation}
U=\frac{1+\frac{\mathrm{i}}{\sqrt{2}F}\phi }{1-\frac{\mathrm{i}}{\sqrt{2}F}%
\phi }=1+2\sum_{n=1}^{\infty }\left( \frac{\mathrm{i}}{\sqrt{2}F}\phi
\right) ^{n},  \label{Cayley}
\end{equation}%
where the Goldstone boson fields are arranged into the hermitian matrix $%
\phi =\phi ^{a}t^{a}$ with $t^{a}$ being the $U(N)$ generators. As described
in Appendix \ref{parametrization_appendix}, representation (\ref{Cayley}) is
a special member of a wide class of parameterizations suited for the
construction of the flavor-stripped Feynman rules. The interrelation between
the field $\phi $ and analogous field $\widetilde{\phi }$ of the more usual
exponential parametrization $U=\exp \left( \frac{\mathrm{i}}{F}\widetilde{%
\phi }\right) $ is through the following admissible nonlinear field
redefinition
\begin{equation}
\phi =2F\tan \left( \frac{\mathrm{i}}{2F}\widetilde{\phi }\right) =%
\widetilde{\phi }+O\left( \widetilde{\phi }^{3}\right) .
\label{exp_to_Cayley}
\end{equation}%
As is shown in Appendix \ref{parametrization_appendix}, the flavor-stripped
Feynman rules for vertices read in the Cayley parametrization%
\begin{eqnarray}
V_{2n+1} &=&0  \notag \\
V_{2n+2} &=&-\frac{(-1)^{n}}{2^{n+1}}\left( \frac{1}{F}\right)
^{2n}\sum_{j=0}^{n}\sum_{i=1}^{2n+2}(p_{i}\cdot p_{i+2j+1})=\frac{(-1)^{n}}{%
2^{n}}\left( \frac{1}{F}\right) ^{2n}\left( \sum_{i=0}^{n}p_{2i+1}\right)
^{2},  \label{stripped_rules_Cayley}
\end{eqnarray}%
where we have used the momentum conservation in the last row. For the first
non-trivial vertex $V_{4}$ we get%
\begin{equation}
V_{4}=-\frac{1}{2F^{2}}(p_{1}+p_{3})^{2}=-\frac{1}{2F^{2}}(p_{2}+p_{4})^{2}
\end{equation}%
and the first two non-trivial semi-on-shell amplitudes read in the Cayley
parametrization
\begin{eqnarray}
J(1,2,3) =\frac{1}{2F^{2}p_{4}^{2}}&& {(p_{1}+p_{3})^{2}}  \label{J_3} \\
J(1,2,3,4,5) =\frac{1}{4F^{4}p_{6}^{2}}&& \left[ \frac{%
(p_{1}+p_{2}+p_{3}+p_{5})(p_{1}+p_{3})^{2}}{(p_{1}+p_{2}+p_{3})^{2}}+\frac{%
(p_{1}+p_{3}+p_{4}+p_{5})^{2}(p_{3}+p_{5})^{2}}{(p_{3}+p_{4}+p_{5})^{2}}%
\right.  \notag \\
&&\left. +\frac{(p_{1}+p_{5})^{2}(p_{2}+p_{4})^{2}}{(p_{2}+p_{3}+p_{4})^{2}}%
-(p_{1}+p_{3}+p_{5})^{2}\right]  \label{J_5}
\end{eqnarray}%
Let us illustrate explicitly the dependence of the semi-on-shell amplitudes
on the parametrization. Using the exponential one we obtain different
amplitude $J(1,2,3)$, namely%
\begin{equation}
J(1,2,3)_{\mathrm{exp}}=-\frac{1}{6F^{2}}\frac{%
(p_{1}+p_{2})^{2}+(p_{2}+p_{3})^{2}-2(p_{1}+p_{3})^{2}}{p_{4}^{2}}.
\label{J3_exp}
\end{equation}%
However, both $J(1,2,3)$ and $J(1,2,3)_{\mathrm{exp}}$ give the same
on-shell amplitude (\ref{4p_tamplitude}).

In the next subsection we will prove additional useful properties of the
semi-on-shell amplitudes.

\subsection{Scaling properties of semi-on-shell amplitudes}

The Cayley parametrization is specific in the sense that the semi-on-shell
amplitudes $J_{n}(p_{1},\ldots ,p_{n})$ in this parametrization obey simple
scaling properties when some subset of the momenta $p_{i}$ are scaled $%
p_{i}\rightarrow tp_{i}$ and the scaling parameter $t$ is then send to zero.
Here we will study two important scaling limits, corresponding to the case
when \emph{all odd} or \emph{all even} on-shell momenta are scaled. As we
shall see in the following section, these two scaling limits are the key
ingredients for the construction of the BCFW-like relations for
semi-on-shell amplitudes in the Cayley parametrization.

We will prove that for $n>1$ and $t\rightarrow 0$%
\begin{equation}
J_{2n+1}(tp_{1},p_{2},tp_{3},p_{4},\ldots ,p_{2r},tp_{2r+1},p_{2r+2},\ldots
,p_{2n},tp_{2n+1})=O(t^{2})  \label{licha_Ot2}
\end{equation}%
and
\begin{equation}
\lim_{t\rightarrow 0}J_{2n+1}(p_{1},tp_{2},p_{3},tp_{4},\ldots
,tp_{2r},p_{2r+1},tp_{2r+2},\ldots ,tp_{2n},p_{2n+1})=\frac{1}{(2F^{2})^{n}}.
\label{suda_Ot0}
\end{equation}%

The general proof of (\ref{licha_Ot2}) and (\ref{suda_Ot0}) is by induction. Let us first verify the base cases.
While the second statement holds already for $n=1$%
\begin{equation}
J_{3}(p_{1},tp_{2},p_{3})=\frac{1}{F^{2}}\frac{(p_{1}\cdot p_{3})}{%
(p_{1}+tp_{2}+p_{3})^{2}}\rightarrow \frac{1}{2F^{2}},
\end{equation}%
the first one is not valid unless $n=2$. Indeed
\begin{equation}
J_{3}(tp_{1},p_{2},tp_{3})=\frac{1}{2F^{2}}\frac{t(p_{1}\cdot p_{3})}{%
(p_{1}\cdot p_{2})+(p_{2}\cdot p_{3})+t(p_{1}\cdot p_{3})}=O(t).
\end{equation}%
On the other hand, using the explicit form of $J_{5}$ (cf. (\ref{J_5})) we
get
\begin{equation}
J_{5}(tp_{1},p_{2},tp_{3},p_{4},tp_{5})=O(t^{2});
\end{equation}%
we can therefore proceed by induction starting at $n=2$.

\begin{figure}[h]
\begin{center}
\epsfig{width=0.8\textwidth,figure=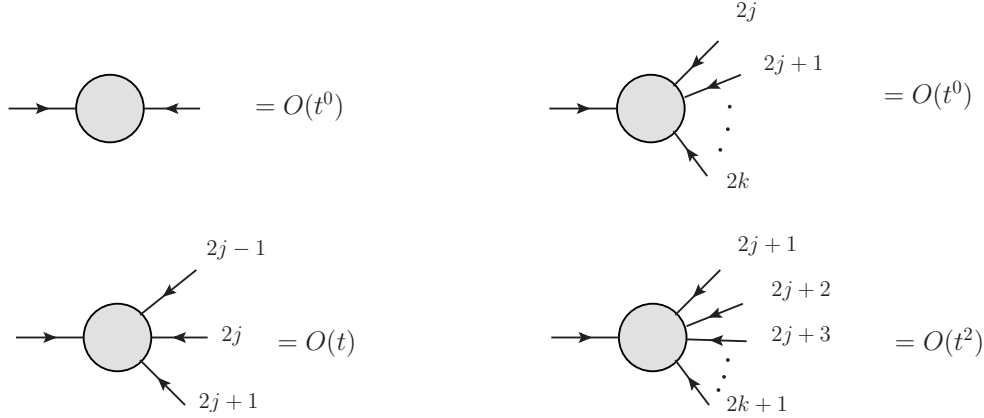}
\end{center}
\caption{Scaling of the building blocks on the right hand hand of the
Berends-Giele recursion relation according to the induction hypothesis when
the odd momenta are scaled.}
\label{blocks}
\end{figure}

\begin{figure}[h]
\begin{center}
\epsfig{width=0.9\textwidth,figure=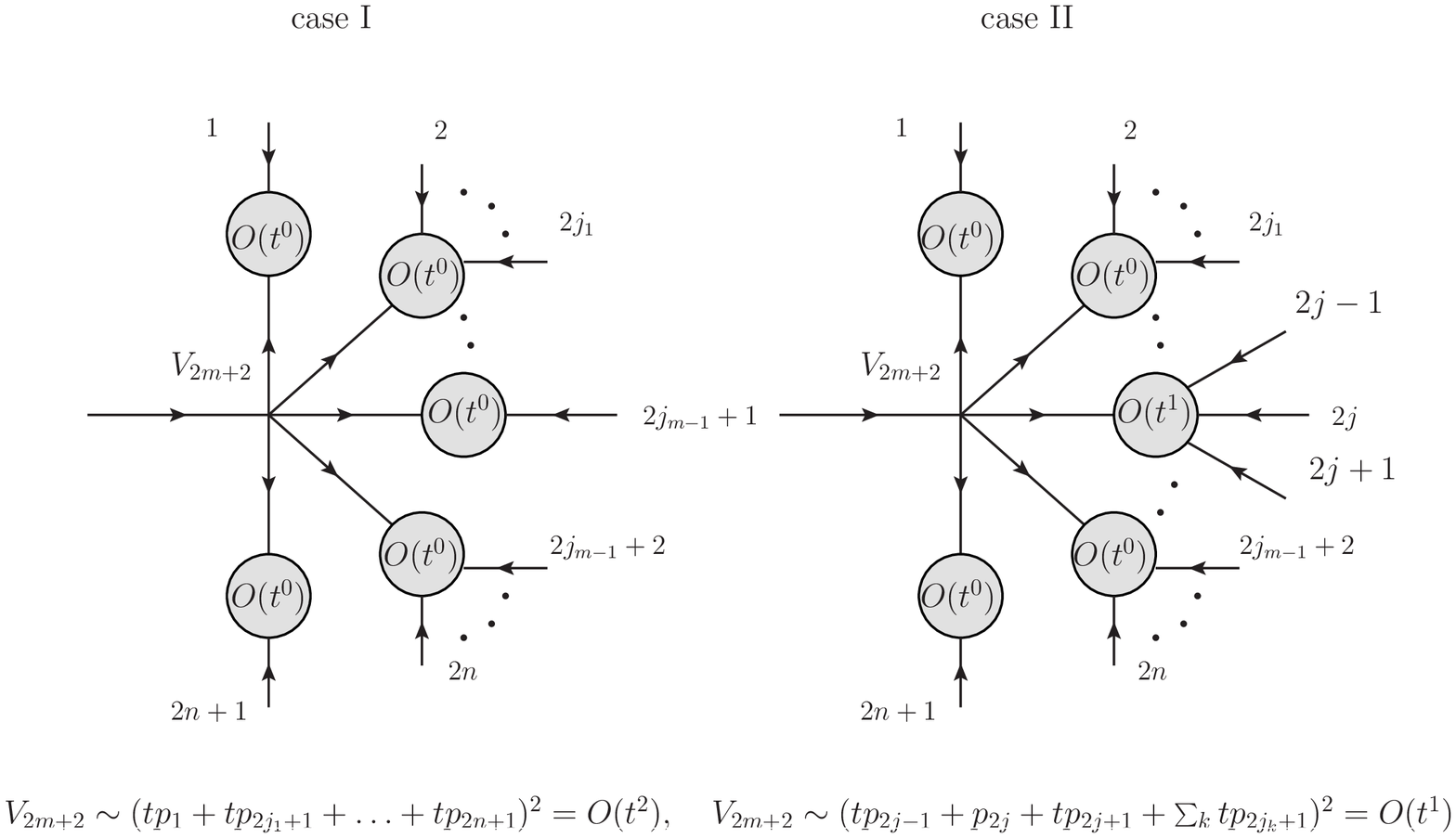}
\end{center}
\caption{Typical terms on the right hand hand of the Berends-Giele recursion
relation to which the induction hypothesis (\protect\ref{licha_Ot2}) cannot
be applied directly. In both cases, to all (case I) or to all but one (case
II) odd lines of the vertex the blocks $J_{1}$ are attached. 
In the case II,
one building block $J_{3}$ is attached to remaining odd line.}
\label{figure_1}
\end{figure}

Let us first prove the scaling property (\ref{licha_Ot2}).
Suppose, that (\ref%
{licha_Ot2}, \ref{suda_Ot0}) holds for all $\bar{n}$, where $1<\bar{n}<n$
and write for the left hand side of (\ref{licha_Ot2}) the Berends-Giele
relation (\ref{Berends_Giele_relations}) expressing $J_{2n+1}$ in terms of $%
J_{2\bar{n}+1}$ with $\bar{n}<n$. After the scaling $p_{2k+1}\rightarrow
tp_{2k+1}$, the $t\rightarrow 0$ behavior of $p_{2n+2}^{2}$ and $V_{m+1}$ is
$O(t^{0})$ and $O(t^{r})$ where $r\geq 0$ respectively. The scaling of the
remaining semi-on-shell amplitudes on the right hand side of (\ref%
{Berends_Giele_relations}) can be deduced from the induction hypothesis.
Note that it depends on the number of the external on-shell legs of $%
J(j_{i-1}+1,\dots ,j_{i})$ as well as on the parity of $j_{i-1}+1$, because
the semi-on-shell amplitude with scaled even or odd momenta scales
differently. Namely, according to the induction hypothesis, the scaling of
these building blocks of the right hand side of (\ref%
{Berends_Giele_relations}) is as follows (see Fig. \ref{blocks})
\begin{eqnarray}
&&J(j)=1=O(t^{0}),~~~~J(2j-1,2j,2j+1)=O(t),~~J(2j,\ldots ,2k)=O(t^{0}),
\notag \\
~~~ &&J(2j+1,\ldots ,2k+1)=O(t^{2})~~\mathrm{for}~~~k-j>1.
\end{eqnarray}%
\begin{figure}[h]
\begin{center}
\epsfig{width=0.9\textwidth,figure=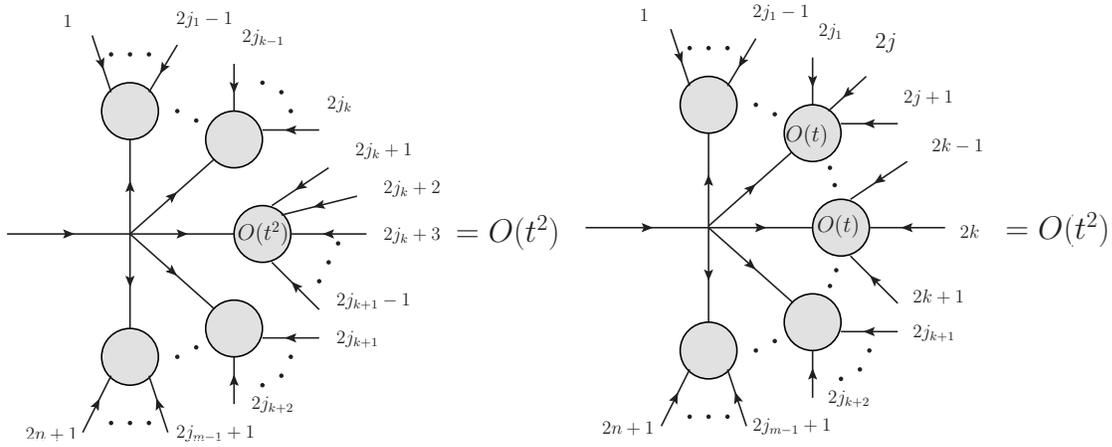}
\end{center}
\caption{The terms on the right hand hand of the Berends-Giele recursion
relation which are automatically $O(t^2)$ using the induction hypothesis
when the odd momenta are scaled.}
\label{inductionfigure}
\end{figure}
This implies, that those terms of Berends-Giele relations which are depicted
in Fig. \ref{inductionfigure}, i.e. those which contain at least one block $%
J(2j+1,\ldots ,2k+1)=O(t^{2})$ with $k-j>1$ or at least two building blocs $%
J(2j-1,2j,2j+1)$ are automatically $O(t^2)$. Therefore, the only dangerous
terms on the right hand side of (\ref{Berends_Giele_relations}) are those
without the buildings block of the type $J(2j+1,\ldots ,2k+1)=O(t^{2})$ with
$k-j>1$ and at the same time without (case I) or with just one (case II)
building block $J(2j-1,2j,2j+1)=O(t)$ (see Fig. \ref{figure_1}). To this
terms the induction hypothesis cannot be applied directly.

In the case I, the odd lines of the corresponding vertex $V_{2m+2}$ are
attached to $J(2j_{k}+1)=1$ and such a vertex is then proportional to the
squared sum of the odd momenta $tp_{2j_{k}+1}$, (cf. (\ref%
{stripped_rules_Cayley}))
\begin{equation}
V_{2m+2}(tp_{1},p_{2,2j_{1}},tp_{2j_{1}+1},\dots ,tp_{2n+1})\sim
(tp_{1}+tp_{2j_{1}+1}+\dots +tp_{2n+1})^{2}
\end{equation}%
which means that it scales as $O(t^{2})$. This is in fact the scaling of the
complete contribution of the terms in the case I, because all the remaining
building blocs are of the order $O(t^{0})$ for $t\rightarrow 0$.

In the case II with exactly one building block $%
J_{3}(tp_{2j-1},p_{2j},tp_{2j+1})=O(t)$ (note that, it has to be attached to
the odd line of the vertex $V_{2m+2}$), all the other odd lines of $V_{2m+2}$
are attached to $J(2j_{k}+1)=1$ and such a vertex is then proportional to
the squared sum of the momenta $tp_{2j_{k}+1}$ and the momentum of the line
which is attached to $J_{3}(tp_{2j-1},p_{2j},tp_{2j+1})$, namely
\begin{equation}
V_{2m+2}\sim \left( tp_{2j-1}+p_{2j}+tp_{2j+1}+\sum_{k}tp_{2j_{k}+1}\right)
^{2}=O(t).
\end{equation}%
Therefore the complete contribution of the dangerous terms in the case II is
in fact $O(t^{2})$ for $t\rightarrow 0$ because both $V_{2m+2}$ and $%
J_{3}(tp_{2j-1},p_{2j},tp_{2j+1})$ scale as $O(t)$ and again all the
remaining building blocks are of the order $O(t^{0})$ for $t\rightarrow 0$.
\ All the other \textquotedblleft non-dangerous'' terms on
the right hand side of the Berends-Giele relations scale at least as $%
O(t^{2})$, which finishes the proof of (\ref{licha_Ot2}).

\begin{figure}[tbh]
\begin{center}
\epsfig{width=0.35\textwidth,figure=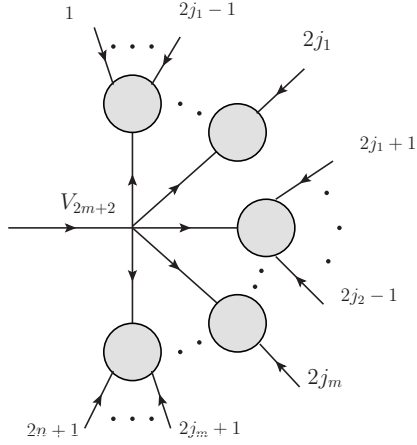}
\end{center}
\caption{Typical terms on the right hand hand of the Berends-Giele recursion
relation which contribute to (\protect\ref{suda_Ot0}). Here to all even
lines of the vertex the blocks $J_1$ are attached. }
\label{figure_2}
\end{figure}

Let us now prove (\ref{suda_Ot0}), i.e. the case when all even momenta are
scaled. Suppose validity of this relation for $\bar{n}<n$ $\ $\ and again
write the Berends-Giele relation for the left hand side of (\ref{suda_Ot0}).
Thanks to the just proven statement (\ref{licha_Ot2}), the terms on the
right hand side of \ (\ref{Berends_Giele_relations}) with at least one
building block $J(j_{k}+1,\ldots ,j_{k+1})$ with odd $j_{k}$ and $%
j_{k+1}-j_{k}>1$ do not contribute in the limit $t\rightarrow 0$. Such a
block can be attached only to the even line of the vertex $V_{m+1}$.
Therefore, the only terms which can contribute in the limit $t\rightarrow 0$
have the form depicted in Fig. \ref{figure_2}, i.e. those with the building
blocks $J_{1}$ attached to all even lines of the vertex.

According to the induction hypothesis and using the explicit form of $%
V_{2k+2}$ this gives for $t\rightarrow 0$

\begin{equation}
-\frac{(-1)^{k}}{2^{k}F^{2k}}\prod\limits_{l=1}^{k+1}\frac{1}{%
(2F^{2})^{j_{l}-j_{l-1}-1}}=-\frac{(-1)^{k}}{2^{n}F^{2n}}
\end{equation}%
where we denote $j_{0}=0$ and $j_{k+1}=n+1.$ Sum of all such contributions is%
\begin{equation}
\sum_{k=1}^{n}\sum_{1\leq j_{1}<j_{2}<\ldots ,j_{k}\leq n}\frac{(-1)^{k-1}}{%
2^{n}F^{2n}}=\frac{1}{2^{n}F^{2n}}\sum_{k=1}^{n}\left(
\begin{array}{c}
n \\
k%
\end{array}%
\right) (-1)^{k-1}=\frac{1}{2^{n}F^{2n}},
\end{equation}
which finishes the proof.

Another independent scaling properties of the semi-on-shell amplitudes $J_{2n+1}$
can be proven using the same strategy. For instance, when all odd momenta
and one additional even momentum (say $p_{2r}$) are scaled, we get%
\begin{equation}
\lim_{t\rightarrow 0}J_{2n+1}(tp_{1},p_{2},tp_{3},p_{4},\ldots
,tp_{2r-1},tp_{2r},tp_{2r+1},\ldots ,p_{2N},tp_{2n+1})=0
\end{equation}%
for $n>1$. We postpone the proof to the Appendix \ref{extra_scaling}.

Let us note that due to the {homogeneity of $J(1,2,\ldots ,2n+1)$ we can
rewrite the relations (\ref{licha_Ot2}) and (\ref{suda_Ot0}) as a statement
on the asymptotic behavior of the scaled amplitudes for }$t\rightarrow
\infty $, namely%
\begin{equation}
\lim_{t\rightarrow \infty }J_{2n+1}(tp_{1},p_{2},\ldots
,p_{2n},tp_{2n+1})=\lim_{t\rightarrow \infty
}J_{2n+1}(p_{1},t^{-1}p_{2},\ldots ,t^{-1}p_{2n},p_{2n+1})=\frac{1}{%
(2F^{2})^{n}}  \label{licha_infty}
\end{equation}%
and
\begin{equation}
J_{2n+1}(p_{1},tp_{2},\ldots
,tp_{2n},p_{2n+1})=J_{2n+1}(t^{-1}p_{1},p_{2},\ldots
,p_{2n},t^{-1}p_{2n+1})=O(t^{-2}).  \label{suda_infty}
\end{equation}

\subsection{BCFW reconstruction}

As we have mentioned in the previous subsection, the standard BCFW-like
deformation of the external momenta $p_{i}$ yields deformed amplitudes which
behave as a non-negative power of $z$ for $z\rightarrow \infty $. As a
result, for the reconstruction of the amplitude from its pole structure we need to
use the general reconstruction formula (\ref{generalized_BCFW}) for which
additional information on the on-shell amplitude (its values at several points) is
necessary. However, such an information is not at our disposal.  We solve this problems by the following trick:
 we
relax some demands placed on the usual BCFW-like deformation and allow more
general ones for which either the reconstruction formula without
subtractions can be applied or additional information on the deformed
amplitudes is accessible. The momentum conservation cannot be evidently
avoided, what remains is the on-shell condition of all the external momenta.
It seems therefore to be natural to relax this constraint and instead of the
on-shell amplitudes ${\mathcal{M}}_{2n+2}$ to use the semi-on-shell
amplitudes $J_{2n+1}$, or the cut semi-on-shell amplitudes $M_{2n+1}$
defined as%
\begin{equation}
M_{2n+1}\left( p_{1},\ldots ,p_{2n+1}\right) =p_{1,2n+1}^{2}J_{2n+1}\left(
p_{1},\ldots ,p_{2n+1}\right).  \label{cut_amplitude}
\end{equation}

Motivated by the results of the previous section let us assume the following
deformation of the semi-on-shell amplitude $M_{2n+1}$ in the Cayley
parametrization%
\begin{equation}
M_{2n+1}(z)\equiv M_{2n+1}(p_{1},zp_{2},p_{3},zp_{4},\ldots
,zp_{2r},p_{2r+1},zp_{2r+2},\ldots ,zp_{2n},p_{2n+1})
\end{equation}%
i.e. all even momenta are scaled by the complex parameter $z$ and the odd
momenta are not deformed
\begin{equation}
p_{2k}(z)=zp_{2k},~~~p_{2k+1}(z)=p_{2k+1}  \label{p2k}
\end{equation}%
Note that in contrast to the standard BCFW shift this deformation is
possible for general number of space-time dimensions $d$. The physical
amplitude corresponds to $z=1$. For $n=1$ we get explicitly%
\begin{equation}
M_{3}(z)=\frac{1}{F^{2}}(p_{1}\cdot p_{3})  \label{J1}
\end{equation}%
For general $n$ let us denote the sums of all odd (even) momenta as
\begin{equation}
p\_=\sum_{k=0}^{n}p_{2k+1},~~p_{+}=\sum_{k=1}^{n}p_{2k}~.
\end{equation}%
Then in general case the function $M_{2n+1}(z)$ has the following important
properties:

\begin{enumerate}
\item {With generic fixed $p_{i}$ it is a meromorphic function of $z$ with
simple poles.}

\item {\ The asymptotics of $M_{2n+1}(z)$ can be deduced form the known
properties of $J_{2n+1}$, namely for $n>1$ we get as a consequence of (\ref%
{suda_infty})}%
\begin{equation}
M_{2n+1}(z)=(p_{+}z+p_{-})^{2}J_{2n+1}(p_{1},zp_{2},\ldots
,zp_{2n},p_{2n+1})=O(z^{0}).  \label{Jinfty}
\end{equation}

\item {\ For $n\geq 1$ we have according to known scaling property (\ref%
{suda_Ot0}) of $J_{2n+1}$
\begin{equation}
\lim_{z\rightarrow 0}M_{2n+1}(z)=\frac{1}{(2F^{2})^{n}}p_{-}^{2}  \label{J0}
\end{equation}%
}
\end{enumerate}

The first two properties allows us to write for $M_{2n+1}(z)$ the
reconstruction formula with one subtraction, i.e. the relation (\ref%
{generalized_BCFW}) with $k=0$. The third property is the key one for the
complete reconstruction and determines both the ``subtraction point`` $%
a_{1}=0$ and the ``subtraction constant`` $M_{2n+1}(a_{1})=p_{-}^{2}/{%
(2F^{2})^{n}}$. The resulting formula reads\footnote{%
Let us note, that we could write analogous reconstruction formula directly
for the currents $J_{2n+1}$ as we did in \cite{Kampf:2012fn}. In such a case we do not need any subtraction.  The price to
pay is that we get two more poles, the residues of which cannot be determined
recursively from unitarity. Fortunately, the relation (\ref{suda_infty}) and
the residue theorem can be used in order to obtain the unknown
residues in terms of the remaining ones. The resulting formula is fully
equivalent to (\ref{master}), however it is a little bit less elegant.}
\begin{equation}
M_{2n+1}(z)=\frac{1}{(2F^{2})^{n}}p_{-}^{2}+\sum_{P}\frac{\mathrm{Res}\left(
M_{2n+1},z_{P}\right) }{z-z_{P}}\frac{z}{z_{P}}  \label{master}
\end{equation}%
where the sum is over the poles $z_{P}$ of $M_{2n+1}(z)$. \ The position of
the poles is known and the corresponding residues can be determined
recursively as in usual BCFW relations, however, there are some subtleties.

The poles $z_P$ of $M_{2n+1}(z)$ correspond to the vanishing denominators of
the deformed propagators $p_{P}^{2}(z)=0$, where%
\begin{equation}
p_{P}^{2}(z)\equiv p_{i,j}(z)^{2}=0,~~\mathrm{for}~~2\leq j-i<2n
\label{pole_equation}
\end{equation}%
and where $j-i$ is even; in this formula $p_{i,j}(z)=zp_{i,j}^{+}+p_{i,j}^{-}
$ with
\begin{equation}
p_{i,j}^{+}=\sum_{i\leq 2k\leq j}p_{2k},~\ \ p_{i,j}^{-}=\sum_{i\leq
2k+1\leq j}p_{2k+1},~
\end{equation}%
i.e. $p_{i,j}^{\pm }$ is a sum of all even (odd) momenta from the ordered
set $p_{i},p_{i+1},\ldots ,p_{j-1},p_{j}$. Explicitly for $j-i>2$%
\begin{equation}
z_{i,j}^{\pm }=\frac{-(p_{i,j}^{+}\cdot \ p_{i,j}^{-})\pm \left(
-G(p_{i,j}^{+},\ p_{i,j}^{-})\right) ^{1/2}}{p_{i,j}^{+2}}  \label{poly}
\end{equation}%
where $G(a,b)=a^{2}b^{2}-(a\cdot b)^{2}$ is the Gram determinant, which is
nonzero for generic momenta $p_{i},\ldots ,p_{j}$. Therefore in the generic
case for $j-i>2$ we deal with doublets of single poles.

The case of three-particle poles corresponding to $j-i=2$ has to be treated
separately. In this case either $p_{i,j}^{+2}=0$ or $p_{i,j}^{-2}=0$ (this
sets in for $p_{i,j}^{+}=p_{i+1}$ or for $p_{i,j}^{-}=p_{i+1}$ respectively;
let us remind that $p_{k}$ are on-shell). In the first case we have only one
pole%
\begin{equation}
z_{2j-1,2j+1}=-\frac{(p_{2j-1}\cdot p_{2j+1})}{p_{2j}\cdot
(p_{2j-1}+p_{2j+1})}  \label{z2jm1}
\end{equation}%
while in the second case we have apparently two poles%
\begin{eqnarray}
z_{2j,2j+2}^{+} &=&0 \\
z_{2j,2j+2}^{-} &\equiv &z_{2j,2j+2}=-\frac{p_{2j+1}\cdot (p_{2j}+p_{2j+2})}{%
(p_{2j}\cdot p_{2j+2})}  \label{z2j2jp2min}
\end{eqnarray}%
However $z_{2j,2j+2}^{+}=0$ cannot be a pole according to (\ref{J0}) and the
corresponding residue has to be zero.

\begin{figure}[t]
\begin{center}
\epsfig{width=0.5\textwidth,figure=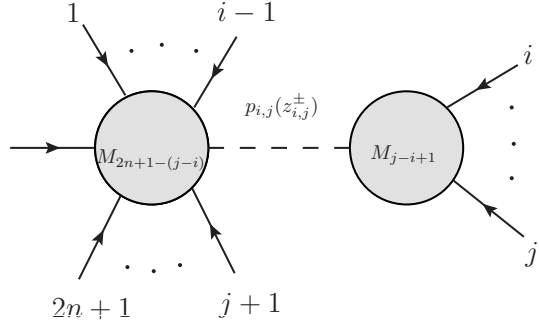}
\end{center}
\caption{Graphical representation of the right hand side of the relation (%
\protect\ref{residua}). }
\label{figure_4}
\end{figure}

The residues of the function $M_{2n+1}(z)$ are dictated by unitarity and at
the poles they factorize (see Fig. \ref{figure_4}). Writing for $j-i>2$
\begin{equation}
(zp_{i,j}^{+}+p_{i,j}^{-})^{2}=p_{i,j}^{+2}(z-z_{i,j}^{+})(z-z_{i,j}^{-})
\label{p_pole}
\end{equation}%
we get for $j-i>2$%
\begin{equation}
\mathrm{Res}\left( M_{2n+1},z_{i,j}^{\pm }\right) =\pm \frac{%
M_{L}^{(i,j)}(z_{i,j}^{\pm })M_{R}^{(i,j)}(z_{i,j}^{\pm })}{%
p_{i,j}^{+2}(z_{i,j}^{+}-z_{i,j}^{-})}  \label{residua}
\end{equation}%
where we denoted
\begin{eqnarray}
M_{L}^{(i,j)}(z_{i,j}^{\pm }) &=&M_{2n+1-(j-i)}(p_{1}(z_{i,j}^{\pm }),\ldots
,p_{i-1}(z_{i,j}^{\pm }),p_{i,j}(z_{i,j}^{\pm }),p_{j+1}(z_{i,j}^{\pm
}),\ldots ,p_{2n+1}(z_{i,j}^{\pm }))  \label{ML} \\
M_{R}^{(i,j)}(z_{i,j}^{\pm }) &=&M_{j-i+1}(p_{i}(z_{i,j}^{\pm
}),p_{i+1}(z_{i,j}^{\pm }),\ldots ,p_{j}(z_{i,j}^{\pm })).  \label{MR}
\end{eqnarray}%
Note that, while the amplitude $M_{L}^{(i,j)}$ remains semi-on-shell, the
amplitude $M_{R}^{(i,j)}$ is fully on-shell, because the deformed momentum $%
p_{i,j}(z)$ is on-shell for $z=z_{i,j}^{\pm }$.

The formula (\ref{residua}) is valid also for the three-particle pole $%
z_{2j,2j+2}$ given by (\ref{z2j2jp2min}). However the pole $z_{2j-1,2j+1}$
deserves a special remark because the corresponding residue is determined by
the formula different from (\ref{residua}), namely%
\begin{equation}
\mathrm{Res}\left( M_{2n+1},z_{2j-1,2j+1}\right) =\frac{%
M_{L}^{(2j-1,2j+1)}(z_{2j-1,2j+1})M_{R}^{(2j-1,2j+1)}(z_{2j-1,2j+1})}{%
2p_{2j-1,2j+1}^{+}\cdot p_{2j-1,2j+1}^{-}}  \label{residua1}
\end{equation}%
where $M_{L,R}^{(2j-1,2j+1)}(z_{2j-1,2j+1})$ are given by (\ref{ML}) and (%
\ref{MR}) with $z_{i,j}^{\pm }$ replaced by $z_{2j-1,2j+1}$.

To summarize, we have found a closed system of recursive BCFW-like relations
for the tree cut semi-on-shell amplitudes $M_{2n+1}$, which consists of the
reconstruction formula (\ref{master}), the pole positions (\ref{poly}), (\ref%
{z2jm1}) and (\ref{z2j2jp2min}) and the residue formulae (\ref{residua}) and
(\ref{residua1}). Note that the initial condition for the recursion (\ref{J1}%
) can be understood as the special case of (\ref{master}) for $n=1$ because
then there is no pole $z_{i,j}$ with $2\leq j-i<2$ and the sum of the
residue contributions is empty. The physical amplitude $M_{2n+1}(p_{1},%
\ldots ,p_{2n+1})$ corresponds to $z=1$
\begin{equation}
M_{2n+1}(p_{1},
\ldots ,p_{2n+1})=\frac{1}{(2F^{2})^{n}}p_{-}^{2}+\sum_{P}\frac{\mathrm{Res}\left(
M_{2n+1},z_{P}\right) }{z_{P}}\frac{1}{1-z_{P}}.  \label{master_phys}
\end{equation}%
As a final result we get then using (\ref{residua}), (\ref{residua1}), (\ref%
{z2jm1}), (\ref{z2j2jp2min}) and (\ref{p_pole})
\begin{equation}
M_{2n+1}(p_{1},
\ldots ,p_{2n+1})=\frac{1}{(2F^{2})^{n}}p_{-}^{2}+\sum_{P}M_{L}^{(P)}(z_{P})\frac{%
R_{P}}{p_{P}^{2}}M_{R}^{(P)}(z_{P}).  \label{master_phys_poles}
\end{equation}%
Note that there is an extra function $R_{P}$ in contrast to the standard
BCFW formula (\ref{phys_general_BCFW}), namely%
\begin{equation}
R_{P}=\left\{
\begin{array}{l}
z_{P}^{-2}~~~~~~~~~\,\,\,\,\,\,\,\,\,\,\,\mathrm{for}~\ z_{P}=z_{2j,2j+2}~
\\
z_{P}^{-1}~~~~~~~~~\,\,\,\,\,\,\,\,\,\,\,\mathrm{for}~\ z_{P}=z_{2j-1,2j+1}
\\
\frac{1}{z_{i,j}^{\pm }-z_{i,j}^{\mp }}\frac{1-z_{i,j}^{\mp }}{z_{i,j}^{\pm }%
}\,\,\,\,\,\mathrm{for}~\ z_{P}=z_{i,j}^{\pm }%
\end{array}%
\right.
\end{equation}%
For further convenience, we rewrite (\ref{master_phys_poles}) with help of (%
\ref{J1}) in the following more explicit form
\begin{eqnarray}
&&M_{2n+1}\left( p_{1},\ldots ,p_{2n+1}\right) =\frac{1}{(2F^{2})^{n}}%
p_{-}^{2}+  \notag \\
&&+\sum_{j=1}^{n-1}M_{L}^{(2j,2j+2)}(z_{2j,2j+2})\frac{1}{p_{2j,2j+2}^{2}}%
\frac{p_{2j}\cdot p_{2j+2}}{F^{2}}  \notag \\
&&-\sum_{j=1}^{n}M_{L}^{(2j-1,2j+1)}(z_{2j-1,2j+1})\frac{1}{p_{2j-1,2j+1}^{2}%
}\frac{p_{2j-1,2j+1}^{+}\cdot p_{2j-1,2j+1}^{-}}{F^{2}}  \notag \\
&&+\sum_{2<j-i<2n}\frac{1}{z_{i,j}^{+}-z_{i,j}^{-}}\left(
M_{L}^{(i,j)}(z_{i,j}^{+})\frac{1}{p_{i,j}^{2}}M_{R}^{(i,j)}(z_{i,j}^{+})%
\frac{1-z_{i,j}^{-}}{z_{i,j}^{+}}-M_{L}^{(i,j)}(z_{i,j}^{-})\frac{1}{%
p_{i,j}^{2}}M_{R}^{(i,j)}(z_{i,j}^{-})\frac{1-z_{i,j}^{+}}{z_{i,j}^{-}}%
\right) .  \notag \\
&&  \label{explicit_BCFW}
\end{eqnarray}%
The on-shell amplitude is then%
\begin{equation}
\mathcal{M}_{2n}(1,2,\ldots ,2n-1;2n)=-\lim_{p_{1,2n-1}^{2}\rightarrow
0}M_{2n-1}(1).  \label{on_shell_LSZ}
\end{equation}

\subsection{Explicit example of application of BCFW relations: 6pt amplitude}

As an illustration let us apply the BCFW-like recursive relations (\ref%
{master}) to the amplitude $M_{5}(z)\equiv
M_{5}(p_{1},zp_{2},p_{3},zp_{4},p_{5})$. In this case we have three poles,
all of them being three-particle, namely%
\begin{equation}
z_{1,3}=1-\frac{s_{1,3}}{s_{1,2}+s_{2,3}},~~~z_{2,4}=\left( 1-\frac{s_{2,4}}{%
s_{2,3}+s_{3,4}}\right) ^{-1},~~~z_{3,5}=1-\frac{s_{3,5}}{s_{3,4}+s_{4,5}}
\label{z13}
\end{equation}%
where the variables $s_{i,j}$ are given by (\ref{s_ij_definition}).The
residues are given by the relations (\ref{residua}) for $z_{2,4}$ and (\ref%
{residua1} ) for $z_{1,3}$ and $z_{3,5}$. After simple algebra using the
explicit form of the poles (\ref{z13}) we get%
\begin{eqnarray}
\frac{\mathrm{Res}\left( M_{5},z_{1,3}\right) }{z_{1,3}} &=&\frac{1}{4F^{4}}%
(1-z_{1,3})(s_{2,5}-s_{2,4}+s_{3,4}-s_{3,5})-\frac{1}{4F^{4}}\left(
s_{1,5}-s_{1,4}-s_{4,5}\right)  \notag \\
\frac{\mathrm{Res}\left( M_{5},z_{3,5}\right) }{z_{3,5}} &=&\frac{1}{4F^{4}}%
(1-z_{3,5})(s_{1,4}-s_{1,3}+s_{2,3}-s_{2,4})-\frac{1}{4F^{4}}%
(s_{1,5}-s_{1,2}-s_{2,5})  \notag \\
\frac{\mathrm{Res}\left( M_{5},z_{2,4}\right) }{z_{2,4}} &=&\frac{1}{4F^{4}}%
\left( s_{1,5}-s_{1,4}+s_{2,4}-s_{2,5}\right) .
\end{eqnarray}%
Note that the potential  unphysical poles $z_{i,j}(p_k)=0$ have canceled completely. We have
also%
\begin{equation}
(1-z_{1,3})^{-1}=\frac{s_{1,2}+s_{2,3}}{s_{1,3}},~~(1-z_{3,5})^{-1}=\frac{%
s_{3,4}+s_{4,5}}{s_{3,5}},~~~~(1-z_{2,4})^{-1}=1-\frac{s_{2,3}+s_{3,4}}{%
s_{2,4}}
\end{equation}%
These factors are responsible for setting of the physical poles in the
resulting amplitude. After inserting this to the formula (\ref{master_phys})
we get for the individual contributions to the semi-on-shell amplitude in
the Cayley parametrization%
\begin{eqnarray}
\frac{\mathrm{Res}\left( M_{5},z_{1,3}\right) }{z_{1,3}(1-z_{1,3})} &=&\frac{%
1}{4F^{2}}\left[ \frac{\left( s_{1,4}+s_{4,5}-s_{1,5}\right)
(s_{1,2}+s_{2,3})}{s_{1,3}}+s_{2,5}-s_{2,4}+s_{3,4}-s_{3,5}\right]  \notag \\
\frac{\mathrm{Res}\left( M_{5},z_{3,5}\right) }{z_{3,5}(1-z_{3,5})} &=&\frac{%
1}{4F^{2}}\left[ \frac{(s_{1,2}+s_{2,5}-s_{1,5})(s_{3,4}+s_{4,5})}{s_{3,5}}%
+s_{1,4}-s_{1,3}+s_{2,3}-s_{2,4}\right]  \notag \\
\frac{\mathrm{Res}\left( M_{5},z_{2,4}\right) }{z_{2,4}(1-z_{2,4})} &=&\frac{%
1}{4F^{2}}\left[ \frac{\left( s_{1,4}+s_{2,5}-s_{1,5}\right)
(s_{2,3}+s_{3,4})}{s_{2,4}}+s_{1,5}-s_{1,4}+s_{2,4}-s_{2,5}-s_{2,3}-s_{3,4}%
\right]  \notag \\
\frac{p_{-}^{2}}{4F^{2}} &=&\frac{1}{4F^{2}}\left[
s_{1,3}-s_{1,2}-s_{2,3}+s_{1,5}-s_{1,4}+s_{2,4}-s_{2,5}+s_{3,5}-s_{3,4}-s_{4,5}%
\right].
\end{eqnarray}%
Finally we get
\begin{eqnarray}
&&4F^{2}M_{5}(1)=  \notag \\
&=&\frac{\left( s_{1,4}+s_{4,5}-s_{1,5}\right) (s_{1,2}+s_{2,3})}{s_{1,3}}+%
\frac{(s_{1,2}+s_{2,5}-s_{1,5})(s_{3,4}+s_{4,5})}{s_{3,5}}+\frac{\left(
s_{1,4}+s_{2,5}-s_{1,5}\right) (s_{2,3}+s_{3,4})}{s_{2,4}}  \notag \\
&&+2s_{1,5}-s_{1,2}-s_{1,4}-s_{2,3}-s_{2,5}-s_{3,4}-s_{4,5}.
\end{eqnarray}%
Taking this amplitude on-shell according to (\ref{on_shell_LSZ}), i.e.
setting $s_{1,5}\rightarrow 0$ and changing the overall sign, we reproduce
the parametrization independent physical amplitude (\ref{6pt_amplitude}).

\section{More properties of stripped semi-on-shell amplitudes}

The BCFW recursive relations provides us with a Lagrangian-free formulation
of the tree-level nonlinear $SU(N)$ sigma model in the Cayley
parametrization. We can use them similarly as the Berends-Giele relations as
a tool for the investigation of further interesting features of the stripped
semi-on-shell amplitudes $M_{2n+1}$ and $J_{2n+1}$. As we have already
mentioned, these features are not universal because of the parametrization
dependence of $M_{2n+1}$ and $J_{2n+1}$, however, their implications for the fully on shell amplitudes
hold universally\footnote{Let us remind that the on-shell amplitudes are parametrization independent.}. In this section we will concentrate on the problem of
single soft limits (Adler zeroes) and double soft limit of the semi-on-shell
amplitudes.

The presence of Adler zeroes for the \emph{on-shell} Goldstone boson
amplitudes $\mathcal{M}^{a_{1}\ldots a_{2n}}(p_{1},\ldots ,p_{2n}) $, i.e.
validity of the limit
\begin{equation}
\lim_{p_{j}\rightarrow 0}\mathcal{M}^{a_{1}a_{2}\ldots
a_{2n}}(p_{1},p_{2},\ldots ,p_{2n})=0,  \label{Ma_zero}
\end{equation}
is a well known consequence of the nonlinearly realized chiral symmetry.
More generally it is an universal (non-perturbative) feature in the theories
with spontaneous breakdown of a global symmetry. In such theories the
amplitudes with one extra Goldstone boson $\pi ^{a}$ in the \emph{out }(or
\emph{in}) state vanishes when the Goldstone boson become soft, e.g.
\begin{equation}
\lim_{p\rightarrow 0}\langle f+\pi ^{a}(p),\mathrm{out}|i,\mathrm{in}\rangle
=0,  \label{Gen_zero}
\end{equation}
provided the $\pi ^{a}$ cannot be emitted from the external lines
corresponding to the states $|i,\mathrm{{in}\rangle }$ or $|f,\mathrm{{out}%
\rangle }$. In the $SU(N)$ nonlinear sigma model the Adler zero is present
also for the stripped on-shell amplitudes $\mathcal{M}_{2n}(p_{1},p_{2},%
\ldots ,p_{2n})$ due to the leading $N$ orthogonality relations (\ref%
{N_orthogonality}) and corresponding uniqueness of the decomposition (\ref%
{M_stripped}). However, this property is not guaranteed automatically for
the semi-on-shell amplitudes $M_{2n+1}$ and the soft Goldstone boson
behavior can depend on the parametrization. For instance using the Cayley
parametrization, we find for the amplitude $M_{3}=(p_{1}\cdot p_{3})/F^{2}$
the Adler zero for soft $p_{1}$ and $p_{3}$, however there is no zero for
soft $p_{2}$ in general when keeping $p_{4}$ off-shell. For the same
amplitude in the exponential parametrization (cf. (\ref{J3_exp})) we have no
Adler zero at all. As we shall show in this section, for the semi-on-shell
amplitudes $M_{2n+1}$ in the Cayley parametrization we can prove, using the
BCFW-like relation, the Adler zero for half of the momenta  (namely for those $%
p_{j}$ with \emph{odd} index $j$).

The double soft limit of the Goldstone boson on-shell amplitudes $\mathcal{M}%
^{a_{1}a_{2}\ldots a_{2n+2}}(p_{1},p_{2},\ldots ,p_{2n+2})$ is more
complicated and has been studied relatively recently in connection with the
regularized action of the broken generators on the $n$ Goldstone boson
states \cite{ArkaniHamed:2008gz}. Motivated by direct inspection of the six
Goldstone boson amplitude in the nonlinear chiral $SU(2)$ sigma model it
was conjectured that provided the two soft momenta are sent to zero with the
same rate, the following limit holds%
\begin{eqnarray}
&&\lim_{t\rightarrow 0}\mathcal{M}^{aba_{1}a_{2}\ldots
a_{2n}}(tp,tq,p_{1},p_{2},\ldots ,p_{2n})  \notag \\
&=&-\frac{1}{2F^{2}}\sum_{i=1}^{n}f^{abc}f^{ca_{i}d}\frac{p_{i}\cdot (p-q)}{%
p_{i}\cdot (p+q)}\mathcal{M}^{a_{1}\ldots a_{i-1}da_{i+1}\ldots
a_{2n}}(p_{1},p_{2},\ldots ,p_{2n}),  \label{AH_double_soft}
\end{eqnarray}%
where $f^{abc}$ are the structure constants. Analogous statement has been
then rigorously proven for the tree-level amplitudes in the $\mathcal{N}=8$
supergravity using BCFW relations. In fact, for the on-shell amplitudes, the
formula (\ref{AH_double_soft}) can be proven non-perturbatively under some
assumptions for the general enough case of the theory with global symmetry
breaking (including the case of chiral nonlinear sigma model with general
chiral group $G$) using the symmetry arguments only (cf. the PCAC soft-pions
theorems \cite{Dashen:1969ez}) . We postpone the details to the Appendix \ref%
{double_soft_appendix} . \

In terms of the stripped \emph{on-shell} amplitudes the relation (\ref%
{AH_double_soft})\ $\ $\ can be rewritten as
\begin{eqnarray}
&&\lim_{t\rightarrow 0}\mathcal{M}_{2n+2}(p_{1},\ldots
,p_{i-1},tp_{i},\ldots ,tp_{j},p_{j+1},\ldots p_{2n+2})  \notag \\
&=&\frac{1}{4F^{2}}\delta _{j,i+1}\left( \frac{p_{i+2}\cdot (p_{i}-p_{i+1})}{%
p_{i+2}\cdot (p_{i}-p_{i+1}}-\frac{p_{i-1}\cdot (p_{i}-p_{i+1})}{%
p_{i-1}\cdot (p_{i}-p_{i+1}}\right) \mathcal{M}_{2n}(p_{1},\ldots
,p_{i-1},p_{i+2},\ldots p_{2n+2}).  \label{stripped_double_soft}
\end{eqnarray}%
In this section we will prove this relation also for the tree-level
semi-on-shell amplitudes $J_{2n+1}$ (and consequently for $M_{2n+1}$) of the
$SU(N)$ nonlinear sigma model in the Cayley parametrization using suitable
form of the generalized BCFW representation.

\subsection{Adler zeroes}

In this subsection we will use the BCFW-like relations (\ref{explicit_BCFW})
derived in the previous section and prove the presence an Adler zero at $%
M_{2n+1}$ when one of the \emph{odd} momenta, say $p_{2l-1}$, is soft, i.e.
we will prove that for $l=1,2,\ldots ,n+1$%
\begin{equation}
\lim_{t\rightarrow 0}M_{2n+1}(p_{1},p_{2},\ldots
,p_{2l-2},tp_{2l-1},p_{2l+1},\ldots ,p_{2n+1})=0.  \label{adler_zero}
\end{equation}%
For the fundamental amplitude $M_{3}(p_{1},p_{2},p_{3})$ we have explicitly%
\footnote{%
Note however that for $t\rightarrow 0$ according to (\ref{suda_Ot0}).%
\begin{equation*}
M_{3}(p_{1},tp_{2},p_{3})\rightarrow \frac{1}{2F^{2}}(p_{1}+p_{3})^{2}
\end{equation*}%
and therefore the statement analogous to (\ref{adler_zero}) for even momenta
does not hold.}
\begin{equation}
M_{3}(tp_{1},p_{2},p_{3})=M_{3}(p_{1},p_{2},tp_{3})=\frac{1}{F^{2}}%
t(p_{1}\cdot p_{3})\rightarrow 0.
\end{equation}%
In the general case the proof of (\ref{adler_zero}) is by induction. Let us
assume validity of (\ref{adler_zero}) for $m<n$. This assumption also means
that, taking the cut semi-on-shell amplitude $M_{2m+1}$ on shell, i.e. for $%
p_{1,2n+1}^{2}\rightarrow 0$, the Adler zero is in fact present at $%
M_{2m+1}|_{\mathrm{on~~shell}}$ $=-\mathcal{M}_{2m+2}$ for all momenta, i.e.
\begin{equation}
\lim_{t\rightarrow 0}M_{2m+1}(p_{1},p_{2},\ldots ,tp_{j},\ldots p_{2m+1})|_{%
\mathrm{on~~shell}}=0  \label{adler_for_M}
\end{equation}%
for all $j=1,\ldots ,2m+1$ due to the cyclicity of $\mathcal{M}_{2m+2}$.

Let us now substitute $p_{2l-1}\rightarrow tp_{2l-1}$ to the right hand side
of (\ref{explicit_BCFW}). Note that, under such substitution, the position
of the poles $z_{2j,2j+2}\allowbreak $, $z_{2j-1,2j+1}$ and $z_{i,j}^{\pm }$
become $t-$dependent. The $t-$ dependence of the right hand side of (\ref%
{explicit_BCFW}) is therefore both explicit (due to the explicit dependence
on $p_{2l-1}$) and implicit (due to the implicit $t-$dependence of the poles
$z_{P}$).

We will now inspect the behavior of the individual terms under the limit $%
t\rightarrow 0$. The first term gives finite limit
\begin{equation}
\frac{1}{(2F^{2})^{n}}p_{-}^{2}\rightarrow \frac{1}{(2F^{2})^{n}}%
p_{-}^{2}|_{p_{2l-1}\rightarrow 0}.  \label{first}
\end{equation}%
\begin{figure}[t]
\begin{center}
\epsfig{width=0.8\textwidth,figure=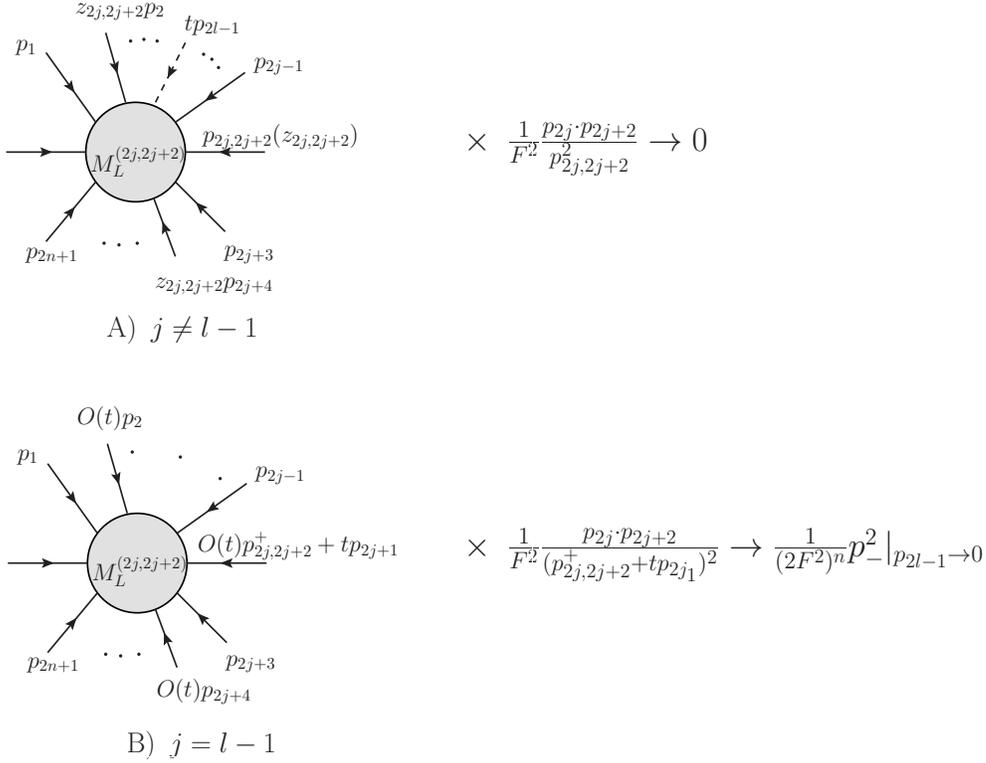}
\end{center}
\caption{Graphical representation of the $t\rightarrow 0$ limit of the
second term on the right hand side of (\protect\ref{explicit_BCFW}). The
soft momentum is denoted by dashed line in the case A. In the case B, $O(t)$
indicates the order of the $t-$dependent $z_{2j,2j+2}$. }
\label{figureAdler}
\end{figure}
As far as the second term is concerned, the individual terms of the sum over
$j$ vanish in this limit unless $j=l-1$. The reason is as follows. For $%
j\neq l-1$ (the case A in the Figure \ref{figureAdler}), the kinematical
factor $p_{2j}\cdot p_{2j+2}/p_{2j,2j+2}^{2}\ $as well as the position of
the pole $z_{2j,2j+2}$ are $t-$independent and because $tp_{2l-1}$ is placed
on the \emph{odd} position in $M_{L}^{(2j,2j+2)}(z_{2j,2j+2})$, we can safely%
\footnote{%
Indeed, in general the momenta $p_{k}(z_{2j,2j+2})$ and $%
p_{2j,2j+2}(z_{2j,2j+2})$ are $t-$independent and nonzero$.$} use the
induction hypothesis to conclude that%
\begin{equation*}
\lim_{t\rightarrow 0}M_{L}^{(2j,2j+2)}(z_{2j,2j+2})|_{p_{2l-1}\rightarrow
0}=0.
\end{equation*}%
For $j=l-1$ (the case B in the Figure \ref{figureAdler}), the kinematical
factor $p_{2j}\cdot p_{2j+2}/p_{2j,2j+2}^{2}$ becomes explicitly $t-$%
dependent and tends to $1/2$ for $t\rightarrow 0$, while $%
M_{L}^{(2j,2j+2)}(z_{2j,2j+2})$ has both explicit (through $%
p_{2j,2j+2}=z_{2j,2j+2}(p_{2j}+p_{2j+2})+tp_{2j+1}$) and implicit $t-$%
dependence. In this case $z_{2j,2j+2}=O(t)$, as can be seen from (\ref%
{z2j2jp2min}). Therefore, all \emph{even} momenta in $%
M_{L}^{(2j,2j+2)}(z_{2j,2j+2})$ are scaled by $O(t)$ factor, in the same way
as in (\ref{suda_Ot0}). We can therefore conclude with help of (\ref%
{suda_Ot0}) that%
\begin{equation}
\lim_{t\rightarrow 0}M_{L}^{(2j,2j+2)}(z_{2j,2j+2})\frac{1}{p_{2j,2j+2}^{2}}%
\frac{p_{2j}\cdot p_{2j+2}}{F^{2}}=\delta _{j,l-1}\frac{1}{(2F^{2})^{n}}%
p_{-}^{2}|_{p_{2l-1}\rightarrow 0}.  \label{second}
\end{equation}%
\begin{figure}[t]
\begin{center}
\epsfig{width=0.8\textwidth,figure=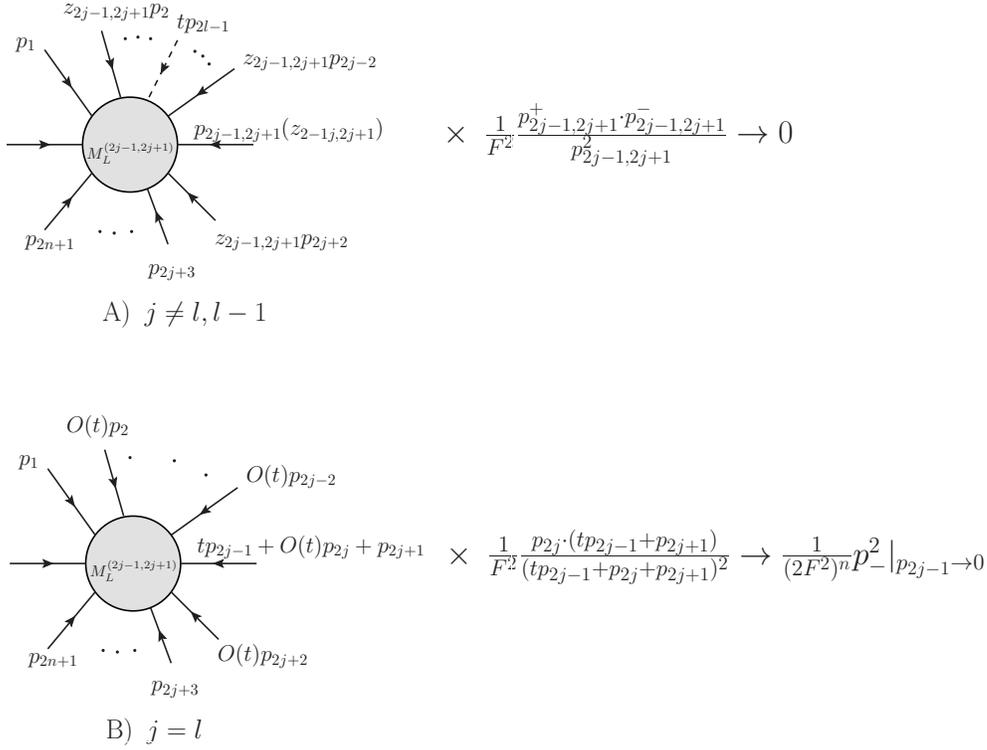}
\end{center}
\caption{Graphical representation of the $t\rightarrow 0$ limit of the third
term on the right hand side of (\protect\ref{explicit_BCFW}). The soft
momentum is denoted by dashed line in the picture A. In the picture B, we
show only the $j=l$ case, the $j=l-1$ case is treated analogously. $O(t)$
indicates the order of the $t-$dependent $z_{2j,2j+2}$. }
\label{figureAdler1}
\end{figure}
The third term on the right hand side of (\ref{explicit_BCFW}) can be
treated exactly in the same way as the second (see Fig. \ref{figureAdler1}).
Also here the individual terms of the sum over $j$ do not contribute with
the only exception of $j=l~$and $j=l-1$ by induction hypothesis applied to $%
M_{L}^{(2j-1,2j+1)}(z_{2j-1,2j+1})$ which has for $j\neq l,l-1$ only
explicit $t-$dependence. In the remaining two cases $j=l~$and $j=l-1$, the
explicitly $t-$dependent kinematical factors $p_{2j-1,2j+1}^{+}\cdot
p_{2j-1,2j+1}^{-}/p_{2j-1,2j+1}^{2}$ tend again to $1/2$ and within $%
M_{L}^{(2j-1,2j+1)}(z_{2j-1,2j+1})$ the even momenta are scaled by $%
z_{2j-1,2j+1}=O(t)$ (see (\ref{z2jm1})) and thus (\ref{suda_Ot0}) can be used%
\footnote{%
Note that, the odd momenta are $t-$idependent with the only exception of $%
p_{2j-1,2j+1}(z_{2j-1,2j+1})|_{p_{2j\mp 1}\rightarrow tp_{2j\mp 1}}$ the
limit of which is $p_{2j\pm 1}$.} to conclude that%
\begin{equation}
\lim_{t\rightarrow 0}M_{L}^{(2j-1,2j+1)}(z_{2j-1,2j+1})\frac{1}{%
p_{2j-1,2j+1}^{2}}\frac{p_{2j-1,2j+1}^{+}\cdot p_{2j-1,2j+1}^{-}}{F^{2}}%
=\left( \delta _{j,l}+\delta _{j,l-1}\right) \frac{1}{(2F^{2})^{n}}%
p_{-}^{2}|_{p_{2l-1}\rightarrow 0}.  \label{third}
\end{equation}%
The fourth term on the right hand side of (\ref{explicit_BCFW}) vanish
completely in the limit $t\rightarrow 0$. \ This is easy to see for those
terms of the sum over $(i,j)$ for which\footnote{%
It is easy to realize that $\lim_{t\rightarrow 0}z_{i,j}^{+}\neq
\lim_{t\rightarrow 0}z_{i,j}^{-}$ for generic $p_{k}$.} $\lim_{t\rightarrow
0}z_{i,j}^{\pm }$ $\neq 0$. In this case either $M_{L}^{(i,j)}(z_{i,j}^{\pm
})$ or $M_{R}^{(i,j)}(z_{i,j}^{\pm })$ have explicit $t-$dependence through $%
tp_{2l-1}$ (which is for $M_{L}^{(i,j)}(z_{i,j}^{+})$ on odd position) and
thus the induction hypothesis in the form (\ref{adler_zero}) or (\ref%
{adler_for_M}) can be used\footnote{%
Let us remind that $M_{R}^{(i,j)}(z_{i,j}^{+})$ is fully on-shell$.$}. By
direct inspection of (\ref{poly}) we find that the only case for which the
above argumentation does not apply is the case $j-i=4$ with $i$ even and $%
i\leq 2l-1\leq j$. Here $\lim_{t\rightarrow 0}z_{i,j}^{-}\neq 0$ and so for
the \textquotedblleft minus'' part of this $(i,j)$ term we
can use the induction hypothesis as above. However, the \textquotedblleft
plus'' part might be problematic because
\begin{equation}
z_{i,j}^{+}=-\frac{(p_{2l-1}\cdot p_{2l-1\pm 2})}{(p_{2l-1\pm 2}\cdot
p_{i,j}^{+})}t+O(t^{2}).
\end{equation}
Using this formula and (\ref{J_5}) we find after some algebra
\begin{equation}
M_{R}^{(i,j)}(z_{i,j}^{+})=M_{5}(p_{i}(z_{i,j}^{t+}),\ldots tp_{2l-1},\ldots
,p_{j}(z_{i,j}^{t+}))=O(t^{2}).
\end{equation}%
which shows that also the \textquotedblleft plus'' part has
vanishing $t\rightarrow 0$ limit.

Putting therefore the only nonzero contributions (\ref{first}), (\ref{second}%
) and (\ref{third}) together we get finally%
\begin{eqnarray*}
&&\lim_{t\rightarrow 0}M_{2n+1}(p_{1},p_{2},\ldots
,p_{2l-2},tp_{2l-1},p_{2l+1},\ldots ,p_{2n+1}) \\
&=&\frac{1}{(2F^{2})^{n}}p_{-}^{2}|_{p_{2l-1}\rightarrow 0}\left(
1+\sum_{j=1}^{n-1}\delta _{j,l-1}-\sum_{j=1}^{n}\left( \delta _{j,l}+\delta
_{j,l-1}\right) \right) =0,
\end{eqnarray*}%
which finishes the proof.

\subsection{Double-soft limit}

Let us now study the behavior of the semi-on-shell amplitude $J_{2n+1}$ in
the Cayley parametrization under the double soft limit, i.e. the case when
two external momenta, say $p_{i}$ and $p_{j}$, are scaled according to $%
p_{i,j}\rightarrow tp_{i,j}$ and $t$ is sent to zero. In this section we
will prove, that for $1<i<j<2n+1$%
\begin{eqnarray}
&&\lim_{t\rightarrow 0}J_{2n+1}(p_{1},\ldots ,p_{2n+1})|_{p_{i}\rightarrow
tp_{i},p_{j}\rightarrow tp_{j}}  \notag \\
&=&\delta _{j,i+1}\frac{1}{2F^{2}}\left( \frac{(p_{i}\cdot p_{i+2})}{%
p_{i+2}\cdot (p_{i+1}+p_{i})}-\frac{(p_{i}\cdot p_{i-1})}{p_{i-1}\cdot
(p_{i+1}+p_{i})}\right) J_{2n-1}(p_{1},\ldots ,p_{i-1},p_{i+2}\ldots
,p_{2n+1}),  \label{dslres}
\end{eqnarray}%
which has an identical form as (\ref{stripped_double_soft})$\footnote{%
Indeed,
\begin{equation*}
\frac{(p_{i}\cdot p_{i+2})}{p_{i+2}\cdot (p_{i+1}+p_{i})}-\frac{(p_{i}\cdot
p_{i-1})}{p_{i-1}\cdot (p_{i+1}+p_{i})}=\frac{1}{2}\left( \frac{p_{i+2}\cdot
(p_{i}-p_{i+1})}{p_{i+2}\cdot (p_{i}-p_{i+1})}-\frac{p_{i-1}\cdot
(p_{i}-p_{i+1})}{p_{i-1}\cdot (p_{i}-p_{i+1})}\right) .
\end{equation*}%
}$. The key ingredient of the proof is the generalized form of the BCFW
representation mentioned in Section \ref{subtracted_BCFW} written for a
suitable two-parameter complex deformation of the amplitude $J_{2n+1}$. Such
a representation allows us to calculate the double soft limit with help of
the known behavior of the poles and corresponding residues in this limit.
Useful information on this behavior can be inferred from the statement (\ref%
{adler_zero}) concerning the Adler zeroes proved in the previous subsection.

The above mentioned deformation of $J_{2n+1}$ can be defined as the
following function of two complex variables $z$ and $t$
\begin{equation}
S_{i,j}^{n}(z,t)=J(p_{1},\ldots ,p_{2n+1})|_{p_{i}\rightarrow
tp_{i},p_{j}\rightarrow zp_{j}},
\end{equation}%
therefore%
\begin{equation}
S_{i,j}^{n}(1,1)=J_{2n+1}(p_{1},\ldots ,p_{2n+1})
\end{equation}%
Various types of the double soft limit correspond then to various ways of
taking the limit $(z,t)\rightarrow (0,0)$ in the double complex plane $(z,t)$%
; the limit (\ref{dslres}) corresponds to $\lim_{t\rightarrow
0}S_{i,j}^{n}(t,t)\equiv S_{i,j}^{n,0} $.

For $z\rightarrow \infty $ and $t>0$ fixed the following asymptotic behavior
holds
\begin{equation}
S_{i,j}^{n}(z,t)=O(z^{0}),
\end{equation}%
as can be easily proved e.g. by induction with help of the Berends-Giele
recursive relations (\ref{Berends_Giele_relations}). We can therefore write
the generalized BCFW relation with one subtraction in the form (\ref%
{generalized_BCFW})
\begin{equation}
S_{i,j}^{n}(z,t)=S_{i,j}^{n}(a,t)+\sum_{k,l}\frac{\mathrm{Res}\left(
S_{i,j}^{n};z_{k,l}(t)\right) }{z-z_{k,l}(t)}\frac{z-a}{z_{k,l}(t)-a}.
\label{S_ij_BCFW}
\end{equation}%
where $a\ne z_{k,l}(t)$ is a priory arbitrary, however, as we shall see in
what follows, appropriate choice of $a$ can simplify the calculation.

The poles $z_{k,l}(t)$ for $k\leq j\leq l$ correspond to the conditions $%
p_{k,l}^{2}|_{p_{i}\rightarrow tp_{i},p_{j}\rightarrow zp_{j}}=0$, or
explicitly%
\begin{equation}
z_{k,l}(t)=-\frac{p_{k,l}^{2}|_{p_{i}\rightarrow tp_{i},p_{j}\rightarrow 0}}{%
2(p_{j}\cdot p_{k,l})|_{p_{i}\rightarrow tp_{i}}}.
\end{equation}%
The residues at the poles $z_{k,l}(t)$ factorize
\begin{eqnarray}
\mathrm{Res}\left( S_{i,j}^{n};z_{k,l}(t)\right) &=&\frac{1}{2(p_{j}\cdot
p_{k,l})|_{p_{i}\rightarrow tp_{i}}}[J_{2n+1-(l-k)}(p_{1},\ldots
,p_{k-1},p_{k,l},p_{l+1},\ldots ,p_{2N+1})  \notag \\
&&\times M_{l-k+1}(p_{k},\ldots ,p_{l})|_{p_{i}\rightarrow
tp_{i},p_{j}\rightarrow zp_{j}}]|_{z\rightarrow z_{k,l}(t)},
\label{S_ij_residua}
\end{eqnarray}%
where $M_{l-k+1}$ is the cut amplitude (\ref{cut_amplitude}). Namely the
latter two formulae along with (\ref{adler_zero}) contain sufficient amount
of information for the calculation of the double soft limit.

Let us first assume $i<j$ where $i$ is odd and $j$ arbitrary. This choice is
a technical one, and as we shall see, the general case can be easily
obtained using the symmetry properties of the amplitude. In what follows we
set $a=1$ in (\ref{S_ij_BCFW}), the double soft limit then simplifies to%
\begin{eqnarray}
S_{i,j}^{n,0} \equiv \lim_{t\rightarrow 0}S_{i,j}^{n}(t,t)
=\lim_{t\rightarrow 0}\sum_{k,l}\frac{\mathrm{Res}\left(
S_{i,j}^{n};z_{k,l}(t)\right) }{t-z_{k,l}(t)}\frac{t-1}{z_{k,l}(t)-1},
\label{dsl}
\end{eqnarray}%
where we have used the existence of the Adler zero for $%
S_{i,j}^{n}(1,t)=J_{2n+1}(p_{1},\ldots ,tp_{i},\ldots ,p_{2n+1})$ and $i$
odd (cf. (\ref{adler_zero})).

\begin{figure}[h]
\begin{center}
\epsfig{width=1.00\textwidth,figure=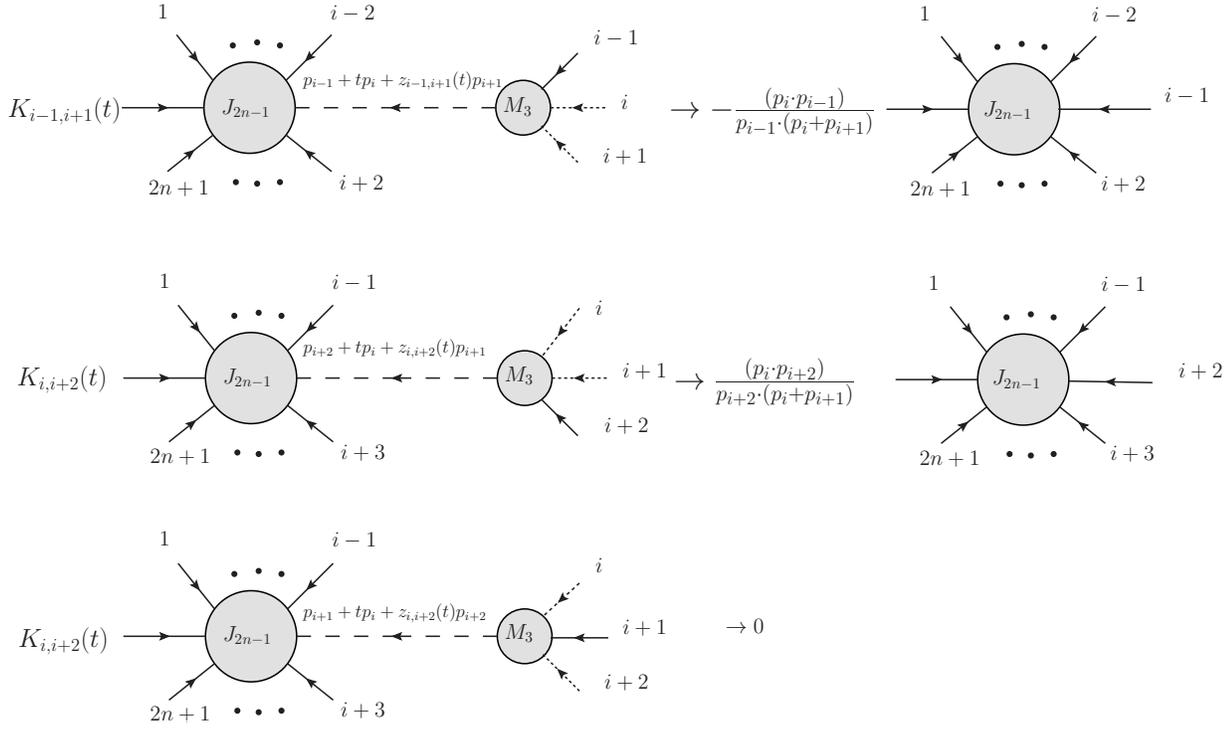}
\end{center}
\caption{Graphical representation of the $t\rightarrow 0$ limit of the three
cases (\protect\ref{adj1b}), (\protect\ref{adj1a}) and (\protect\ref{semiadj}%
) for which $z_{k,l}(t)\rightarrow 0$. The soft momenta are denoted by
dotted lines. The multiplicative factors $K_{k,l}(t)$ stays for $%
(t-1)/(t-z_{k,l}(t))(z_{k,l}(t)-1)$.}
\label{figuredsoft}
\end{figure}

For generic $p_{r}$ there exist a finite limit
\begin{equation}
z_{k,l}(0)=\lim_{t\rightarrow 0}z_{k,l}(t)\ne 1
\end{equation}%
In fact the only nonzero contributions to the right hand side of (\ref{dsl})
stem from the cases for which $z_{k,l}(0)=0$. Indeed, for $z_{k,l}(0)\ne 0$
we get for the corresponding contribution%
\begin{equation}
\frac{1 }{z_{k,l}(0)(z_{k,l}(0)-1)}\lim_{t\rightarrow 0}\mathrm{Res}\left(
S_{i,j}^{n};z_{k,l}(t)\right),
\end{equation}%
and, according to (\ref{adler_zero}), on the right hand side of (\ref%
{S_ij_residua}) we get either
\begin{equation}
\lim_{t\rightarrow 0}[M_{l-k+1}(p_{k},\ldots ,p_{l})|_{p_{i}\rightarrow
tp_{i},p_{j}\rightarrow zp_{j}}]|_{z\rightarrow z_{k,l}(t)}=0
\end{equation}%
for $k\leq i<j\leq l$ or
\begin{equation}
\lim_{t\rightarrow 0}J_{2n+1-(l-k)}(p_{1},\ldots ,tp_{i},\ldots
,p(k,l)(t),p_{k+1},\ldots ,p_{2n+1})=0
\end{equation}%
for $i<k<j\leq l$. In both cases the complementary factor has finite limit
and therefore
\begin{equation}
\lim_{t\rightarrow 0}\mathrm{Res}\left( S_{i,j}^{n};z_{k,l}(t)\right) =0.
\end{equation}%
Let us therefore discuss the contributions form the poles for which $%
z_{k,l}(0)=0$. Note that, for generic $p_r$ such a pole does not exist
provided $j>i+2$. We can therefore immediately conclude
\begin{equation}
S_{i,j}^{n,0}=0~~~\mathrm{for}~~~j>i+2.
\end{equation}
What remains are the following two alternatives for which the three-particle
poles $z_{k,l}(t)$ with $l=k+2$ can vanish in the limit $t\rightarrow 0$
(see Fig. \ref{figuredsoft})
\begin{enumerate}
\item $j=i+1$ and either $k=i$ or $k=i-1$. In this case either%
\begin{equation}
~p_{i-1,i+1}^{2}|_{p_{i}\rightarrow tp_{i},p_{j}\rightarrow 0}\rightarrow
p_{i-1}^{2}=0  \label{adj1b}
\end{equation}
or%
\begin{equation}
p_{i,i+2}^{2}|_{p_{i}\rightarrow tp_{i},p_{j}\rightarrow 0}\rightarrow
p_{i+2}^{2}=0  \label{adj1a}
\end{equation}
\item $j=i+2$ and $k=i$, in this case%
\begin{equation}
p_{i,i+2}^{2}|_{p_{i}\rightarrow tp_{i},p_{j}\rightarrow 0}=p_{i+1}^{2}=0.
\label{semiadj}
\end{equation}
\end{enumerate}
In what follows we will discuss separately the cases $j=i+1$ and $j=i+2$.
Let us first study the double soft limit of two adjacent momenta, i.e. $%
j=i+1 $ where $i$ is odd. We will investigate the contributions of
individual poles $z_{k,l}(t)$ on the right hand side of (\ref{dsl})
separately. In this case we get for $i>1$ only two potentially nonzero
contributions (i.e. (\ref{adj1a}) and (\ref{adj1b})) to the right hand side
of (\ref{dsl}), namely
\begin{equation}
S_{i,i+1}^{n,0}=\lim_{t\rightarrow 0}\frac{\mathrm{Res}\left(
S_{i,i+1}^{n};z_{i-1,i+1}(t)\right) }{t-z_{i-1,i+1}(t)}\frac{t-1}{%
z_{i-1,i+1}(t)-1}+\lim_{t\rightarrow 0}\frac{\mathrm{Res}\left(
S_{i,i+1}^{n};z_{i,i+2}(t)\right) }{t-z_{i,i+2}(t)}\frac{t-1}{z_{i,i+2}(t)-1}%
.  \label{ij_adjacent}
\end{equation}%
We get for the poles $z_{i-1,i+1}(t)$ and $z_{i,i+2}(t)$
\begin{equation}
z_{k,k+2}(t)=-\frac{p_{k,k+2}^{2}|_{p_{i}\rightarrow tp_{i},p_{j}\rightarrow
0}}{2(p_{j}\cdot p_{k,k+2})|_{p_{i}\rightarrow tp_{i}}}=-t\frac{(p_{i}\cdot
p_{r})}{(p_{j}\cdot p_{r})}+O(t^{2}),  \label{z_kl_t}
\end{equation}%
where either $r=i+2$ (for $k=i$) or $r=i-1$ (for $k=i-1$), and as a
consequence,%
\begin{equation}
\frac{1}{t-z_{k,k+2}(t)}\frac{t-1}{z_{k,k+2}(t)-1}=\frac{1}{t}\frac{%
(p_{j}\cdot p_{r})}{p_{r}\cdot (p_{j}+p_{i})}(1+O(t)).  \label{1_t_z}
\end{equation}%
We have further
\begin{equation}
p_{k,k+2}(t)=tp_{i}+z_{k,k+2}(t)p_{j}+p_{r}\rightarrow p_{r}\neq 0
\label{p_kl_t}
\end{equation}%
and therefore in both cases%
\begin{equation}
\lim_{t\rightarrow 0}J_{2n-1}(p_{1},\ldots
,p_{k-1},p_{k,k+2}(t),p_{k+3},\ldots ,p_{2n+1})=J_{2n-1}(p_{1},\ldots
,p_{i-2},p_{i-1},p_{i+2},\ldots ,p_{2n+1}).  \label{J_t}
\end{equation}%
For the remaining ingredients of the formula (\ref{S_ij_residua}) we get
\begin{eqnarray}
M_{3}(tp_{i},z_{i,i+2}(t)p_{i+1},p_{i+2}) &=&\frac{1}{F^{2}}t(p_{i}\cdot
p_{i+2}) \\
M_{3}(p_{i-1},tp_{i},z_{i-1,i+1}(t)p_{i+1}) &=&\frac{1}{F^{2}}%
z_{i-1,i+1}(t)(p_{i-1}\cdot p_{i+1})=-t\frac{1}{F^{2}}(p_{i}\cdot
p_{i-1})(1+O(t)).
\end{eqnarray}%
Inserting this into the formulae (\ref{S_ij_residua}) and (\ref{ij_adjacent}%
) get finally for $i>1$
\begin{equation}
S_{i,i+1}^{n,0}=\frac{1}{2F^{2}}\left( \frac{(p_{i}\cdot p_{i+2})}{%
p_{i+2}\cdot (p_{i+1}+p_{i})}-\frac{(p_{i}\cdot p_{i-1})}{p_{i-1}\cdot
(p_{i+1}+p_{i})}\right) J_{2n-1}(p_{1},\ldots
,p_{i-2},p_{i-1},p_{i+2},\ldots ,p_{2n+1}).  \label{rezult_dsl}
\end{equation}%
In the same way, for $i=1$ only the first term on the right hand side of (%
\ref{rezult_dsl}) contributes.

Let us proceed to the case 2. when $j=i+2$ and $z_{i,i+2}(t)\rightarrow 0$
for $t\rightarrow 0$ is the only pole which can give nonzero contribution to
(\ref{dsl}). In this case we have

\begin{equation}
S_{i,i+2}^{n,0}=\lim_{t\rightarrow 0}\frac{\mathrm{Res}\left(
S_{i,i+2}^{n};z_{i,i+2}(t)\right) }{t-z_{i,i+2}(t)}\frac{t-1}{z_{i,i+2}(t)-1}%
.
\end{equation}%
The formulae (\ref{z_kl_t}, \ref{1_t_z}, \ref{p_kl_t}, \ref{J_t}) are still
valid with $r=i+1$, but now we have%
\begin{equation}
M_{3}(tp_{i},p_{i+1},z_{i,i+2}(t)p_{i+2})=\frac{1}{F^{2}}tz_{i,i+2}(t)(p_{i}%
\cdot p_{i+2})=O(t^{2}).
\end{equation}%
which implies $S_{i,i+2}^{n,0}=0$.

To summarize, we have for $k>0$%
\begin{eqnarray}
&&\lim_{t\rightarrow 0}J_{2n+1}(p_{1},\ldots ,p_{2k},tp_{2k+1},\ldots
,tp_{j},\ldots ,p_{2n+1})=  \notag  \label{imj} \\
&=&\delta _{j,2k+2}\frac{1}{2F^{2}}J_{2n-1}(p_{1},\ldots
,p_{2k},p_{2k+3},\ldots ,p_{2n+1})\left( \frac{(p_{2k+1}\cdot p_{2k+3})}{%
p_{2k+3}\cdot (p_{2k+2}+p_{2k+1})}-\frac{(p_{2k+1}\cdot p_{2k})}{p_{2k}\cdot
(p_{2k+2}+p_{2k+1})}\right)  \notag \\
&&
\end{eqnarray}%
and for $k=0$
\begin{equation*}
\lim_{t\rightarrow 0}J_{2n+1}(tp_{1},\ldots ,tp_{j},\ldots ,p_{2n+1})=\delta
_{j,2}\frac{1}{2F^{2}}\frac{(p_{1}\cdot p_{3})}{(p_{2}\cdot
p_{3})+(p_{1}\cdot p_{3})}J_{2n-1}(p_{3},\ldots ,p_{2n+1}).
\end{equation*}%
As it is clear from the above discussion, the \textquotedblleft
asymmetry'' of the latter result stems from the fact that $%
p_{2n+2}$ is off-shell and therefore the three-particle pole corresponding
to $(p_{3}+p_{4}+\ldots +p_{2n+1})^{2}=(p_{2n+2}-p_{1}-p_{2})^{2}\rightarrow
p_{2n+2}^{2}\neq 0$ does not contribute.

Because%
\begin{equation}
J(1,2,\ldots ,2n+1)=J(2n+1,2n,\ldots ,2,1),
\end{equation}%
we get for $j<2k+1$%
\begin{equation}
J_{2n+1}(p_{1},\ldots ,tp_{j}\ldots ,p_{2k},tp_{2k+1},\ldots
,,p_{2n+1})=J_{2n+1}(p_{2n+1,}\ldots ,tp_{2k+1},p_{2k},\ldots ,tp_{j},\ldots
,p_{1}).
\end{equation}%
On the right hand side of this identity the momentum $p_{2k+1}$ stays on the
odd position and thus%
\begin{eqnarray}  \label{jmi}
&&\lim_{t\rightarrow 0}J_{2n+1}(p_{1},\ldots ,tp_{j}\ldots
,p_{2k},tp_{2k+1},\ldots ,p_{2n+1})  \notag \\
&=&\delta _{j,2k}\frac{1}{2F^{2}}J_{2n-1}(p_{1}\ldots
,p_{2k-1},p_{2k+2},\ldots ,p_{2n+1})\left( -\frac{(p_{2k}\cdot p_{2k-1})}{%
p_{2k-1}\cdot (p_{2k}+p_{2k+1})}+\frac{(p_{2k}\cdot p_{2k+2})}{p_{2k+2}\cdot
(p_{2k}+p_{2k+1})}\right)  \notag \\
\end{eqnarray}%
Putting (\ref{imj}) and (\ref{jmi}) together the final result (\ref{dslres})
follows.

\section{Summary and conclusion}

We have studied various aspects of the $SU(N)$ chiral nonlinear sigma model
which describes the low-energy dynamics of the Goldstone bosons
corresponding to the spontaneous chiral symmetry breaking $SU(N)\times
SU(N)\rightarrow SU(N)$. As we have shown, the tree-level scattering
amplitudes of the Goldstone bosons can be constructed from the stripped
amplitudes, which are identical as those of the $U(N)$ chiral nonlinear
sigma model. It is therefore possible to use this correspondence and to
investigate both the $SU(N)$ and $U(N)$ cases on the same footing.
Especially we are allowed to choose any parametrization (field redefinition)
of the chiral unitary matrix $U(x)$ entering the Lagrangian from the wide
class of parametrizations admissible for the extended $U(N)$ case, because
the fully on-shell stripped amplitudes do not depend on the parametrization.
For the direct calculation of the flavor ordered Feynman graphs, the most
convenient choice proved to be the minimal parametrization (\ref{minimal_parametrization_text}), which we have
chosen in order to calculate the on-shell amplitudes up to 10 Goldstone
bosons.

The proliferation of the Feynman graphs with increasing number of the
Goldstone bosons call for alternative methods of calculation. The more
efficient method is based on the Berends-Giele recursive relations for the
semi-on-shell amplitudes, but due to the infinite number of the interaction
vertices in the Lagrangian of the nonlinear sigma model, the number of
terms necessary to evaluate the $n-$point amplitude grows much faster (exponentially) with $%
n $ than for the case of the power-counting renormalizable theories (where the growth is polynomial).

The BCFW recursive relations could make the calculation of the on-shell
stripped amplitude as effective as for the renormalizable theories at least
as far as the number of terms (which is in both cases related to the number of factorization channels) is concerned. However, the standard way of the
BCFW reconstruction is not directly applicable for the nonlinear sigma
model because of the bad behavior of the BCFW deformed amplitudes at
infinity. We have therefore proposed an alternative deformation of the
semi-on-shell amplitudes based on the scaling of all odd or all even
momenta, for which we were able to prove exact results concerning the
behavior of the semi-on-shell amplitudes when the scaling parameter tended
to zero. Using the Berends-Giele recursive relations we were able to prove
this scaling properties for general $n-$point amplitude. An essential
ingredient of the proof was the fact that the semi-on-shell amplitudes (unlike the on-shell ones) are
parametrization dependent and we could therefore make an appropriate choice
of the parametrization (the Cayley one). We have then used these exact
scaling properties for a generalized BCFW reconstruction formula (with one
subtraction) which determines fully all the semi-on-shell amplitudes in the
Cayley parametrization including the basic four-point one.  Putting then the semi-on-shell amplitudes on-shell we reconstruct simply the parametrization independent on-shell amplitudes. In contrast to
the standard BCFW relations our procedure is not restricted to $d\geq 4$
space-time dimensions.

The BCFW recursive relation are also a suitable tool for investigation of
the properties of the amplitudes. We have illustrated this in two cases,
namely we have proved the presence of the Adler zero and established the
general form of the double soft limit for the semi-on-shell amplitudes in
the Cayley parametrization.

The existence of BCFW recursion relations for power-counting
non-renormalizable effective theory as the $SU(N)$ chiral nonlinear sigma
model gives an evidence that the on-shell methods can be used for much
larger classes of theories than has been considered so far. It also
indicates that the $SU(N)$ chiral nonlinear sigma model is rather special
and deeper understanding of all its properties is desirable. For future
directions, it would be interesting to see whether the construction can be
re-formulated purely in terms of on-shell scattering amplitudes not using
the semi-on-shell ones. Next possibility is to focus on loop amplitudes. As
was shown in \cite{ArkaniHamed:2010kv} the loop integrand can be also in
certain cases constructed using BCFW recursion relations, it would be
spectacular if the similar construction can be applied for effective field
theories.

\section*{Acknowledgement}

We would like to thank Nima Arkani-Hamed and David McGady for useful
discussions and comments on
the manuscript. JT is supported by NSF grant PHY-0756966. This work is
supported in part by projects MSM0021620859 of Ministry of Education of the
Czech Republic and GAUK- 514412.

\appendix

\section{General parametrization\label{parametrization_appendix}}

In this Appendix we will discuss a very general class of parameterizations
of the $U(N)$ sigma model originally studied in \cite{Cronin:1967jq}, which
is suited for a derivation of the stripped Feynman rules. Within this class
the field $U(x)\in U(N)$ is expressed in the form
\begin{equation}
U=\sum_{k=0}^{\infty }a_{k}\left( \sqrt{2}\frac{\mathrm{i}}{F}\phi \right)
^{k}  \label{general_U}
\end{equation}%
where $\phi =t^{a}\phi ^{a}$ , $\phi ^{a}$ are the Goldstone boson fields, $%
t^{a}$ are the $U(N)$ generators normalized according to $\langle
t^{a}t^{b}\rangle =\delta ^{ab}$ and $a_{k}$ are real coefficients. These
coefficients are not completely arbitrary, because the unitarity condition $%
U^{+}U=1$ implies the following constraint
\begin{equation}
\sum_{k=0}^{n}a_{k}a_{n-k}(-1)^{k}=\delta _{n,0}.  \label{unitarita_a}
\end{equation}%
For $n=0$ we get $a_{0}^{2}=1$ and without lose of generality we can set $%
a_{0}=1$. In order to preserve the correct normalization of the kinetic term
and to keep the interpretation of $F$ as the decay constant for the fields $%
\phi ^{a}$ we have to fix also $a_{1}=1$.

For $n$ odd the relations (\ref{unitarita_a}) are satisfied automatically
while for $n=2k$ we can solve them for $a_{2k}$ and get a recurrent formula
for the even coefficients expressed in terms of the odd ones

\begin{equation}
a_{2k}=-\frac{(-1)^{k}}{2}a_{k}^{2}-\sum_{j=1}^{k-1}(-1)^{j}a_{j}a_{2k-j}.
\label{a_recursive}
\end{equation}%
This gives up to $k=3$
\begin{eqnarray}
a_{2} &=&\frac{1}{2}a_{1}^{2}=\frac{1}{2}  \notag \\
a_{4} &=&-\frac{1}{2}a_{2}^{2}+a_{1}a_{3}=-\frac{1}{8}+a_{3}  \notag \\
a_{6} &=&\frac{1}{2}a_{3}^{2}+a_{1}a_{5}-a_{2}a_{4}=\frac{1}{16}-\frac{1}{2}%
a_{3}+\frac{1}{2}a_{3}^{2}+a_{5}
\end{eqnarray}

The explicit solution of the recurrent relations (\ref{a_recursive}) to all
orders can be easily found by means of the following trick. Let us introduce
the generating function $f(x)$ of the above coefficients $a_{k}$
\begin{equation}
f(x)=\sum_{k=0}^{\infty }a_{k}x^{k}.
\end{equation}%
The relations of unitarity with the initial conditions $a_{0}=a_{1}=1$ are
then equivalent to
\begin{equation}
f(-x)f(x)=1,~f(0)=1,~f^{^{\prime }}(0)=1  \label{ff_unitarita}
\end{equation}%
which represents a functional equations for the generating functions $f(x)$.
Let us define $f_{\pm }(x)$ to be the even and odd part of $f(x)$, i.e. $%
f_{\pm }(x)=\left( f(x)\pm f(-x)\right) /2$. From (\ref{ff_unitarita}) we get
then
\begin{equation}
f_{+}(x)^{2}-f_{-}(x)^{2}=1
\end{equation}%
or finally%
\begin{equation}
f_{+}(x)=\sqrt{1+f_{-}(x)^{2}}.
\end{equation}%
The formal series expansion of both sides of the last equation at $x=0$
gives the solution of the recurrent relations (\ref{a_recursive}), i.e. the
explicit expressions for $a_{2k}$ in terms of an infinite number of free
parameters $a_{2k+1}$. The general solution of the functional equation (\ref%
{ff_unitarita}) is then%
\begin{equation}
f(x)=f_{-}(x)+\sqrt{1+f_{-}(x)^{2}}
\end{equation}%
where $f_{-}(x)$ is arbitrary odd real function analytic for $x=0$
satisfying $f^{\prime }(0)=1$. The minimal parameter-free solution
corresponds to the choice $a_{2k+1}=0$ for $k>0$, i.e. $f_{-}^{\min }(x)=x$
and%
\begin{equation}
f_{\min }(x)=x+\sqrt{1+x^{2}}
\end{equation}%
i.e. for $k\geq 1$%
\begin{equation}
a_{2k}^{\min }=\frac{(-1)^{k+1}}{2^{2k-1}}C_{k-1},
\end{equation}%
where
\begin{equation}
C_{n}=\frac{1}{n+1}\left(
\begin{array}{c}
2n \\
n%
\end{array}%
\right)  \label{catalan}
\end{equation}%
are the Catalan numbers.

Another frequently used choices are the exponential and Cayley
parameterizations corresponding to $f_{\exp }(x)$ and $f_{\mathrm{Cayley}%
}(x) $ respectively, where
\begin{eqnarray}
f_{\exp }(x) &=&\mathrm{e}^{x} \\
f_{\mathrm{Cayley}}(x) &=&\frac{1+(x/2)}{1-(x/2)},
\end{eqnarray}%
or in terms of the coefficients $a_{k}$
\begin{eqnarray}
a_{k}^{\exp } &=&\frac{1}{k!}  \label{exp_a} \\
a_{k}^{\mathrm{Cayley}} &=&\frac{1}{1+\delta _{k,0}}\frac{1}{2^{k-1}}.
\label{cayley_a}
\end{eqnarray}%
These two parameterizations can be understood as minimal parameter-free
variants with respect to other two possible forms of the general solutions
of the functional equation (\ref{ff_unitarita}), namely%
\begin{equation}
f(x)=\exp g(x)
\end{equation}%
and
\begin{equation}
f(x)=\frac{h(x)}{h(-x)}
\end{equation}%
where $g(x)$ and $h(x)$ are arbitrary real functions analytic for $x=0$ for
which
\begin{eqnarray}
g(x) &=&-g(-x), \\
g(0) &=&0,~~g^{\prime }(0)=1
\end{eqnarray}%
and
\begin{equation}
h^{\prime }(0)=\frac{1}{2}h(0)\neq 0.
\end{equation}

As was proved in \cite{Cronin:1967jq}, for $N>2$ the only parametrization
from the class (\ref{general_U}) admissible also for $SU(N)$ sigma model is
the exponential one. The reason is that, under the general axial $SU(N)$
transformation%
\begin{equation}
U(x)^{\prime }=\sum_{k=0}^{\infty }a_{k}\left( \sqrt{2}\frac{\mathrm{i}}{F}%
\phi ^{\prime }\right) ^{k}=U_{A}\sum_{k=0}^{\infty }a_{k}\left( \sqrt{2}%
\frac{\mathrm{i}}{F}\phi \right) ^{k}U_{A}
\end{equation}%
which defines corresponding nonlinear transformation of the matrix of the
Goldstone boson fields $\phi =\sum_{a=1}^{N^{2}-1}\phi ^{a}t^{a}$ the $SU(N)$
condition for the trace $\langle \phi ^{\prime }\rangle =0$ is not preserved
unless $a_{k}=1/k!$. Of course, in the case  $N>2$ we can use different admissible parameterizations of $SU(N)$ which, however, do not belong to the class  (\ref{general_U}) (see e.g. \cite{Bijnens:2013yca}).

Let us now find the stripped Feynman rules. Using the general
parametrization (\ref{general_U}) we can write the Lagrangian of the
nonlinear $U(N)$ sigma model in the expanded form

\begin{equation}
\mathcal{L}^{(2)}=\frac{F^{2}}{4}\langle \partial U\cdot \partial
U^{+}\rangle =\sum_{n,m=0}^{\infty }v_{n,m}\langle \partial \phi \phi
^{n}\cdot \partial \phi \phi ^{m}\rangle .
\end{equation}%
where we get for $v_{n,m}$ after some algebra (and using the unitarity
condition (\ref{unitarita_a}))%
\begin{equation}
v_{n,m}=(1+(-1)^{n+m})\frac{(-\mathrm{i})^{n+m}}{4F^{n+m}}%
\sum_{k=0}^{m}a_{k}a_{m+n+2-k}(-1)^{k+1}(k-1-m)  \label{v_NM}
\end{equation}%
Therefore only the terms with even number of fields survive, explicitly%
\begin{equation}
\mathcal{L}^{(2)}=\sum_{n=0}^{\infty }\mathcal{L}_{2n+2}^{(2)}
\end{equation}%
where

\begin{equation}
\mathcal{L}_{2n+2}^{(2)}=\sum_{k=0}^{2n}v_{k,2n-k}\langle \partial \phi \phi
^{k}\cdot \partial \phi \phi ^{2n-k}\rangle
\end{equation}%
The usual Feynman rules for the vertices can be easily obtained as a sum
over permutations
\begin{eqnarray}
V_{2n+2}^{a_{1},\ldots ,a_{2n+2}}(p_{1},p_{2},\ldots ,p_{2n+1};p_{2n+2})
&=&-2^{n+1}\sum_{\sigma \in S_{2n+2}}\langle t^{a_{\sigma (1)}}\ldots
t^{a_{\sigma (2n+2)}}\rangle  \notag \\
&&\times \sum_{k=0}^{2n}v_{k,2n-k}(p_{\sigma (1)}\cdot p_{\sigma (1)+k+1})
\end{eqnarray}%
The stripped Feynman rule then follows in the form%
\begin{equation}
V_{2n+2}(p_{1},p_{2},\ldots
,p_{2n+1};p_{2n+2})=-2^{n+1}\sum_{k=0}^{2n}\sum_{i=1}^{2n+2}v_{k,2n-k}(p_{i}%
\cdot p_{i+k+1})
\end{equation}%
Inserting (\ref{exp_a}) into (\ref{v_NM}) we get after some algebra for the
exponential parametrization%
\begin{equation}
v_{k,2n-k}^{\exp }=\frac{(-1)^{n}}{2F^{2n}}\frac{(-1)^{k}}{(2n+2)!}\left(
\begin{array}{c}
2n \\
k%
\end{array}%
\right) .
\end{equation}%
while for the Cayley parametrization we have $v_{2k+1,2n-2k-1}^{\mathrm{%
Cayley}} =0$ and%
\begin{eqnarray}
v_{2k,2n-2k}^{\mathrm{Cayley}} &=&\frac{(-1)^{n}}{2F^{2n}}\frac{1}{2^{2n+1}}
.
\end{eqnarray}%
Similar calculations can be made also for the minimal parametrization, but
the result is much more lengthy and we will not need it explicitly. Instead
we will rewrite the Feynman rules for the vertex $V_{2n+2}$ with $2n+2$
external legs in terms of the variables%
\begin{equation}
s_{i,j}~=p_{i,j}^{2}
\end{equation}%
where $1\leq i<j\leq 2n+1$ and%
\begin{equation}
p_{i,j}=\sum_{k=i}^{j}p_{k}
\end{equation}%
Here we identify%
\begin{eqnarray}
s_{2n+2,2n+2+k} &=&s_{k+1,2n+1} \\
s_{i,2n+2+k} &=&s_{k+1,i-1}.
\end{eqnarray}%
The scalar products $(p_{i}\cdot p_{j})$ can be then expressed as
\begin{eqnarray}
(p_{i}\cdot p_{i}) &=&s_{i.i} \\
(p_{i}\cdot p_{i+1}) &=&\frac{1}{2}(s_{i,i+1}-s_{i,i}-s_{i+1,i+1})
\end{eqnarray}%
and for $k\geq 2$%
\begin{equation}
(p_{i}\cdot p_{i+k})=\frac{1}{2}%
(s_{i,i+k}-s_{i,i+k-1}+s_{i+1,i+k-1}-s_{i+1,i+k}).
\end{equation}%
On-shell we get $s_{i,i} =0 $ and $s_{1,2n+1}=0$. The stripped Feynman rule
in these variables can be written in the form valid for $n\geq 1$%
\begin{equation}
V_{2n+2}(s_{i,j})=(-1)^{n}\left( \frac{2}{F^{2}}\right)
^{n}\sum_{k=0}^{n}w_{k,n}\sum_{i=1}^{2n+2}s_{i,i+k}
\end{equation}%
where
\begin{eqnarray}
w_{0,n} &=&(-1)^{n}2F^{2n}\left( 2v_{0,2n}-v_{1,2n-1}\right) \\
w_{k,n} &=&(-1)^{n}2F^{2n}\left(
2v_{k,2n-k}-v_{k-1,2n+1-k}-v_{k+1,2n-1-k}\right) ~~\mathrm{for}~~k<n \\
w_{n,n} &=&(-1)^{n}2F^{2n}(v_{n,n}-v_{n-1,n+1}).
\end{eqnarray}%
Within the general parametrization we get from (\ref{v_NM}) and (\ref%
{unitarita_a}) after some algebra%
\begin{equation}
w_{k,n}=\frac{(-1)^{k}}{1+\delta _{kn}}a_{k+1}a_{2n+1-k}.
\end{equation}%
For the above special cases this reads for $N\geq 1$%
\begin{eqnarray}
w_{k,n}^{\exp } &=&\frac{(-1)^{k}}{1+\delta _{kn}}\frac{1}{(2n+2)!}\left(
\begin{array}{c}
2n+2 \\
k+1%
\end{array}%
\right) \\
w_{k,n}^{\mathrm{Cayley}} &=&\frac{(-1)^{k}}{1+\delta _{kn}}\frac{1}{2^{2n}}
\\
w_{0,n}^{\min } &=&w_{2k,n}^{\min }=0 \\
w_{2k+1,n}^{\min } &=&\frac{1}{1+\delta _{2k+1,n}}\frac{(-1)^{n}}{2^{2n}}%
C_{k}C_{n-k-1}.
\end{eqnarray}%
Note that, for the minimal parametrization the coefficients $w_{0,n}^{\min }$
at $s_{i,i}=p_{i}^{2}$ vanish, therefore the stripped Feynman rules for
vertices do not depend on the off-shellness of the momenta in this case.
This fact has been observed already in \cite{Ellis:1970nt} without
calculating the explicit Feynman rules.

\section{More examples of amplitudes\label{explicit_amplitudes_appendix}}

\begin{figure}[h]
\begin{center}
\epsfig{width=0.4\textwidth,figure=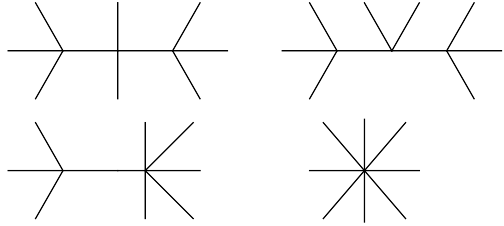}
\end{center}
\caption{Graphical representation of the 8-point amplitude (\protect\ref%
{m8pt}) with cycling tacitly assumed.}
\label{figure_8pt}
\end{figure}

\begin{figure}[t]
\begin{center}
\epsfig{width=0.6\textwidth,figure=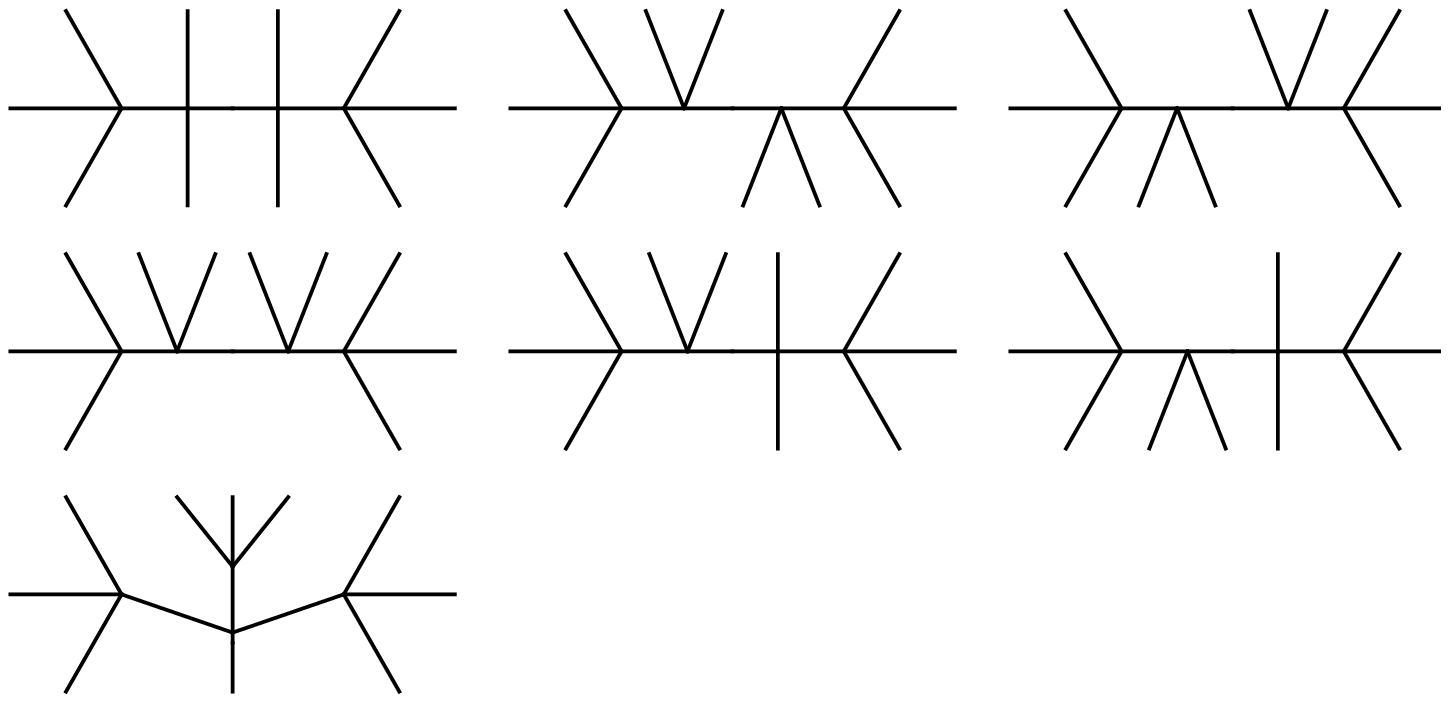}\vspace{0.4cm} %
\epsfig{width=0.6\textwidth,figure=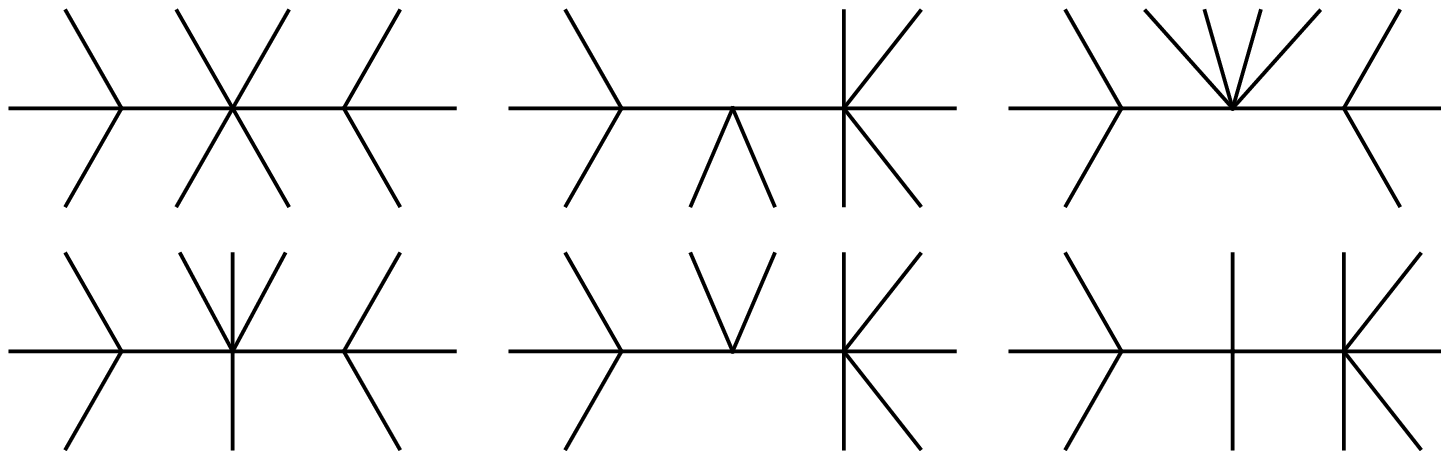}\vspace{0.4cm} %
\epsfig{width=0.6\textwidth,figure=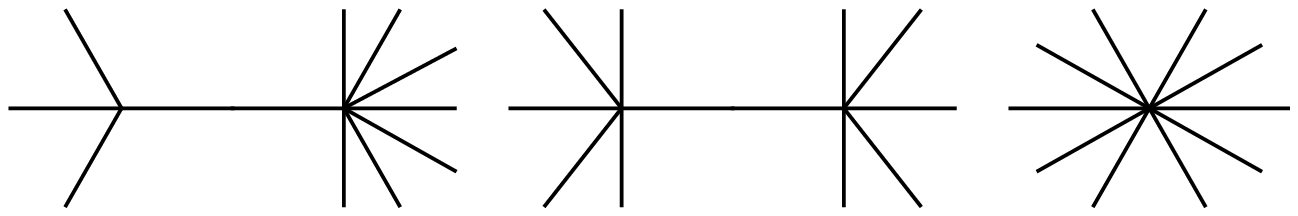}\vspace*{-0.4cm}
\end{center}
\caption{Graphical representation of the 10-point amplitude (\protect\ref%
{m10pt}) with cycling tacitly assumed.}
\label{figure_10pt}
\end{figure}

The eight-point amplitude is
\begin{align}
& 8F^{6}\mathcal{M}(1,2,3,4,5,6,7,8)=  \notag \\
& =\frac{1}{2}\frac{(s_{1,2}+s_{2,3})(s_{1,4}+s_{4,7})(s_{5,6}+s_{6,7})}{%
s_{1,3}s_{5,7}}+\frac{(s_{1,2}+s_{2,3})(s_{1,4}+s_{4,5})(s_{6,7}+s_{7,8})}{%
s_{1,3}s_{6,8}}  \notag \\
& \phantom{=}\,\,-\frac{%
(s_{1,2}+s_{2,3})(s_{4,5}+s_{4,7}+s_{5,6}+s_{5,8}+s_{6,7}+s_{7,8})}{s_{1,3}}%
+2s_{1,2}+\frac{1}{2}s_{1,4}+\text{cycl}  \label{m8pt}
\end{align}%
and graphically in Fig.~\ref{figure_8pt}.
Finally the ten-point amplitude is given by
\begin{align}
& 16F^{8}\mathcal{M}(1,2,3,4,5,6,7,8,9,10)=-\frac{s_{1,2}+s_{2,3}}{s_{1,3}}%
\Bigl\{  \notag \\
& \phantom{=+}\,\,\frac{1}{2}\frac{%
(s_{1,4}+s_{4,9})(s_{5,8}+s_{6,9})(s_{6,7}+s_{7,8})}{s_{5,9}s_{6,8}}+\frac{1%
}{2}\frac{(s_{1,4}+s_{4,5})(s_{1,8}+s_{6,9})(s_{6,7}+s_{7,8})}{s_{1,5}s_{6,8}%
}  \notag \\
& \phantom{=}\,\,+\frac{1}{2}\frac{%
(s_{1,8}+s_{4,9})(s_{4,5}+s_{5,8})(s_{6,7}+s_{7,8})}{s_{4,8}s_{6,8}}+\frac{%
(s_{1,4}+s_{4,5})(s_{1,6}+s_{6,7})(s_{1,8}+s_{8,9})}{s_{1,5}s_{1,7}}  \notag
\\
& \phantom{=}\,\,+\frac{(s_{1,4}+s_{4,5})(s_{1,6}+s_{6,9})(s_{7,8}+s_{8,9})}{%
s_{1,5}s_{7,9}}+\frac{(s_{1,8}+s_{4,9})(s_{4,7}+s_{5,8})(s_{5,6}+s_{6,7})}{%
s_{4,8}s_{5,7}}  \notag \\
& \phantom{=}\,\,+\frac{(s_{1,6}+s_{4,9})(s_{4,5}+s_{5,6})(s_{7,8}+s_{8,9})}{%
s_{4,6}s_{7,9}}-\frac{1}{2}\frac{%
(s_{1,4}+s_{1,8}+s_{4,5}+s_{4,9}+s_{5,8}+s_{6,9})(s_{6,7}+s_{7,8})}{s_{6,8}}
\notag \\
& \phantom{=}\,\,-\frac{%
(s_{1,8}+s_{4,9})(s_{4,5}+s_{4,7}+s_{5,6}+s_{5,8}+s_{6,7}+s_{7,8})}{s_{4,8}}
\notag \\
& \phantom{=}\,\,-\frac{%
(s_{1,4}+s_{1,6}+s_{4,5}+s_{4,7}+s_{5,6}+s_{6,7})(s_{1,8}+s_{8,9})}{s_{1,7}}
\notag \\
& \phantom{=}\,\,-\frac{%
(s_{1,4}+s_{1,6}+s_{4,5}+s_{4,9}+s_{5,6}+s_{6,9})(s_{7,8}+s_{8,9})}{s_{7,9}}
\notag \\
& \phantom{=}\,\,-\frac{%
(s_{1,4}+s_{4,5})(s_{1,6}+s_{1,8}+s_{6,7}+s_{6,9}+s_{7,8}+s_{8,9})}{s_{1,5}}
\notag \\
& \phantom{=}\,\,-\frac{%
(s_{1,4}+s_{4,9})(s_{5,6}+s_{5,8}+s_{6,7}+s_{6,9}+s_{7,8}+s_{8,9})}{s_{5,9}}
\notag \\
& \phantom{=}\,%
\,+2s_{1,4}+s_{1,6}+2s_{1,8}+2s_{4,5}+s_{4,7}+2s_{4,9}+2s_{5,6}+s_{5,8}+2s_{6,7}+s_{6,9}+2s_{7,8}+2s_{8,9}%
\Bigr\}  \notag \\
& \phantom{=}\,\,-\frac{1}{2}\frac{%
(s_{1,2}+s_{1,4}+s_{2,3}+s_{2,5}+s_{3,4}+s_{4,5})(s_{1,6}+s_{1,8}+s_{6,7}+s_{6,9}+s_{7,8}+s_{8,9})%
}{s_{1,5}}  \notag \\
& \phantom{=}\,\,+5s_{1,2}+2s_{1,4}+\text{cycl}  \label{m10pt}
\end{align}%
with one-to-one correspondence with Fig.~\ref{figure_10pt}

\section{Relative efficiency of  Feynman diagrams and Berends-Giele relations\label{graph_numbers}}

In this appendix we review the solution of several types of recursive
relations which count the number of ordered Feynman graphs needed for the
semi-on-shell amplitude $J(1,2,\dots ,n)$ in the nonlinear sigma model and
related toy models.

\subsection{Number of the Feynman graphs}

Let us start with the case of nonlinear sigma model, i.e. with the case
with infinite number of vertices in the interaction Lagrangian. The above
recursive relations, which determine the number $f(2n+1)$ of the (flavor
ordered) Feynman graphs which contribute to $J(1,2,\ldots ,2n+1)$, are
tightly related to the Berends-Giele relations (\ref{Berends_Giele_relations}%
). Indeed, after making the following substitution to (\ref%
{Berends_Giele_relations})
\begin{equation}
J(1,2,\ldots ,2n+1)\rightarrow f(2n+1),~~\frac{\mathrm{i}}{p_{2n+2}^{2}}%
\rightarrow 1,~~\mathrm{i}V_{2k+1}\rightarrow 0,~~\mathrm{i}V_{2k+2}=1,
\end{equation}%
the individual terms on the right hand side just count the number of Feynman
graphs generated from these terms by the iterations of the recursive
procedure. As a result we get for $f(2n+1)$ the following recursive relation%
\begin{equation}
f(2n+1)=\sum_{k=1}^{n}\sum_{\{n_{i}\}}\prod\limits_{i=1}^{2k+1}f(2n_{i}+1),
\label{kombinatorics1}
\end{equation}%
with the initial condition $f(1)=1$. In the above formula the sum over $%
\{n_{i}\}$ is constrained by the requirement%
\begin{equation}
\sum_{i=1}^{2k+1}(2n_{i}+1)=2n+1\Leftrightarrow \sum_{i=1}^{2k+1}n_{i}=n-k
\end{equation}%
i.e. it corresponds to the sum over all possible decompositions of ordered
set of $2n+1$ momenta to non-empty clusters with odd number of momenta in
each cluster (cf. (\ref{odd_zero}) and Fig. \ref{figure_0}), i.e. more
explicitly
\begin{equation}
f(2n+1)=\sum_{k=1}^{n}\sum_{\sum_{i}n_{i}=n-k}\prod%
\limits_{i=1}^{2k+1}f(2n_{i}+1),~~~f(1)=1.  \label{rec_a}
\end{equation}%
Standard method for solution of this type of recursive relation is based on
the generating function defined as%
\begin{equation}
A(x)=\sum_{n=0}^{\infty }f(2n+1)x^{n}.
\end{equation}%
The recursive formula (\ref{rec_a}) implies the following equation for $A(x)$%
\begin{equation}
A=1+\sum_{k=1}^{\infty }x^{k}A^{2k+1}=1+\frac{xA^{3}}{1-xA^{2}}
\end{equation}%
or%
\begin{equation}
x=\frac{B}{\left( B+1\right) ^{2}\left( 2B+1\right) }\equiv \frac{B}{g(B)}
\end{equation}%
where $B=A-1$ and $g(z)=(z+1)^{2}(2z+1)$.\ In this form, the problem is
prepared for the application of the Lagrange--B\"{u}rmann \ inversion formula%
\begin{equation}
B(x)=\sum_{n=0}^{\infty }\frac{x^{n}}{n!}\frac{\mathrm{d}^{n-1}}{\mathrm{d}%
z^{n-1}}g(z)^{n}|_{z=0}=\sum_{n=1}^{\infty }\frac{x^{n}}{n!}\frac{\mathrm{d}%
^{n-1}}{\mathrm{d}z^{n-1}}(z+1)^{2n}(2z+1)^{n}|_{z=0}.
\end{equation}%
After straightforward algebra with help of Leibnitz rule we get for $n\geq 1$%
\begin{equation}
f(2n+1)=\frac{2^{n-1}}{n}\sum_{k=0}^{n-1}\left(
\begin{array}{c}
n \\
k+1%
\end{array}%
\right) \left(
\begin{array}{c}
2n \\
k%
\end{array}%
\right) 2^{-k}=2^{n-1}~_{2}F_{1}\left( 1-n,-2n,2;\frac{1}{2}\right) ,
\end{equation}%
where $_{2}F_{1}(\alpha .\beta ,\gamma ;z)$ is the hypergeometric function.
In the same way one can solve the recurrence relations for the number of
ordered Feynman graphs for the semi-on-shell amplitudes $J(1,2,\ldots ,n)$
in the cases when only quadrilinear vertices (\textquotedblleft $\phi ^{4}$
theory''), only trilinear vertices (\textquotedblleft $\phi
^{3}$ theory'') or both trilinear and quadrilinear vertices
(\textquotedblleft $\phi ^{3}+\phi ^{4}$ theory'') are
present in the Lagrangian. In the first case, similarly to the nonlinear
sigma model, only $J(1,2,\ldots ,n)$ with $n$ odd can be nonzero, while in
the remaining two cases $J(1,2,\ldots ,n)$ both parities of $n$ are
generally allowed. Let us denote the number of the Feynman graphs for $%
J(1,2,\ldots ,n)$ as $f_{4}(n)$, $f_{2}(n)$ and $f_{3+4}(n)$ respectively.
We get the following recurrence relations
\begin{eqnarray}
f_{4}(2n+1) &=&\sum_{n_{1}+n_{2}+n_{3}=n-1,~n_{i}\geq
0}f_{4}(2n_{1}+1)f_{4}(2n_{2}+1)f_{4}(2n_{3}+1) \\
f_{3}(n) &=&\sum_{n_{1}+n_{2}=n,~n_{i}\geq 1}f_{3}(n_{1})f_{3}(n_{2}) \\
f_{3+4}(n) &=&\sum_{n_{1}+n_{2}=n,~n_{i}\geq 1}f_{3+4}(n_{1})f_{3+4}(n_{2})
\\
&&+\sum_{n_{1}+n_{2}+n_{3}=n,~n_{i}\geq
1}f_{3+4}(n_{1})f_{3+4}(n_{2})f_{3+4}(n_{3})
\end{eqnarray}%
with initial conditions $f_{j}(1)=1$, $j=3,4,3+4$. The corresponding
generating functions
\begin{equation}
A_{4}(x) =\sum_{n=0}^{\infty }f_{4}(2n+1)x^{n},\,\,\,\,\,\,\, A_{3,3+4}(x)
=\sum_{n=1}^{\infty }f_{3,3+4}(n)x^{n}
\end{equation}%
then satisfy
\begin{equation}
A_{4} =1+xA_{4}^{3},\,\,\,\,\, A_{3} =x+A_{3}^{2},\,\,\,\,\, A_{3+4}
=x+A_{3+4}^{2}+A_{3+4}^{3}.
\end{equation}%
In the second case we get%
\begin{equation}
A_{3}(x)=\frac{1-\sqrt{1-4x}}{2}=\frac{1}{2}\left( 1-\sum_{n=0}^{\infty
}\left(
\begin{array}{c}
1/2 \\
k%
\end{array}%
\right) (-4x)^{k}\right)
\end{equation}%
and therefore%
\begin{equation}
f_{3}(n)=\frac{1}{n}\left(
\begin{array}{c}
2(n-1) \\
n-1%
\end{array}%
\right) =C_{n-1}
\end{equation}%
where $C_{n}$ are the Catalan numbers.
\begin{table}[tbp]
{\small {\
\begin{tabular}{|l|r|r|r|r|r|r|r|r|r|r|}
\hline
$n$ & 2 & 3 & 4 & 5 & 6 & 7 & 8 & 9 & 10 & 11 \\ \hline
$f_{3}(n)$ & 1 & 2 & 5 & 14 & 42 & 132 & 429 & 1 430 & 4 862 & 16 796 \\
$f_{3+4}(n)$ & 1 & 3 & 10 & 38 & 154 & 654 & 2 871 & 12 925 & 59 345 & 276
835 \\
$f_{4}(2n+1)$ & 3 & 12 & 55 & 273 & 1 428 & 7 752 & 43 263 & 246 675 & 1 430
715 & 8 414 640 \\
$f(2n+1)$ & 4 & 21 & 126 & 818 & 5 594 & 39 693 & 289 510 & 2 157 150 & 16
348 960 & 125 642 146 \\ \hline
\end{tabular}%
}}
\caption{Number of flavor ordered Feynman graphs for $J(1,\dots ,n)$ and $%
J(1,\dots ,2n+1)$ in the models of the type $\protect\phi ^{3}$, $\protect%
\phi ^{3}+\protect\phi ^{4}$, $\protect\phi ^{4}$ and nonlinear sigma
model. }
\label{tab2}
\end{table}
In the first case, writing%
\begin{equation}
x=\frac{A_{4}-1}{A_{4}^{3}}=\frac{B_{4}}{(B_{4}+1)^{3}}
\end{equation}%
and using the Lagrange--B\"{u}rmann \ inversion formula we get for $n>0$%
\begin{equation}
f_{4}(2n+1)=\frac{1}{n!}\frac{\mathrm{d}^{n-1}}{\mathrm{d}z^{n-1}}%
(z+1)^{3n}|_{z=0}=\frac{1}{2n+1}\left(
\begin{array}{c}
3n \\
n%
\end{array}%
\right) .
\end{equation}%
In the third case, we get from
\begin{equation}
x=A_{3+4}\left( 1-A_{3+4}-A_{3}^{2}\right)
\end{equation}%
and using the Lagrange--B\"{u}rmann \ inversion formula%
\begin{eqnarray}
f_{3+4}(n) &=&\frac{1}{n!}\frac{\mathrm{d}^{n-1}}{\mathrm{d}z^{n-1}}\left(
\frac{1}{1-z-z^{2}}\right) ^{n}|_{z=0} =\frac{(-1)^{n}}{n!}\frac{\mathrm{d}%
^{n-1}}{\mathrm{d}z^{n-1}}\left( \frac{1}{z_{1}-z}\right) ^{n}\left( \frac{1%
}{z_{2}-z}\right) ^{n}|_{z=0}
\end{eqnarray}%
(where $z_{1}=-\phi $, $z_{2}=\phi $ $-1$ and $\phi =(1+\sqrt{5})/2$ is the
Golden ratio) the result
\begin{eqnarray}
f_{3+4}(n) &=&(-1)^{n+1}\frac{\phi ^{1-n}}{n}\sum_{k=0}^{n-1}\left(
\begin{array}{c}
n-1+k \\
k%
\end{array}%
\right) \left(
\begin{array}{c}
2(n-1)-k \\
n-1%
\end{array}%
\right) \left( \frac{\phi }{1-\phi }\right) ^{k} \nonumber \\
&=&\left( -\frac{4}{\phi }\right) ^{n-1}\Gamma \left( n-\frac{1}{2}\right)
~_{2}F_{1}\left( 1-n,n,2-2n;\frac{\phi }{1-\phi }\right) .
\end{eqnarray}%
The first twelve members of the above sequences are illustrated in the Table %
\ref{tab2}.

\subsection{Efficiency of the Berends-Giele relations}

We can compare this with the number of terms generated by Berends-Giele
recursion. For the nonlinear sigma model, the number of terms on the right
hand side of \ (\ref{Berends_Giele_relations}) is just%
\begin{equation}
t(2n+1)=\sum_{k=1}^{n}\sum_{\{n_{i}\}}1=\sum_{k=1}^{n}\left(
\begin{array}{c}
n+k \\
n-k%
\end{array}%
\right) =F_{2n+1}-1
\end{equation}%
where
\begin{equation}
F_{n}=\frac{1}{\sqrt{5}}\left( \phi ^{n}-(\phi -1)^{n}\right)
\end{equation}%
are the Fibonacci numbers and $\phi =(1+\sqrt{5})/2$ is the Golden ratio.
Therefore, using the known results for $J(1,2,\ldots ,2m+1)$ with $m<n$ at
each step, we need to evaluate altogether%
\begin{equation}
b(2n+1)=\sum_{m=1}^{n}t(2m+1)=\frac{1}{\sqrt{5}}\left( \phi ^{3}\frac{\phi
^{2n}-1}{\phi ^{2}-1}-(\phi -1)^{3}\frac{(\phi -1)^{2n}-1}{(\phi -1)^{2}-1}%
\right) -n
\end{equation}%
terms in order to calculate $J(1,2,\ldots ,2n+1)$ using the Berends-Giele
recursion. We show the sequences $t(2n+1)$ and $b(2n+1)$ in the first and
second row of Tab.1 respectively.

In the same way we can calculate analogous numbers $t_{j}(n)$ and $b_{j}(n)$
for $j=3,4,3+4$, i.e. for \textquotedblleft $\phi ^{3}$ theory%
'' , \textquotedblleft $\phi ^{3}$ theory''
or \textquotedblleft $\phi ^{3}+\phi ^{4}$ theory''. For
instance, for $t_{4}(2n+1)$ we have (see Tab. 1 for numerical values)%
\begin{equation}
t_{4}(2n+1)=\left(
\begin{array}{c}
n+1 \\
2%
\end{array}%
\right) ,~~~b_{4}(2n+1)=\sum_{m=1}^{n}t_{4}(2m+1)=\frac{1}{6}n(n+1)(n+2)
\end{equation}
Note the exponential growth of $t(2n+1)$ and $b(2n+1)$ with increasing $n$
in contrast to the only polynomial growth of $t_4(2n+1)$ and $b_4(2n+1)$ .

\section{Other example of scaling properties of the semi-on-shell amplitudes\label%
{extra_scaling}}

In this appendix we prove the following scaling limit
\begin{equation}
\lim_{t\rightarrow 0}J_{2n+1}(tp_{1},p_{2},tp_{3},p_{4},\ldots
,tp_{2r-1},tp_{2r},tp_{2r+1},\ldots ,p_{2N},tp_{2n+1})=0
\label{all_odd_one_even}
\end{equation}%
which is valid for for $n>1$. Let us note, however, that%
\begin{equation}
J_{3}(tp_{1},tp_{2},tp_{3})=J_{3}(p_{1},p_{2},p_{3})\neq 0.  \label{all4}
\end{equation}%
On the other hand, for $N=2$ \ we get by direct calculation
\begin{equation}
\lim_{t\rightarrow
0}J_{5}(tp_{1},tp_{2},tp_{3},p_{4},tp_{5})=\lim_{t\rightarrow
0}J_{5}(tp_{1},p_{2},tp_{3},tp_{4},tp_{5})=0
\end{equation}%
and we can therefore proceed by induction based on Berends-Giele relations
almost exactly as in the case of the proof of (\ref{licha_Ot2}). The only
modification here is that, along with the \textquotedblleft
dangerous'' contributions without blocks $J(j_{k}+1,\ldots
,j_{k+1})$ where $j_{k}$ is even and $j_{k+1}-j_{k}>1$ attached to the odd
line of the vertex $V_{m+1}$(provided at least one such a block is present,
the contribution vanish either by the induction hypothesis or by (\ref%
{licha_Ot2}) ) we have to discuss separately new type of \textquotedblleft
dangerous'' terms with building block $%
J(p_{2r-1},p_{2r},p_{2r+1})$ (this block does not vanish due to (\ref{all4}%
)). The \textquotedblleft old'' dangerous terms do not in
fact contribute as was already discussed within the proof of (\ref{licha_Ot2}%
). The \textquotedblleft new'' dangerous terms have the
following general form form
\begin{eqnarray}
&&\frac{\mathrm{i}}{p_{2N+2}^{2}}\mathrm{i}%
V_{2k+2}(p_{1},p_{2,2j_{1}},p_{2j_{1}+1},\ldots
,p_{2j_{l}+2,2r-2},p_{2r-1,2r+1},p_{2r+2,2j_{l+1}}\ldots \notag\\
&&\ldots ,p_{2j_{k-1},2n},p_{2n+1},-p_{1,2n+1}) \notag\\
&&\times J(p_{1})J(2,\ldots ,2j_{1})J(p_{2j_{1}+1})\ldots J(2j_{l}+2,\ldots
,2r-2) \notag\\
&&\times J(p_{2r-1},p_{2r},p_{2r+1})J(2r+2,\ldots ,2j_{l+1})\cdots
J(2j_{k-1},\ldots ,2n)J(p_{2n+1}).
\end{eqnarray}%
Note that, $p_{2r-1,2r+1}$ is attached to the odd line of the vertex $%
V_{2k+2}$ and scales as
\begin{equation}
p_{2r-1,2r+1}\rightarrow tp_{2r-1,2r+1}
\end{equation}%
i.e. in the same way as the remaining momenta attached to the odd lines of
the vertex. The vertex being proportional the squared sum of the odd line
momenta scales therefore as $O(t^{2})$, and the contribution of the
\textquotedblleft new'' dangerous terms vanish. This
finishes the proof.

\section{Double soft limit of Goldstone boson amplitudes\label%
{double_soft_appendix}}

In this appendix we will discuss the  properties of the on-shell scattering amplitudes of the Goldstone bosons, which are dictated by the symmetry, namely the limits of the amplitudes for soft external momenta.  Some
of these properties have been obtained in the special case of pions by PCAC methods in the late
sixties (see e.g.\cite{Dashen:1969ez}). Here we enlarge and reformulate them in a more general form appropriate for our purposes with stress on the proof of the double soft limit discussed recently for pions and ${\cal{N}}=8$ supergravity in \cite{ArkaniHamed:2008gz}.

Let us assume a general theory with spontaneous symmetry breaking according
to the pattern $G\rightarrow H$ where the homogeneous space $G/H$ is a
symmetric space, i.e. the vacuum little group $H$ is the maximal subgroup
invariant with respect to some involutive automorphism of $G$
(\textquotedblleft parity''). This implies the following
structure of the Lie algebra of $G$
\begin{eqnarray}
\lbrack T^{a},T^{b}] &=&\mathrm{i}f_{T}^{abc}T^{c}  \notag \\
\lbrack T^{a},X^{b}] &=&\mathrm{i}f_{X}^{abc}X^{c}  \notag \\
\lbrack X^{a},X^{b}] &=&\mathrm{i}F^{abc}T^{c}.
\end{eqnarray}%
Here $T^{a}$ and $X^{a}$ are the unbroken and broken generators respectively
and $f_{T}^{abc}$, $f_{X}^{abc}$ and $F^{abc}$ are the structure constants.
The chiral nonlinear sigma model is a special case for which $%
f_{T}^{abc}=f_{X}^{abc}=F^{abc}=f^{abc}$.

The invariance of the theory with respect to the group $G$ can be expressed
in terms of the Ward identities for the correlators in the general form
\begin{eqnarray}
\mathrm{i}p^{\mu }\langle \widetilde{V}_{\mu }^{a}(p)\widetilde{O}%
_{1}(p_{1})\ldots \widetilde{O}_{n}(p_{n})\rangle &=&-\sum_{i=1}^{n}\mathrm{i%
}\langle \widetilde{O}_{1}(p_{1})\ldots \delta _{T}^{a}\widetilde{O}%
_{i}(p_{i}+p)\ldots \widetilde{O}_{n}(p_{n})\rangle  \label{vectioWI} \\
\mathrm{i}p^{\mu }\langle \widetilde{A}_{\mu }^{a}(p)\widetilde{O}%
_{1}(p_{1})\ldots \widetilde{O}_{n}(p_{n})\rangle &=&-\sum_{i=1}^{n}\mathrm{i%
}\langle \widetilde{O}_{1}(p_{1})\ldots \delta _{X}^{a}\widetilde{O}%
_{i}(p_{i}+p)\ldots \widetilde{O}_{n}(p_{n})\rangle .  \label{axialWI}
\end{eqnarray}%
Here $V_{\mu }^{a}(x)$ and $A_{\mu }^{a}(x)$ are the Noether currents
corresponding to the generators $T^{a}$ and $X^{a}$ respectively (in analogy
with the chiral theories we will call them \emph{vector} and \emph{axial}
currents in what follows and to the Ward identities (\ref{vectioWI}) and (%
\ref{axialWI}) we will refer to the \emph{vector} and \emph{axial} WI) , $%
O_{i}(x)$ are (generally composite) local operators, $\delta
_{T}^{a}O_{i}(x) $ and $\delta _{X}^{a}O_{i}(x)$ are their infinitesimal
transforms with respect to the generators $T^{a}$ and $X^{a}$ . The tilde
means the Fourier transform%
\begin{equation}
\widetilde{O}_{i}(p)=\int \mathrm{d}^{4}xe^{\mathrm{i}p\cdot x}O_{i}(x).
\end{equation}

According to the Goldstone theorem the spectrum of the theory contains as
many Goldstone bosons $\pi ^{a}$ as the broken generators $X^{a}$ for which
the currents $A_{\mu }^{a}(x)$ play the role of the interpolating fields,
i.e.
\begin{equation}
\langle 0|A_{\mu }^{a}(0)|\pi ^{b}(p)\rangle =\mathrm{i}p_{\mu }F\delta
^{ab}.  \label{F}
\end{equation}%
where $F$ is the Goldstone boson decay constant. Let as denote $%
M^{a_{1}\ldots a_{n}}(p_{1},\ldots ,p_{n})$ the on-shell scattering
amplitude of the Goldstone bosons $\pi ^{a_{1}}(p_{1}),\ldots ,\pi
^{a_{n}}(p_{n})$. In what follows we will concentrate on the properties of $%
M^{a_{1}\ldots a_{n}}(p_{1},\ldots ,p_{n})$ dictated by the symmetry, i.e.
those which are encoded in the WI (\ref{vectioWI}) and (\ref{axialWI}).

\subsection{Vector WI and symmetry with respect to $H$}

The invariance with respect to the unbroken subgroup $H$ implies
\begin{equation}
\sum_{i=1}^{n}f_{X}^{aa_{i}b}M^{a_{1}\ldots a_{i-1}ba_{i+1}\ldots
a_{n}}(p_{1},\ldots ,p_{n})=0.  \label{H_invariance}
\end{equation}%
This can be understood as the consequence of the vector WI of the form
\begin{equation}
-\mathrm{i}p^{\mu }\langle \widetilde{V}_{\mu }^{a}(p)\widetilde{A}_{\mu
_{1}}^{a_{1}}(p_{1})\ldots \widetilde{A}_{\mu _{n}}^{a_{n}}(p_{n})\rangle
=-\sum_{i=1}^{n}\mathrm{i}\langle \widetilde{A}_{\mu
_{1}}^{a_{1}}(p_{1})\ldots \delta _{T}^{a}\widetilde{A}_{\mu
_{i}}^{a_{i}}(p+p_{i})\ldots \widetilde{A}_{\mu _{n}}^{a_{n}}(p_{n})\rangle
\label{WIvector1}
\end{equation}%
Note that the infinitesimal transformations $\delta ^{a}V_{\nu }^{b}$ and $%
\delta ^{a}A_{\nu }^{b}$ of these currents with respect to the generator $%
T^{a}$ of the unbroken subgroup $H$ are as follows
\begin{eqnarray}
\delta _{T}^{a}A_{\nu }^{b} &=&-f_{X}^{abc}A_{\nu }^{c} \\
\delta _{T}^{a}V_{\nu }^{b} &=&-f_{T}^{abc}V_{\nu }^{c}.
\end{eqnarray}%
Because there is no pole for $p\rightarrow 0$ in the correlator on the left
hand side of (\ref{WIvector1}), we get in this limit%
\begin{equation}
\sum_{i=1}^{n}f^{aa_{i}b}\langle \widetilde{A}_{\mu
_{1}}^{a_{1}}(p_{1})\ldots \widetilde{A}_{\mu _{i}}^{b}(p_{i})\ldots
\widetilde{A}_{\mu _{n}}^{a_{n}}(p_{n})\rangle =0.
\end{equation}%
Using the LSZ formula we get according to (\ref{F})
\begin{equation}
\langle \widetilde{A}_{\mu _{1}}^{a_{1}}(p_{1})\ldots \widetilde{A}_{\mu
_{n}}^{a_{n}}(p_{n})\rangle =\left( \prod_{i=1}^{n}\frac{\mathrm{i}}{%
p_{i}^{2}}Z_{\mu _{i}}\right) M^{a_{1}\ldots a_{n}}(p_{1},\ldots
,p_{n})+R_{\mu _{1}\ldots }^{a_{1}\ldots }  \label{LSZ}
\end{equation}%
where $Z_{\mu _{i}}=\mathrm{i}Fp_{i\mu _{i}}$ and the remnant $R_{\mu
_{1}\ldots }^{a_{1}\ldots }$ is regular on shell in the sense that
\begin{equation}
\lim_{p_{i}^{2}\rightarrow 0}\left( \prod\limits_{i=1}^{n}p_{i}^{2}\right)
R_{\mu _{1}\ldots }^{a_{1}\ldots }=0.
\end{equation}%
which implies (\ref{H_invariance}) for the on-shell amplitude $%
M^{a_{1}\ldots a_{n}}(p_{1},\ldots ,p_{n})$.

\subsection{Soft vector current singularity}

Let us assume now the following matrix element%
\begin{equation}
\langle \widetilde{V}_{\mu }^{a}(p)|\pi ^{a_{1}}(p_{1})\ldots \pi
^{a_{i}}(p_{i})\ldots \pi ^{a_{n}}(p_{n})\rangle .  \label{Vmatrix}
\end{equation}%
In what follows we will discuss the behavior of this object in the limit $%
p\rightarrow 0$. \ On the level of the Feynman graphs, the only
singularities in the soft limit $p\rightarrow 0$ are those which stem from
the one-Goldstone-boson-reducible graphs for which the vector current $%
\widetilde{V}_{\mu }^{a}(p)$ is attached to the external Goldstone boson
line. The potential singularities are therefore of the form (see Fig. \ref%
{softV} )%
\begin{figure}[t]
\begin{center}
\epsfig{width=0.45\textwidth,figure=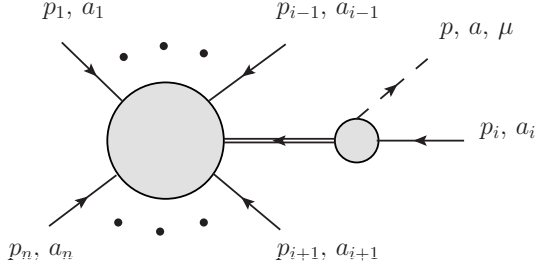}
\end{center}
\caption{Graphical representation of the singular contributions to the
matrix element (\protect\ref{Vmatrix}). }
\label{softV}
\end{figure}
\begin{equation}
\langle \widetilde{V}_{\mu }^{a}(p)\phi ^{a_{j}}(0)|\pi
^{a_{i}}(p_{1})\rangle _{1PI}\mathrm{i}\Delta ^{a_{j}a_{k}}(\left(
p-p_{i}\right) ^{2})\langle \phi ^{a_{k}}(0)|\pi ^{a_{1}}(p_{1})\ldots
\widehat{\pi ^{a_{i}}(p_{i})}\ldots \pi ^{a_{n}}(p_{n})\rangle _{1PI}
\end{equation}%
where the subscript $1PI$ means one-Goldstone-boson-irreducible block, the
hat means omitting of the corresponding particle, $\phi ^{a}(x)$ is the
Goldstone boson interpolating field normalized as%
\begin{equation}
\langle 0|\phi ^{a}(0)|\pi ^{b}(p)\rangle =\delta ^{ab}
\end{equation}%
and $\Delta ^{a_{j}a_{k}}(q^{2})$ is a Goldstone boson propagator. For $%
q^{2}\rightarrow 0$ we have%
\begin{equation}
\Delta ^{a_{j}a_{k}}(q^{2})=\frac{\delta ^{a_{j}a_{k}}}{q^{2}}\left(
1+O(q^{2})\right) .
\end{equation}%
As a consequence of the Lorentz invariance, invariance with respect to $H$
and LSZ formulae we have
\begin{equation}
\langle \widetilde{V}_{\mu }^{a}(p)\phi ^{a_{j}}(0)|\pi
^{a_{i}}(p_{i})\rangle _{1PI}=\mathrm{i}%
f_{X}^{aa_{i}a_{j}}F_{V}(p^{2})(2p_{i}-p)_{\mu }+O(\left( p-p_{i}\right)
^{2})
\end{equation}%
where $F_{V}(p^{2})$ is the on-shell vector form-factor defined as\footnote{%
The form of the right hand side is dictated by $H$-invariance, Bose and
crossing symmetry.}
\begin{equation}
\langle \pi ^{a_{j}}(p-p_{i})|\widetilde{V}_{\mu }^{a}(p)|\pi
^{a_{i}}(p_{i})\rangle =\mathrm{i}%
f_{X}^{aa_{i}a_{j}}F_{V}(p^{2})(2p_{i}-p)_{\mu }.
\end{equation}%
We can fix the normalization of the vector currents $V_{\mu }^{a}$ in such a
way that%
\begin{equation}
F_{V}(p^{2})=1+O(p^{2}).
\end{equation}%
Analogously we have
\begin{equation}
\langle \phi ^{a_{k}}(0)|\pi ^{a_{1}}(p_{1})\ldots \widehat{\pi
^{a_{i}}(p_{i})}\ldots \pi ^{a_{n}}(p_{n})\rangle _{1PI}=M^{a_{1}\ldots
a_{i-1}a_{k}a_{i+1}\ldots a_{n}}(p_{1},\ldots ,p_{n})+O(\left(
p-p_{i}\right) ^{2}).
\end{equation}%
Using $\left( p-p_{i}\right) ^{2}=-2(p\cdot p_{i})+p^{2}$ and putting all
the ingredients together we get for $p\rightarrow 0$
\begin{eqnarray}
\langle \widetilde{V}_{\mu }^{a}(p)|\pi ^{a_{1}}(p_{1})\ldots \pi
^{a_{i}}(p_{i})\ldots \pi ^{a_{n}}(p_{n})\rangle
&=&\sum_{i=1}^{n}f_{X}^{aa_{i}d}\frac{(2p_{i}-p)_{\mu }}{2(p\cdot p_{i})}%
M^{a_{1}\ldots a_{i-1}da_{i+1}\ldots a_{n}}(p_{1},\ldots ,p_{n})+O(1)  \notag
\label{soft_vector} \\
&&
\end{eqnarray}

\subsection{Axial WI and Adler zero}

To illustrate the method which we will use in the next subsection, let us
briefly recapitulate the textbook example of the derivation of the Adler
zero for the amplitude $M^{a_{1}\ldots a_{n}}(p_{1},\ldots ,p_{n})$ (see
e.g. \cite{Weinberg:1996kr}). Let us start with the axial WI in the form
\begin{equation}
-\mathrm{i}p^{\mu }\langle \widetilde{A}_{\mu }^{a}(p)\widetilde{A}_{\mu
_{1}}^{a_{1}}(p_{1})\ldots \widetilde{A}_{\mu _{n}}^{a_{n}}(p_{n})\rangle
=-\sum_{i=1}^{n}\mathrm{i}\langle \widetilde{A}_{\mu
_{1}}^{a_{1}}(p_{1})\ldots \delta _{X}^{a}\widetilde{A}_{\mu
_{i}}^{a_{i}}(p+p_{i})\ldots \widetilde{A}_{\mu _{n}}^{a_{n}}(p_{n})\rangle
\label{WIaxial}
\end{equation}%
where now%
\begin{eqnarray}
\delta _{X}^{a}A_{\nu }^{b} &=&-F^{abc}V_{\nu }^{c}  \notag \\
\delta _{X}^{a}V_{\nu }^{b} &=&-f_{X}^{abc}A_{\nu }^{c}.
\label{axial_variation_VA}
\end{eqnarray}%
Applying on both sides of (\ref{WIaxial}) the LSZ reduction to all but one
axial currents, we get the conservation of the axial current in terms of the
transversality of the matrix element of $A_{\mu }^{a}$ between the initial
and final states $|i\rangle $ and $\langle f|$
\begin{equation}
-\mathrm{i}p^{\mu }\langle f|\widetilde{A}_{\mu }^{a}(p)|i\rangle =0.
\label{Aconservation}
\end{equation}%
On the other hand from (\ref{LSZ}) we get the Goldstone boson pole dominance
for $p^{2}\rightarrow 0$
\begin{equation}
-\mathrm{i}p^{\mu }\langle f|\widetilde{A}_{\mu }^{a}(p)|i\rangle =\frac{1}{%
p^{2}}p^{\mu }Z_{\mu }\langle f+\pi ^{a}(p)|i\rangle -\mathrm{i}p^{\mu
}R_{\mu ,fi}^{a}  \label{GBpoledominance}
\end{equation}%
where $Z_{\mu }=\mathrm{i}Fp_{\mu }$ and the remnant $R_{\mu ,fi}^{a}$ is
regular in this limit
\begin{equation}
\lim_{p^{2}\rightarrow 0}p^{2}R_{\mu ,fi}^{a}=0.  \label{Rregularity}
\end{equation}%
Putting (\ref{Aconservation}) and (\ref{GBpoledominance}) together we get
for the amplitude with emition of the Goldstone boson $\pi ^{a}(p)$ in the
final state%
\begin{equation}
\langle f+\pi ^{a}(p)|i\rangle =\frac{1}{F}p^{\mu }R_{\mu ,fi}^{a}.
\label{softGBreduction}
\end{equation}%
Provided the following stronger regularity condition holds%
\begin{equation}
\lim_{p\rightarrow 0}p^{\mu }R_{\mu ,fi}^{a}=0,
\label{pR}
\end{equation}%
we get
\begin{equation}
\langle f+\pi ^{a}(0)|i\rangle =0,
\end{equation}%
i.e. the Adler zero for $p\rightarrow 0$.

An useful off-shell generalization of the formula (\ref{GBpoledominance})
reads
\begin{equation}
-\mathrm{i}p^{\mu }\langle \widetilde{A}_{\mu }^{a}(p)\widetilde{A}_{\mu
_{1}}^{a_{1}}(p_{1})\ldots \widetilde{A}_{\mu _{n}}^{a_{n}}(p_{n})\rangle =%
\mathrm{i}F\langle \pi ^{a}(p)|\widetilde{A}_{\mu _{1}}^{a_{1}}(p_{1})\ldots
\widetilde{A}_{\mu _{n}}^{a_{n}}(p_{n})\rangle -\mathrm{i}p^{\mu }R_{\mu
,\mu _{1}\ldots }^{a,a_{1}\ldots }
\end{equation}%
where%
\begin{equation}
\lim_{p^{2}\rightarrow 0}p^{2}R_{\mu ,\mu _{1}\ldots }^{a,a_{1}\ldots }=0.
\end{equation}%
and using the Ward identity (\ref{WIaxial}) and (\ref{axial_variation_VA})
we get%
\begin{eqnarray}
&&F\langle \pi ^{a}(p)|\widetilde{A}_{\mu _{1}}^{a_{1}}(p_{1})\ldots
\widetilde{A}_{\mu _{n}}^{a_{n}}(p_{n})\rangle \notag\\
&=&p^{\mu }R_{\mu ,\mu _{1}\ldots }^{a,a_{1}\ldots
}+\sum_{i=1}^{n}F^{aa_{i}c}\langle \widetilde{A}_{\mu
_{1}}^{a_{1}}(p_{1})\ldots \widetilde{V}_{\mu _{i}}^{c}(p+p_{i})\ldots
\widetilde{A}_{\mu _{n}}^{a_{n}}(p_{n})\rangle .
\end{eqnarray}

\subsection{Double soft limit}

Our starting point is the axial WI (\ref{WIaxial}) rewritten in the form%
\begin{eqnarray}
-\mathrm{i}p^{\mu }\langle \widetilde{A}_{\mu }^{a}(p)\widetilde{A}_{\nu
}^{b}(q)\widetilde{A}_{\mu _{1}}^{a_{1}}(p_{1})\ldots \widetilde{A}_{\mu
_{n}}^{a_{n}}(p_{n})\rangle &=&-\mathrm{i}\langle \delta _{X}^{a}\widetilde{A%
}_{\nu }^{b}(p+q)\widetilde{A}_{\mu _{1}}^{a_{1}}(p_{1})\ldots \widetilde{A}%
_{\mu _{n}}^{a_{n}}(p_{n})\rangle \\
&&-\sum_{i=1}^{n}\mathrm{i}\langle \widetilde{A}_{\nu }^{b}(q)\widetilde{A}%
_{\mu _{1}}^{a_{1}}(p_{1})\ldots \delta _{X}^{a}\widetilde{A}_{\mu
_{i}}^{a_{i}}(p+p_{i})\ldots \widetilde{A}_{\mu _{n}}^{a_{n}}(p_{n})\rangle
\notag
\end{eqnarray}%
Multiplying then both sides by $-\mathrm{i}q^{\nu }$ and using the axial WI (%
\ref{WIaxial}) once again we get
\begin{eqnarray}
&&-p^{\mu }q^{\nu }\langle \widetilde{A}_{\mu }^{a}(p)\widetilde{A}_{\nu
}^{b}(q)\widetilde{A}_{\mu _{1}}^{a_{1}}(p_{1})\ldots \widetilde{A}_{\mu
_{n}}^{a_{n}}(p_{n})\rangle  \notag \\
&=&q^{\nu }F^{abc}\langle \widetilde{V}_{\nu }^{c}(p+q)\widetilde{A}_{\mu
_{1}}^{a_{1}}(p_{1})\ldots \widetilde{A}_{\mu _{n}}^{a_{n}}(p_{n})\rangle
\notag \\
&+&\sum_{i\neq j;i,j=1}^{n}F^{aa_{j}c}F^{ba_{i}d}\langle \widetilde{A}_{\mu
_{1}}^{a_{1}}(p_{1})\ldots \widetilde{V}_{\mu _{i}}^{d}(p_{i}+q)\ldots
\widetilde{V}_{\mu _{j}}^{c}(p+p_{j})\ldots \widetilde{A}_{\mu
_{n}}^{a_{n}}(p_{n})\rangle  \notag \\
&&+\sum_{i=1}^{n}F^{aa_{i}c}f_{X}^{bcd}\langle \widetilde{A}_{\mu
_{1}}^{a_{1}}(p_{1})\ldots \widetilde{A}_{\mu _{i}}^{d}(p+q+p_{i})\ldots
\widetilde{A}_{\mu _{n}}^{a_{n}}(p_{n})\rangle .  \label{WI1}
\end{eqnarray}%
The left hand side of (\ref{WI1}) is symmetric with respect to the
interchange of $(p,a)\leftrightarrow (q,b)$; its right hand side can be
therefore rewritten in the manifestly symmetric form%
\begin{eqnarray}
&&-p^{\mu }q^{\nu }\langle \widetilde{A}_{\mu }^{a}(p)\widetilde{A}_{\nu
}^{b}(q)\widetilde{A}_{\mu _{1}}^{a_{1}}(p_{1})\ldots \widetilde{A}_{\mu
_{n}}^{a_{n}}(p_{n})\rangle  \notag \\
&=&-\frac{1}{2}(p-q)^{\nu }F^{abc}\langle \widetilde{V}_{\nu }^{c}(p+q)%
\widetilde{A}_{\mu _{1}}^{a_{1}}(p_{1})\ldots \widetilde{A}_{\mu
_{n}}^{a_{n}}(p_{n})\rangle  \notag \\
&+&\sum_{i\neq j;i,j=1}^{n}F^{aa_{j}c}F^{ba_{i}d}\langle \widetilde{A}_{\mu
_{1}}^{a_{1}}(p_{1})\ldots \widetilde{V}_{\mu _{i}}^{d}(p_{i}+q)\ldots
\widetilde{V}_{\mu _{j}}^{c}(p+p_{j})\ldots \widetilde{A}_{\mu
_{n}}^{a_{n}}(p_{n})\rangle  \notag \\
&&+\frac{1}{2}\sum_{i=1}^{n}\left(
F^{aa_{i}c}f_{X}^{bcd}+F^{ba_{i}c}f_{X}^{acd}\right) \langle \widetilde{A}%
_{\mu _{1}}^{a_{1}}(p_{1})\ldots \widetilde{A}_{\mu
_{i}}^{d}(p+q+p_{i})\ldots \widetilde{A}_{\mu _{n}}^{a_{n}}(p_{n})\rangle .
\label{doubleWI}
\end{eqnarray}%
On the other hand, the LSZ formula gives for $p^{2},q^{2}\rightarrow 0$
\begin{eqnarray}
\langle \widetilde{A}_{\mu }^{a}(p)\widetilde{A}_{\nu }^{b}(q)\widetilde{A}%
_{\mu _{1}}^{a_{1}}(p_{1})\ldots \widetilde{A}_{\mu
_{n}}^{a_{n}}(p_{n})\rangle &=&\sum_{c,d}\frac{\mathrm{i}}{p^{2}}\langle
0|A_{\mu }^{a}|\pi ^{c}(p)\rangle \frac{\mathrm{i}}{q^{2}}\langle 0|A_{\nu
}^{b}|\pi ^{d}(q)\rangle \notag\\
&&\times \langle \pi ^{c}(p)\pi ^{d}(q)|\widetilde{A}_{\mu
_{1}}^{a_{1}}(p_{1})\ldots \widetilde{A}_{\mu _{n}}^{a_{n}}(p_{n})\rangle
+R_{\mu \nu }^{ab,\ldots }
\end{eqnarray}%
where the regular remnant satisfies
\begin{equation}
\lim_{p^{2},q^{2}\rightarrow 0}p^{2}q^{2}R_{\mu \nu }^{ab,\ldots }=0.
\label{p2q2R}
\end{equation}%
Therefore, using (\ref{F}) we get%
\begin{eqnarray}
&&-p^{\mu }q^{\nu }\langle \widetilde{A}_{\mu }^{a}(p)\widetilde{A}_{\nu
}^{b}(q)\widetilde{A}_{\mu _{1}}^{a_{1}}(p_{1})\ldots \widetilde{A}_{\mu
_{n}}^{a_{n}}(p_{n})\rangle  \notag \\
&=&F^{2}\langle \pi ^{a}(p)\pi ^{b}(q)|\widetilde{A}_{\mu
_{1}}^{a_{1}}(p_{1})\ldots \widetilde{A}_{\mu _{n}}^{a_{n}}(p_{n})\rangle
-p^{\mu }q^{\nu }R_{\mu \nu }^{ab,\ldots }  \label{amp}
\end{eqnarray}%
On the other hand applying the LSZ reduction to (\ref{WI1}, \ref{doubleWI})
(let us note that only the first terms on the right hand side has the
appropriate poles at $p^{2},q^{2}\rightarrow 0$) we get
\begin{eqnarray}
&&p^{\mu }q^{\nu }\langle \widetilde{A}_{\mu }^{a}(p)\widetilde{A}_{\nu
}^{b}(q)|\pi ^{a_{1}}(p_{1})\ldots \pi ^{d}(p_{i})\ldots \pi
^{a_{n}}(p_{n})\rangle \notag\\
&=&q^{\nu }F^{abc}\langle \widetilde{V}_{\nu }^{c}(p+q)|\pi
^{a_{1}}(p_{1})\ldots \pi ^{d}(p_{i})\ldots \pi ^{a_{n}}(p_{n})\rangle \notag\\
&=&p^{\mu }F^{bac}\langle \widetilde{V}_{\mu }^{c}(p+q)|\pi
^{a_{1}}(p_{1})\ldots \pi ^{d}(p_{i})\ldots \pi ^{a_{n}}(p_{n})\rangle \notag\\
&=&-\frac{1}{2}F^{abc}(p-q)^{\mu }\langle \widetilde{V}_{\mu }^{c}(p+q)|\pi
^{a_{1}}(p_{1})\ldots \pi ^{d}(p_{i})\ldots \pi ^{a_{n}}(p_{n})\rangle
\end{eqnarray}%
and as a consequence of LSZ reduction of (\ref{amp})
\begin{eqnarray}
&&F^{2}\langle \pi ^{a}(p)\pi ^{b}(q)|\pi ^{a_{1}}(p_{1})\ldots \pi
^{a_{i}}(p_{i})\ldots \pi ^{a_{n}}(p_{n})\rangle  \notag \\
&=&-\frac{1}{2}F^{abc}(p-q)^{\mu }\langle \widetilde{V}_{\mu }^{c}(p+q)|\pi
^{a_{1}}(p_{1})\ldots \pi ^{d}(p_{i})\ldots \pi ^{a_{n}}(p_{n})\rangle
+p^{\mu }q^{\nu }R_{\mu \nu }^{ab,\ldots }|_{LSZ}.  \label{rell}
\end{eqnarray}%
According to (\ref{soft_vector}) we have for $p,q\rightarrow 0$%
\begin{eqnarray}
&&-\frac{1}{2}F^{abc}(p-q)^{\mu }\langle \widetilde{V}_{\mu }^{c}(p+q)|\pi
^{a_{1}}(p_{1})\ldots \pi ^{d}(p_{i})\ldots \pi ^{a_{n}}(p_{n})\rangle \notag\\
&=&-\frac{1}{2}\sum_{i=1}^{n}F^{abc}f_{X}^{ca_{i}d}\frac{(2p_{i}-p-q)\cdot
(p-q)}{2((p+q)\cdot p_{i})}\langle \pi ^{a_{1}}(p_{1})\ldots \pi
^{d}(p_{i})\ldots \pi ^{a_{n}}(p_{n})\rangle +O(p-q) \notag\\
&=&-\frac{1}{2}\sum_{i=1}^{n}F^{abc}f_{X}^{ca_{i}d}\frac{p_{i}\cdot (p-q)}{%
p_{i}\cdot (p+q)}\langle \pi ^{a_{1}}(p_{1})\ldots \pi ^{d}(p_{i})\ldots \pi
^{a_{n}}(p_{n})\rangle \notag\\
&&+O\left( p-q,\frac{p^{2}-q^{2}}{p_{i}\cdot (p+q)}\right)
\end{eqnarray}

For $p^{2}=q^{2}=0$ we finally get%
\begin{eqnarray}
&&F_{0}^{2}\langle \pi ^{a}(p)\pi ^{b}(q)|\pi ^{a_{1}}(p_{1})\ldots \pi
^{a_{i}}(p_{i})\ldots \pi ^{a_{n}}(p_{n})\rangle \notag\\
&=&-\frac{1}{2}\sum_{i=1}^{n}F^{abc}f_{X}^{ca_{i}d}\frac{p_{i}\cdot (p-q)}{%
p_{i}\cdot (p+q)}\langle \pi ^{a_{1}}(p_{1})\ldots \pi ^{d}(p_{i})\ldots \pi
^{a_{n}}(p_{n})\rangle \notag\\
&&+p^{\mu }q^{\nu }R_{\mu \nu }^{ab,\ldots }|_{LSZ}+O\left( p-q\right).
\end{eqnarray}
Provided condition stronger than (\ref{p2q2R}) holds, namely $%
\lim_{p,q\rightarrow 0}p^{\mu }q^{\nu }R_{\mu \nu }^{ab,\ldots }|_{LSZ}=0$
(cf. (\ref{pR})), we get as a result
\begin{eqnarray}
&&\lim_{t\rightarrow 0}F_{0}^{2}\langle \pi ^{a}(tp)\pi ^{b}(tq)|\pi
^{a_{1}}(p_{1})\ldots \pi ^{a_{i}}(p_{i})\ldots \pi ^{a_{n}}(p_{n})\rangle \notag\\
&=&-\frac{1}{2}\sum_{i=1}^{n}F^{abc}f_{X}^{ca_{i}d}\frac{p_{i}\cdot (p-q)}{%
p_{i}\cdot (p+q)}\langle \pi ^{a_{1}}(p_{1})\ldots \pi ^{d}(p_{i})\ldots \pi
^{a_{n}}(p_{n})\rangle .
\end{eqnarray}
For the chiral nonlinear sigma model corresponding to the symmetry breaking $G\times G\rightarrow G$, we have $F^{abc}=f_X^{abc}=f_T^{abc}$ and we get the formula (\ref{AH_double_soft}) as a special case.

\end{document}